\documentclass[%
 reprint,
superscriptaddress,
nofootinbib,
 amsmath,amssymb,
 aps,
]{revtex4-1}

\usepackage{lipsum}
\usepackage{amsmath}
\usepackage{braket}
\usepackage{graphicx}
\usepackage{dcolumn}
\usepackage{bm}
\usepackage{xcolor}
\definecolor{darkspringgreen}{rgb}{0.09, 0.45, 0.27}
\newcommand{\mcite}[1]{\mbox{\cite{#1}}}

\usepackage{hyperref}
\hypersetup{
     colorlinks=true,
     linkcolor=darkspringgreen,
     filecolor=darkspringgreen,
     citecolor =darkspringgreen,      
     urlcolor=black,
     }

\def	\beq	{\begin{equation}}
\def	\eeq	{\end{equation}}

\begin{document}

\title{A JT supergravity as a double-cut matrix model}

\author{Clifford V. Johnson}
\email{johnson1@usc.edu}
\affiliation{%
Department of Physics and Astronomy, University of Southern California,\\ 
Los Angeles, California 90089-0484, USA
}%

\author{Felipe Rosso}
\email{felipero@usc.edu}
\affiliation{%
Department of Physics and Astronomy, University of Southern California,\\ 
Los Angeles, California 90089-0484, USA
}%

\author{Andrew Svesko}
\email{a.svesko@ucl.ac.uk}
\affiliation{%
Department of Physics and Astronomy,
University College London,\\
Gower Street, London,
WC1E 6BT, UK
}%

\begin{abstract}
We study a Jackiw-Teitelboim (JT) supergravity theory, defined as an Euclidean path integral  over orientable supermanifolds with constant negative curvature, that was argued by  Stanford and Witten to be captured by a random matrix model in the $\boldsymbol{\beta}{=}2$ Dyson-Wigner class. We show that   the theory  is a double-cut  matrix model tuned to a critical point where the two cuts coalesce. Our  formulation  is fully non-perturbative and manifestly stable, providing for explicit unambiguous computation of observables beyond the perturbative recursion relations derivable from loop equations.  Our construction  shows that this JT supergravity theory may be regarded as a particular combination of  certain type 0B minimal string theories, and is hence a natural counterpart to another family of JT supergravity theories recently shown to be built from type~0A minimal strings. We conjecture that certain other JT supergravities can be similarly defined in terms of  double-cut matrix models.
\end{abstract}

\maketitle

\section{Introduction}
\label{sec:1}

Jackiw-Teitelboim (JT) gravity~\cite{Jackiw:1984je,Teitelboim:1983ux}, a theory of 2D  dilaton gravity, has emerged as one of the simplest models for studying non-trivial  problems in quantum gravity.  Describing the low-energy dynamics of a wide class of  near-extremal black holes and branes~\cite{Achucarro:1993fd,Fabbri:2000xh,Nayak:2018qej,Kolekar:2018sba,Ghosh:2019rcj}, JT gravity also features in a precise realization of holographic AdS$_{2}$/CFT$_{1}$, arising as the low-energy gravitational dual to the Sachdev-Ye-Kitaev (SYK) 1D quantum mechanical system~{\cite{Sachdev:1992fk,Kitaev:talks,Jensen:2016pah,Maldacena:2016hyu,Maldacena:2016upp,Engelsoy:2016xyb}}. It has also been a crucial ingredient in recent work that has  yielded  new insights into how  the black hole information puzzle may be  resolved~\cite{Penington:2019npb,Almheiri:2019psf,Almheiri:2019qdq,Penington:2019kki}.

The partition function of the theory $Z(\beta)$  has two natural parameters: The length, $\beta$, of a boundary, which defines the temperature {\it via} $\beta{=}1/T$ and the quantity $S_0$ (which can be thought of as the extremal entropy of the parent black hole), which defines a coupling $\hbar{=}e^{-S_0}$. Surfaces of Euler character $\chi{=}2(1{-}g){-}n$ (there are $g$  handles and $n$ boundaries) contribute with a factor $\hbar^{-\chi}$ to the path integral.  It is vitally important to be able to fully formulate JT gravity as a complete theory of quantum gravity in its own right, not just perturbatively in the parameters, but beyond. 
A landmark discovery in this regard was the remarkable  demonstration by Saad, Shenker and Stanford~\cite{Saad:2019lba} that JT gravity can be formulated as  certain  random Hermitian matrix models, in a double-scaling~\cite{Brezin:1990rb,Douglas:1990ve,Gross:1990vs,Gross:1990aw} limit:
\begin{equation}\label{eq:gravMMequiv}
Z(\beta_1,\dots,\beta_n)
\quad \longleftrightarrow \quad
\big\langle 
{\rm Tr}\,e^{-\beta_1H}
\dots
{\rm Tr}\,e^{-\beta_nH}
\big\rangle_c\ .
\end{equation}

On the left hand side $Z(\beta_{1},...,\beta_{n})$ is the JT gravity path integral, computed as an Euclidean path integral over (connected) Riemann surfaces with constant negative curvature and $n$ asymptotic boundaries of 
lengths~$\beta_{i}$. On the right  hand side is the connected correlation function  of insertions of the operator $\text{Tr}\,e^{-\beta_{i}H}$, with $\langle\, \dots\rangle_{\text{c}}$ implying an ensemble average of the Hermitian matrix~$H$.  The  correspondence was established in ref.~\cite{Saad:2019lba} {\it via} powerful recursion relations, matching the volumes of moduli spaces of Riemann surfaces~\cite{Mirzakhani:2006fta} contributing to $Z(\beta_{1},\dots,\beta_{n})$ to a genus expansion of a Hermitian matrix integral~\cite{Eynard:2004mh,Eynard:2007kz}. While equation~(\ref{eq:gravMMequiv}) is understood to be order by order in genus $g$ and then summed, a matrix model definition can {\it in principle} supply non-perturbative completions that go beyond the sum over genus. Non-perturbative physics is extremely important, especially for studying the low energy regime, and so it is of great interest to fully understand and characterize matrix model formulations, as a means of  extracting it.\footnote{While ref.~\cite{Saad:2019lba}'s Hermitian matrix model definition has a problematic non-perturbative definition, a complete non-perturbative completion of ordinary JT gravity that is  naturally rooted in matrix models can be found, as shown in ref.~\cite{Johnson:2019eik}.}

The correspondence was broadened rather elegantly by Stanford and Witten~\cite{Stanford:2019vob} to include wider classes of JT gravity, incorporating models with fermions and/or time-reversal symmetry. Such models were shown to be classified according to one of the ten types of random matrix ensemble: The  three Dyson-Wigner $\boldsymbol{\beta}$ ensembles and the seven Altland-Zirnbauer $(\boldsymbol{\alpha},\boldsymbol{\beta})$ ensembles~\cite{Dyson:1962es,Altland:1997zz}.   
Many detailed gravity computations (now involving the moduli space of orientable and/or unorientable (super) Riemann surfaces) were shown to   be captured by detailed matrix model integrals, establishing  that   expression~(\ref{eq:gravMMequiv}) still holds (for an appropriate choice of $H$).  In the case of ${\cal N}{=}1$ JT supergravity, which will be the focus of this paper,  the Hermitian matrix is written $H{=}Q^2$, where  it is the supercharge   $Q$ that is (on general grounds) classified in the ten-fold way, depending upon the number,~${\rm N}$, of fermions the effective SYK-like ``boundary'' dual has, and the action of $(-1)^{\textsf F}$ and~${\textsf T}$, where ${\textsf F}$  and~${\textsf T}$ are fermion number and the time reversal operator, respectively \cite{Fu:2016vas,Stanford:2017thb,Li:2017hdt,Kanazawa:2017dpd,Sun:2019yqp,Forste:2017kwy,Murugan:2017eto,Sun:2019yqp,Stanford:2019vob}. 

Symmetry under ${\textsf T}$ results in the inclusion of unorientable surfaces in the path integral, but this paper will feature JT supergravity theories models that do not preserve ${\textsf T}$. So all that is left is to consider whether $(-1)^{\textsf F}$ is a symmetry or not---{\it i.e.,} whether  ${\rm N}$ is even (let's call this Case A) or odd (Case B). 
In Case A, since the supercharge  anti-commutes with $(-1)^{\textsf F}$, it can be written (in a basis where $(-1)^{\textsf F}$ is block diagonal) as: 
\begin{equation}
Q=\begin{pmatrix} 
0 & M\cr M^\dagger & 0
 \end{pmatrix}\ ,
\end{equation}
where $M$ is a complex matrix, and $M^\dagger$ its Hermitian conjugate (see Section 2.6.2 in ref. \cite{Stanford:2019vob}), and the natural combination the matrix model~(\ref{eq:gravMMequiv}) cares about is $M^\dagger M$.  In Case B, there is no grading due to $(-1)^{\textsf F}$, and so $Q$ is itself just an Hermitian matrix. This case should therefore fall into the $\boldsymbol{\beta}{=}2$ Dyson-Wigner class.

Recently, it was  shown in refs.~\cite{Johnson:2020heh,Johnson:2020exp,Johnson:2020mwi} how to use double--scaled random complex matrix models~\cite{Morris:1991cq,Dalley:1992qg,Dalley:1992br,Dalley:1992vr,Dalley:1992yi}  with a potential of the form $V(M^\dagger M)$  to yield  a complete perturbative and non-perturbative definition of the  Case~A supergravity above. The formulation was shown to be equivalent to combining together (in a precise sense reviewed later) an infinite family of models known~\cite{Klebanov:2003wg} to be equivalent to type 0A minimal string theories.\footnote{In fact, the same framework includes some ${\textsf T}$-invariant models too. They all fall into the $(\boldsymbol{\alpha},\boldsymbol{\beta})=(2\Gamma{+}1,2)$  Altland-Zirnbauer series, for $\Gamma=0,+\frac12,-\frac12$. Here,  integer $\Gamma$  breaks ${\textsf T}$.} For other studies of JT gravity and supergravity using minimal models, see refs. \cite{Mertens:2020hbs,Mertens:2019tcm,Turiaci:2020fjj,Mertens:2020pfe,Betzios:2020nry}.

In retrospect, the involvement of type 0A minimal strings is entirely natural. The worldsheet physics of minimal string theories are themselves theories of two dimensional quantum gravity. When fermions are present, there is  a spin structure on the surfaces that enter the path integral, and how these are treated defines different types of string theories~\cite{Polchinski:1998rr}. Type 0A string theories sum over such spin structures and keep track of them by including a weight factor of $(-1)^\zeta$ in the sum, where $\zeta$  (0 or 1 mod 2) is the parity of the spin structure. On the other hand, in the JT classifications above, the boundary $(-1)^{\textsf F}$ directly correlates with the bulk $(-1)^\zeta$. Hence with hindsight it is natural that, if minimal strings are going to be relevant at all (and that is thrust upon us by the presence of double-scaled matrix models in both settings) it is indeed type 0A minimal strings that relate to Case A above. Indeed, many special properties of the non-perturbative ``string equations'' that describe type 0A strings, when deployed appropriately (as demonstrated in ref.~\cite{Johnson:2020heh}) yield many of the remarkable properties  of the supergravity models noticed  in ref.~\cite{Stanford:2019vob}.

This paper answers the natural question as to whether the remaining ${\textsf T}$--breaking JT supergravity,  Case B above, can be given an analogous treatment, yielding  a fully computable stable non-perturbative definition of the theory's observables. In light  of the previous paragraph there is a natural guess: Since it does not respect $(-1)^{\textsf F}$ and hence  results in a sum over bulk geometries that simply {\it ignores} $(-1)^\zeta$, it ought to have a connection to type~0B minimal strings, which themselves ignore that weight factor by definition. The answer will turn out to be yes, and the 0B connection (anticipated in ref.~\cite{Stanford:2019vob}) will prove to be correct. 

Two more clues help lead to these answers, and they can be found in the leading form of the spectral density for $Q$, denoted here $\rho_0(q)$. Since $H{=}Q^2$, this is  determined  by the Laplace transform of the super--Schwarzian partition function, which yields~\cite{Stanford:2017thb} the density ${\rho}^{\rm SJT}_0(E)$  of $H$ (with energies $E{\geq}0$). This gives:
\begin{equation}
\label{eq:classical_densities}
{\rho}^{\rm SJT}_0(E)=\frac{\sqrt{2}\cosh(2\pi\sqrt{E})}{\pi\hbar\sqrt{E}} \quad\! \Longrightarrow\! \quad \rho_0(q)=\frac{\cosh(2\pi q)}{\pi\hbar}\ ,
\end{equation}
since $E{=}q^2$, and after taking into account the Jacobian $dE=2qdq$ in the Laplace transform integral, where the $2$ is absorbed by integrating $q$ over $\mathbb{R}$. (A translation factor of $\gamma^{-1}{=}1/\sqrt{2}$
has also been used to go from the gravity
density to the matrix model density, as in ref. \cite{Stanford:2019vob}.)

\begin{figure}[ht]
\centering
\includegraphics[scale=0.26]{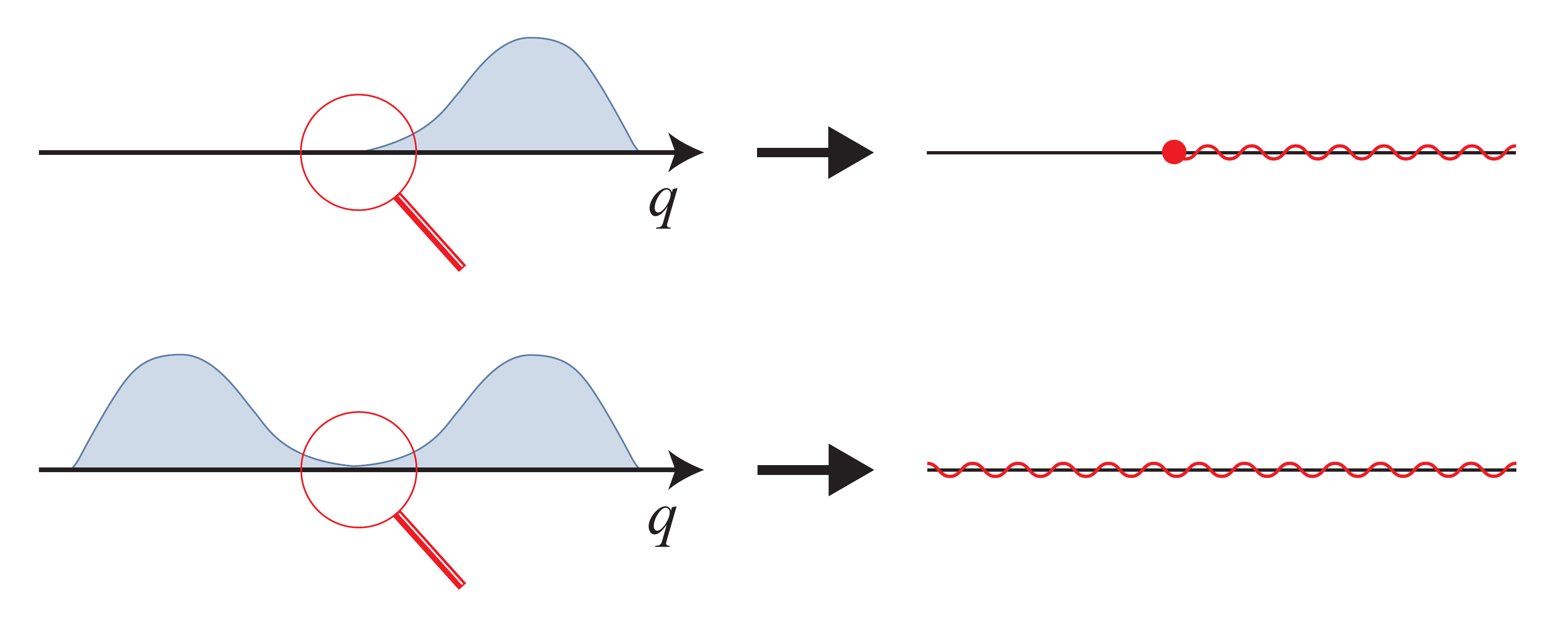}
\caption{The double-scaling limit as a ``zoom in'' to the neighbourhood of the endpoints of spectral densities, resulting in either a semi-infinite cut starting from a single cut case, or resulting in an infinite cut starting from a  double-cut case that has a merger.}
\label{fig:cuts}
\end{figure}

As  noticed in ref.~\cite{Stanford:2019vob}, the Hermitian matrix  model for~$Q$  therefore has (after double--scaling to match gravity) a spectral density that naturally spreads over the entire  real line. This is our first clue: The most commonly studied cases of double-scaled Hermitian matrix models  usually have support on the half--line, resulting from the fact that the double-scaling limit ``zooms in'' to capture the universal physics to be found at one critical endpoint of the (unscaled) spectral density, as shown in the upper part of Figure~\ref{fig:cuts} (see {\it e.g.} ref.~\cite{Ginsparg:1993is} for a review).  

\begin{figure}[ht]
\centering
\includegraphics[scale=0.70]{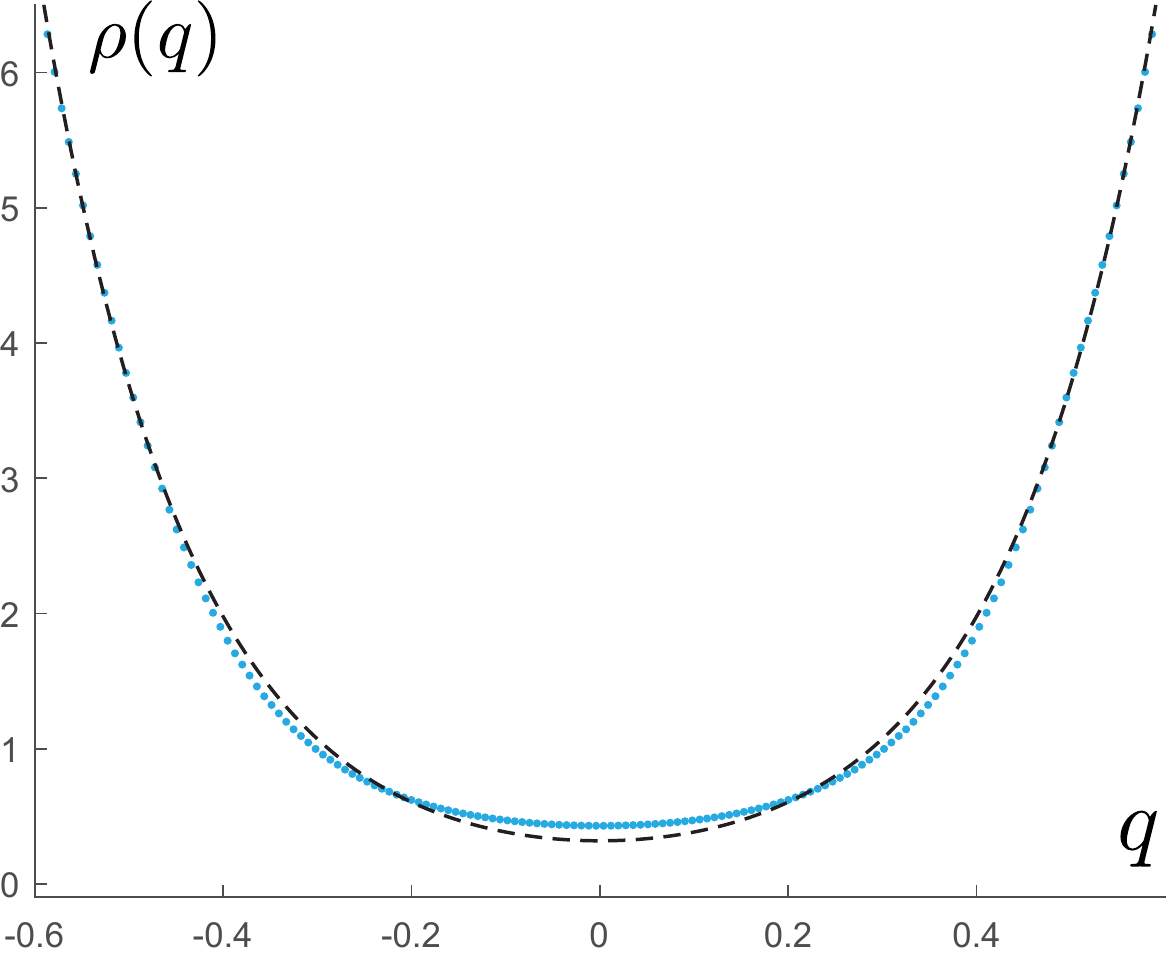}
\caption{Full spectral density for the double-cut model relevant for JT supergravity, obtained from summing the two $s=\pm$ contributions shown in Figure \ref{fig:8.7}. The dashed curve corresponds to the answer to all orders in perturbation theory (\ref{eq:classical_densities}), so that their difference is entirely generated by non-perturbative contributions.}\label{fig:7}
\end{figure}

The simplest way  of arriving at  a density that spreads over the whole real line from double-scaling is to ``zoom in'' on two endpoints that are colliding with each other, as can happen with critical double-cut matrix models (see the lower part of Figure~\ref{fig:cuts}). The second clue is that~$\rho_0(q)$  is symmetric on the real line. As will be reviewed later, the string equations that arise from double-scaling symmetric two-cut  models (pioneered in refs.~\cite{Periwal:1990gf,Periwal:1990qb,Nappi:1990bi,Crnkovic:1990mr,Hollowood:1991xq,Crnkovic:1992wd}) have already been identified in ref.~\cite{Klebanov:2003wg} as capturing type 0B minimal string physics. Moreover, there is a curious universal feature of the string equations that arises in the symmetric sector  (the Painlev\'e~II heirarchy) that, as we will show,  reproduces certain characteristic features of observables in the JT supergravity. Combining together an infinite family of 0B minimal string models in a way precisely analogous to what was done in refs.~\mcite{Johnson:2020heh,Johnson:2020exp,Johnson:2020mwi} for the other JT supergravities provides a {\it fully non-perturbative} definition of the theory. As an example of what can be computed from it, Figure~\ref{fig:7}  displays (blue dots)  the full non-perturbative spectral density for~$Q$. The black dashed line is the perturbative result of equation~(\ref{eq:classical_densities}). Other important quantities can be computed in this way, such as the spectral form factor displayed later in Figure~\ref{fig:6}.

The rest of this paper  is organized as follows. Section~\ref{sec:2}  reviews the construction of double-cut Hermitian matrix models and their double scaling limit. Section~\ref{sec:3}  defines the particular model that  describes the Dyson-Wigner $\boldsymbol{\beta}{=}2$ JT supergravity, showing that it reproduces all of the characteristic perturbative results for this model derived in ref.~\cite{Stanford:2019vob}. Non-perturbative contributions of the matrix model are studied in Section \ref{sec:4}, where the full spectral density and spectral form factor are obtained by combining analytic and numerical techniques.  We finish in Section \ref{sec:6} with some  discussion. 
There are a number of appendices included that expand some computations, and help keep this work somewhat self-contained without diluting the narrative flow. Appendices \ref{zapp:1} and~\ref{zapp:2} are devoted to computing multi-correlators at leading genus and higher genus corrections to the 1-point function $\langle \text{Tr}\,e^{-\beta Q^{2}}\rangle$, respectively. The normalization of  eigenfunctions is discussed in Appendix \ref{zapp:3}. 

\section{Double-cut matrix models}
\label{sec:2}

Very soon after the discovery of the double-scaling limit in the context of single-cut Hermitian matrix models in the 90s~\cite{Brezin:1990rb,Douglas:1990ve,Gross:1990vs,Gross:1990aw}, the study of the double--scaled limit of multi-cut  Hermitian matrix models  was initiated in a series of works~\cite{Douglas:1990xv,Nappi:1990bi,Crnkovic:1990mr,Hollowood:1991xq,Crnkovic:1992wd}, with most of the focus on two-cut models.  Closely allied to the latter are the double-scaled unitary matrix models, discovered slightly earlier in refs.~\cite{Periwal:1990gf,Periwal:1990qb}, which yield a subset of the same physics.\footnote{Transitions of the kind that interest us, involving mergers of cuts, have an older history from before double-scaling and gravity applications (see \emph{e.g.} refs.~\cite{Gross:1980he,Cicuta:1986pu}).} We shall see that in the end, the physics that we need will be accessible from either system, but Hermitian double-cut matrix model language is perhaps the most natural here given the considerations of the previous Section.  They are built from an $N{\times}N$  Hermitian matrix~$Q$,\footnote{Note that the symbol $N$,  the dimension of the square matrix $Q$,  should not  be confused with the $\text{N}$ describing the number of SYK fermions, as used  in Section~\ref{sec:1}.} so that the expectation value of a matrix observable $\mathcal{O}$ is:
\begin{equation}\label{eq:64}
\langle \mathcal{O} \rangle=
\frac{1}{\mathcal{Z}}
\int dQ\,\mathcal{O}\,e^{-N\,{\rm Tr}\,V(Q)}\ ,
\end{equation}
where $\mathcal{Z}=\int dQ\,e^{-N\,{\rm Tr}\,V(Q)}$ is the matrix partition function. The probability measure is determined by the potential $V(Q)$, that is typically taken to be  a polynomial. See refs.~\cite{Akemann:2011csh,DiFrancesco:1993cyw,Ginsparg:1993is} for useful sources of information about random matrix theory and applications of the double-scaling limit.

A central observable that determines the average distribution of eigenvalues $\lambda_i\in\mathbb{R}$ of the matrix $Q$ is the spectral density $\rho(q)$, defined as
\begin{equation}\label{eq:62}
\rho(q)=
\Big\langle 
\frac{1}{N} \sum_{i=1}^N\delta(q-\lambda_i) 
\Big\rangle\ .
\end{equation}
Depending on the particular number and location of the minima of $V(Q)$, the spectral density in the large $N$ limit becomes a smooth function of $q$ supported on a finite number of disjoint intervals in $q\in \mathbb{R}$. These intervals ultimately become cuts in the complex $q$--plane when using complex analysis to  analyze the large $N$ limit. When $\rho(q)$ is supported on more than one interval, the matrix model is said to be ``multi-cut''.

\begin{figure}
\centering
\includegraphics[scale=0.50]{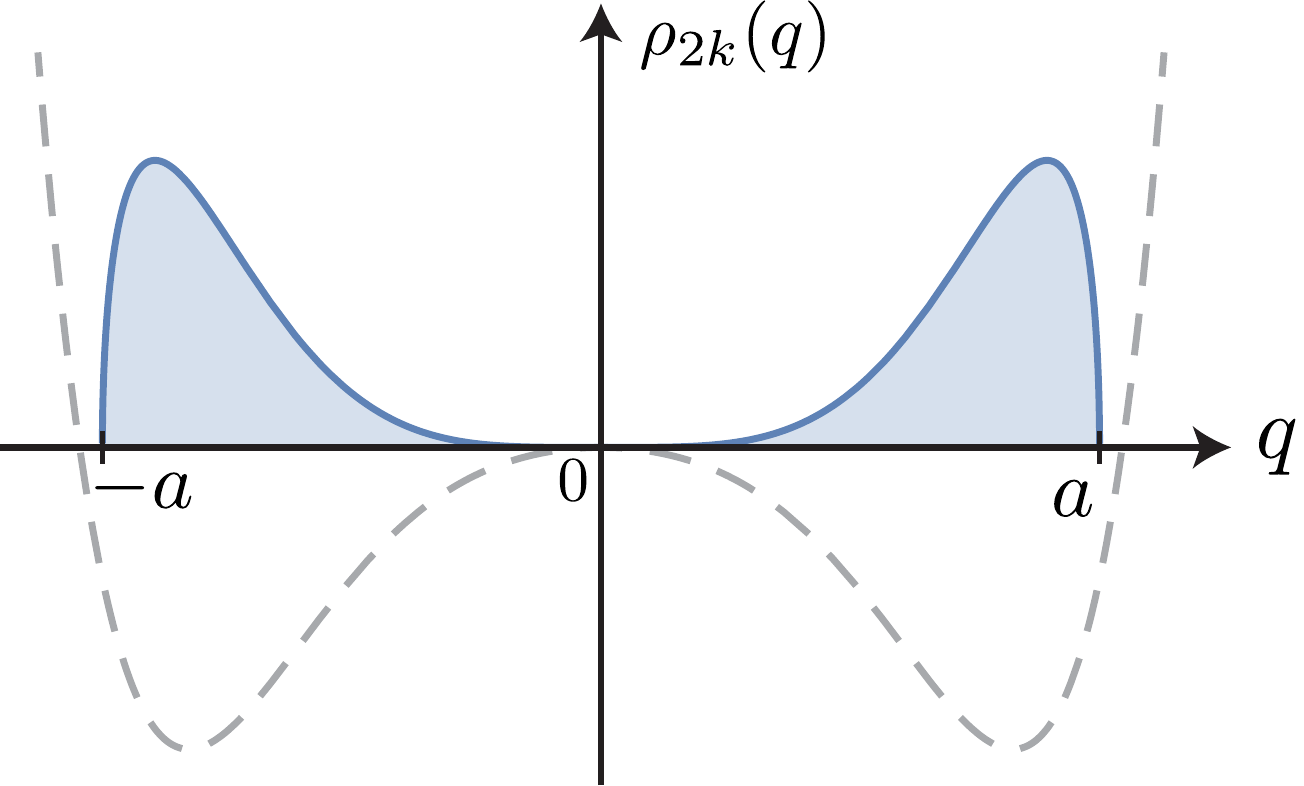}
\caption{In blue we plot the singular spectral density (\ref{eq:12}) with $k=2$, where the spectral density vanishes at the origin were the two cuts meet $\lim_{N\rightarrow \infty}\rho_{2k}(q)\propto q^4$. The dashed curve corresponds to the critical potential (\ref{eq:11}), with the two minima symmetrically located around $q{=}0$.}\label{fig:3}
\end{figure}

The focus here is on systems that transition between  double- and single-cut  phases. Precisely at the transition, the corresponding potential $V(Q)$ is said to be ``critical", with different critical systems characterized~\cite{Kazakov:1989bc} by the rate at which the spectral density vanishes where the two cuts meet at $q{=}0$, \emph{i.e.}, $\lim_{N\rightarrow \infty}\rho_m(q)\sim q^m$ with $m \in \mathbb{N}$. For even potentials we have $m{=}2k$ with $k\in \mathbb{N}$, and the large~$N$ spectral density of a critical system can be written as
\begin{equation}\label{eq:12}
\lim_{N\rightarrow \infty}
\rho_{2k}(q)=
\frac{b_k}{2\pi}
\left(\frac{q}{a}\right)^{2k}
\sqrt{\frac{a^2-q^2}{a^2}}\ ,
\end{equation}
where $a>0$ and $b_k={2^{2k+1}(k+1)!(k-1)!}/{a(2k-1)!}$ is a normalization constant. This spectral density is supported on $q\in (-a,0)\cup (0,a)$ and arises from a two-cut spectral density transitioning to a single-cut phase. Using standard methods (see {\it e.g.,} refs.~\cite{Eynard:2015aea,Anninos:2020ccj}),  the appropriate potential needed to produce this spectral density can be easily worked out to be:
\begin{equation}\label{eq:11}
V'_{2k}(q)=b_k\sum_{n=0}^k
\binom{1/2}{n}
(-1)^n
\left(\frac{q}{a}\right)^{2(k-n)+1}\ ,
\end{equation}
which defines the critical model for arbitrary $N$. The even potential $V_{2k}(q)$ contains two minima symmetrically located in the interval $|q|<a$. Figure \ref{fig:3} is a plot of the potential and spectral density  for $k=2$. Note that it is possible to construct models whose spectral density behaves like $\lim_{N\rightarrow \infty}\rho(q)\sim q^{2k+1}$, but it is more subtle, as these models by themselves are not well defined since the resulting (unscaled) densities are not positive definite. Instead, they can be introduced as perturbations of the even model potentials, as described in refs.~\cite{Neuberger:1990vu,Crnkovic:1992wd}.

\subsection{The double scaling limit}
\label{sec:double-scaling}

The double scaling limit involves approaching the critical potential~(\ref{eq:11}) while simultaneously taking the large~$N$ limit in a  way that  captures only certain universal physics associated to the merging of the two cuts. The leading spectral density in the double scaled model corresponds to the behavior of~$\rho(q)$ at~$q{\sim}0$. For even potentials, this is easily obtained from equation~(\ref{eq:12}) by taking~$a{\rightarrow}\infty$, which gives~$\rho_0(q){=}\lim_{N\rightarrow \infty}\rho_{2k}(q){\sim}q^{2k}$. There will be $1/N$ (topological) perturbative corrections to this leading large~$N$ behaviour, and non-perturbative contributions too. There is a powerful formulation (the orthogonal polynomial methods~\cite{Brezin:1978sv,Bessis:1980ss}) that allows for them to be efficiently extracted. These methods are   reviewed in {\it e.g.,} refs.~\cite{Akemann:2011csh,DiFrancesco:1993cyw,Ginsparg:1993is,Eynard:2015aea}.

A  brief summary of the logic is as follows. After writing the matrix model in terms of integrals over the $N$ eigenvalues $\lambda_n$, a family of $N$  polynomials $P_n(\lambda)=\lambda^n+\cdots$ that are orthogonal with respect to the matrix model measure $d\lambda\, e^{-N V(\lambda)}$ are introduced. A simple argument shows the different polynomials are related according to
\begin{equation}
\lambda P_n(\lambda)= \sqrt{R_{n+1}}P_{n+1}(\lambda)+S_n P_n(\lambda)+\sqrt{R_{n}}P_{n-1}(\lambda)\ ,
\end{equation}
where $R_n$ and $S_n$ depend on the particular theory. The matrix model can be used to derive a family of recursion relations for $R_{n}$ and $S_n$, and solving the model is equivalent to finding them.  The $P_n(\lambda)$ themselves supply a Hilbert space description of the system, on which  matrix model observables become operators. This will be very useful shortly.  At large $N$,  the  $R_n$ and $S_n$ define smooth functions $R(X)$ and $S(X)$ of the variable $X{=}n/N\in(0,1)$, a smooth coordinate along the spectral density.    The scaling limit around the critical behavior at an endpoint is done by introducing a parameter $b$ which goes to zero as $N\to\infty$, and the rate at which it does so is  set by: $\hbar b^{2+1/m}{=}1/N$ (with $m{=}2k$), determining a scaled topological expansion parameter $\hbar$. The rate at which various quantities in the model approach their critical values at the $X{=}1$ endpoint is controlled by a power of $b$ {\it e.g.,} $R(X){=}R_c+(-1)^nb^{1/m}f(x)+\cdots$ where $x{\in}\mathbb{R}$ is the scaling part of $X{=}1{-}b^2(x{-}\mu)$, and the parameter~$\mu$ will be fixed later. A different scaling function $g(x)$ arises from $S(X)$. Inserting all these relations into the recursion relations and taking the limit yields non-trivial coupled ordinary differential equations for $f(x)$ and $g(x)$. These are the so-called ``string equations" for the system indexed by $m$. 

More generally, the critical potentials~(\ref{eq:11}) can be summed together with coefficients $t_m$, and, defining the alternative set of functions $r(x)$ and $\alpha(x)$ by $f(x){=}r(x)\cosh(\alpha(x))$ and $g(x){=}r(x)\sinh(\alpha(x))$ the differential equations can be conveniently written in the conventions of ref.~\cite{Klebanov:2003wg} as:
\begin{equation}\label{eq:15}
\begin{aligned}
&
\sum_{m=1}^{\infty}t_m\,K_m+xK_0=0\ ,\\
&
\sum_{m=1}^{\infty}t_m\,H_m+xH_0+\eta=0\ ,
\end{aligned}
\end{equation}
where $\eta$ is a parameter of the model, arising as an integration constant (see ref.~\cite{Klebanov:2003wg}) and $\lbrace K_m,H_m \rbrace$ are polynomials of $\lbrace r(x),\alpha(x) \rbrace$ and their derivatives determined from the following recursion relations:
\begin{equation}\label{eq:21}
\begin{aligned}
\,\,K_{m+1}&=
\hbar\alpha' K_m+rH_m-\hbar^2(H'_m/r)'\ , \\[4pt]
H'_{m+1}&=\hbar\alpha' H'_m-rK'_m\ ,
\end{aligned}
\end{equation}
where $(K_0,H_0)=(r(x),0)$ and primes are derivatives with respect to $x$ (see ref.~\cite{Klebanov:2003wg} for explicit expressions for the first few values of $m$).\footnote{The string equations for the multi-cut matrix models were first derived in full generality in refs.~\cite{Crnkovic:1990mr,Crnkovic:1992wd}. We are following the conventions in ref.~\cite{Klebanov:2003wg}, slightly changing some of the notation in order to avoid confusion:~$\alpha_{\rm here}{=}\beta_{\rm there}$, $K_{m}^{\rm here}{=}R_{m}^{\rm there}$ and~$\eta_{\rm here}{=}q_{\rm there}$. As pointed out in ref.~\cite{Klebanov:2003wg}, $q_{\rm there}$ arises as an integration constant, and counts the amount of R-R flux in a minimal string interpretation.\label{foot:1}} 

The parameter $t_m$ controls the double scaling limit of the critical model with $\rho_0(q)\propto q^{m}$, \emph{e.g.} the double scaled $2k$th model associated to the critical potential in equation~(\ref{eq:11}) is obtained by setting all $t_m{=}0$ except for $t_{2k}{=}1$.  As differential equations, they encode not just the corrections to the leading large $N$ behavior, but non-perturbative information too, as we shall see. They can be derived from the Zakharov-Shabat (ZS) integrable hierarchy~\cite{zakharov1974scheme}, in an analogue of the manner in which the equations of type~0A minimal strings can be derived from the Korteweg-de Vries (KdV) integrable hierarchy. These models were identified as  type~0B minimal strings in ref.~\cite{Klebanov:2003wg}.

Strictly speaking, we have written here a large family of string equations labeled by {\it any} integer $m$.  For our purposes, it will be enough to restrict ourselves to the~$m$ even models with $\alpha'(x)=0$.\footnote{Note the actual constant value $\alpha(x)=\alpha_0$ does not appear in the string equations and is therefore irrelevant.}  Using the recursion relations~(\ref{eq:21}) it is straightforward to show that $H_{2k}=0$, so that the second string equation  (\ref{eq:15}) is automatically satisfied after fixing $\eta=0$. The $K_{2k}$ are polynomials in $r(x)$ and its derivatives, computed from the following closed recursion relation obtained from equation~(\ref{eq:21})
\begin{equation}\label{eq:40}
K_{2k}=
\frac{2k}{2k-1}
\left[
r(x)\int^xd\bar{x}\,r(\bar{x})K'_{2(k-1)}
-\frac{1}{4} \hbar^2K''_{2(k-1)}
\right]\ ,
\end{equation}
where we have rescaled $\hbar\rightarrow \hbar/2$ as well as $K_{2k}$ so that they are normalized as $K_{2k}=r(x)^{2k+1}+\mathcal{O}(\hbar^2)$.\footnote{The rescaling of $K_{2k}$ can be applied at the level of the string equation by shifting $t_{2k}\rightarrow t_{2k}/P_{2k}(0)$, where $P_{2k}(z)$ is the Legendre polynomial.} The first few  are easily computed and given by:
\begin{equation}
\label{eq:40b}
\begin{split}
&K_0=r(x)\ , \\
&K_2=r(x)^3-\frac{1}{2}\hbar^2r''(x) \ , \\
&K_4=r(x)^5
-\frac{5}{6}\hbar^2r(x)
(r(x)^2)''
+\frac{1}{6}\hbar^4r^{(4)}(x) \ ,\\
&\,\,\,\vdots\\
&K_{2k}=r(x)^{2k+1}+\dots+
\frac{(-1)^kk!(k-1)!}{2(2k-1)!}
\hbar^{2k}r^{(2k)}(x) \ ,
\end{split}
\end{equation}
where $r^{(2k)}(x)$ corresponds to the $2k$ $x$-derivatives acting on $r(x)$.

Finally, the full string equation for this class of even double-cut models is given by:
\begin{equation}\label{eq:13}
\sum_{k=1}^{\infty}t_{2k}K_{2k}+r(x)x=0\ .
\end{equation}
Restricting  to the even $m$ models and turning off $\alpha(x)$ reduces the two string equations (\ref{eq:15}), to this simpler single equation. It is in fact the Painlev\'e~II hierarchy of ODEs,  related to the modified KdV (mKdV) integrable hierarchy, and was first discovered for double--scaled unitary matrix models~\cite{Periwal:1990gf,Periwal:1990qb}. We shall see that equation~(\ref{eq:13}) is enough for describing the JT supergravity theory in which we are interested. 

\subsection{Observables} 

We shall need to compute certain observables from the function $r(x)$ that solves the string equation (\ref{eq:13}). The expectation value of any single trace observable can be computed from a certain  ``macroscopic loop'' formula derived in ref.~\cite{Crnkovic:1992wd}. In our conventions, it is written as \cite{toappear}
\begin{equation}\label{eq:8}
\big\langle 
{\rm Tr}\big(e^{-\beta Q^2}\big)
\big\rangle=
\sum_{s=\pm}
\int_{-\infty}^{\mu}dx
\braket{x|
e^{-\beta \mathcal{H}_s}
|x}\ ,
\end{equation}
where
\begin{equation}\label{eq:34}
\mathcal{H}_s=\hat{p}^2+\left[
r(\hat{x})^2-s\hbar r'(\hat{x})
\right]\ ,
\end{equation}
with $\hat{p}{=}{-}i\hbar \partial_x$. The quantum mechanical system spanned by the position eigenstates $\ket{x}$ arises from the continuum limit of the orthogonal polynomial system described in the previous Subsection, \textit{i.e.} $P_n(\lambda)\sim \ket{n}$ is promoted to a continuous variable $x$~\cite{Gross:1990aw,Banks:1989df}.~\footnote{See refs.~\cite{DiFrancesco:1993cyw,Ginsparg:1993is} for detailed reviews on how this effective description arises and refs. \cite{Johnson:2020lns,Johnson:2020mwi,Johnson:2020heh,Johnson:2019eik} for more recent discussions in the JT supergravity context.} The integral comes from the continuum limit, so that a sum  over $n$ of orthogonal polynomial quantities, becomes an~$X$ integral from $0$ to $1$. What remains is the part that survived in the double-scaling limit,  which zooms into the neighborhood of the $X{=}1$ endpoint defined by $X{=}1{-}b^2(x{-}\mu)$, giving $-\infty$ and~$\mu$ as the  limits on the~$x$ integral. The value of $\mu$ will be fixed later by comparison to the supergravity theory. Compared to the single cut models in refs.~\cite{Banks:1989df,Gross:1989aw}, there is an additional sum over $s=\pm$ that comes from the fact the orthogonal polynomials have a different behavior for even and odd $n$ \cite{Crnkovic:1990mr,toappear}.

The right hand side is computed in  one-dimensional quantum mechanics in the usual way with $\braket{p|x}=e^{ipx/\hbar}/\sqrt{2\pi \hbar}$. This provides an extremely concrete formalism that allows us to compute arbitrary single trace observables to all orders in $\hbar$, and non-perturbatively. As explained and shown in Appendix \ref{zapp:1}, analogous formulae can be derived for higher number of insertions of the matrix operator ${\rm Tr}\,e^{-\beta Q^2}$. This will be useful when we compare results of the multi-cut matrix model to the JT supergravity predictions.


This is all a beautiful counterpart to what occurred for the other JT supergravity model (Case A in the Introduction) and its cousins in refs.~\cite{Johnson:2020lns,Johnson:2020mwi,Johnson:2020heh}. There, the natural system that arose from double-scaled complex matrix models produced single a Schr\"odinger Hamiltonian $\mathcal{H}=\hat{p}^2+u(\hat{x})$, where $u(x)$ solved a different string equation. All of the techniques applied there can be brought to this system in order to fully solve this JT supergravity.\footnote{In the discussion of Section~\ref{sec:6}, there are  further comments about the similarities, and crucial differences, between these two systems.}

For a start, we  can use equation~(\ref{eq:8}) to write a formula for the spectral density $\rho(q)$ using the relation $\langle {\rm Tr}\,e^{-\beta Q^2}\rangle=\int_{-\infty}^{+\infty}dq\rho(q)e^{-\beta q^2}$. Defining the wavefunctions $\psi_{\mathcal{E},s}(x)$ of the operator $\mathcal{H}_s$ (\ref{eq:34})
\begin{equation}\label{eq:35}
\mathcal{H}_s\psi_{\mathcal{E},s}(x)=
\mathcal{E}\psi_{\mathcal{E},s}(x)\ ,
\end{equation}
with eigenvalue $\mathcal{E}$, a simple calculation yields
\begin{equation}\label{eq:38}
\rho(q)=|q|\sum_{s=\pm}
\int_{-\infty}^{\mu}dx\,
|\psi_{q^2,s}(x)|^2\ .
\end{equation}
Now we are ready to apply all of this technology to the JT supergravity of interest.

\section{Perturbative Physics}
\label{sec:3}

An important test is to see if  the double-cut double-scaled model reproduces the perturbative JT supergravity results obtained in Section 5.2.1 of ref.~\cite{Stanford:2019vob}. Recall that this is a supergravity theory with contributions from orientable supermanifolds of constant negative curvature with both even and odd spin structures weighted equally. 
As a reminder, we denote as $Z(\beta_1,\dots,\beta_n)$ the partition function with $n$ asymptotic boundaries of renormalized length $\beta_i$, that includes all possible (connected) genus surfaces weighted by $\hbar^{2(g-1)+n}$, where $\hbar{=} e^{-S_{0}}$. These observables were computed in ref.~\cite{Stanford:2019vob} to all orders in perturbation theory in $\hbar$. They were shown to vanish (see ref.~\cite{Stanford:2019vob}'s Appendices A and D), except for  the cases $n{=}1,2$,  which are  given by:
\begin{equation}\label{eq:17}
\frac{Z(\beta)}{\gamma}=
\frac{e^{\pi^2/\beta}}{\hbar\sqrt{\pi \beta}}\ ,
\quad \quad
\frac{Z(\beta_1,\beta_2)}{\gamma^2}=
\frac{\sqrt{\beta_1\beta_2}}{\pi(\beta_1+\beta_2)}\ .
\end{equation}
Included with each supergravity observable is an inverse factor of $\gamma{=}\sqrt{2}$ for each boundary. These are  in place in preparation for comparison to matrix model quantities. As explained in ref. \cite{Stanford:2019vob},   a factor $\gamma$ is needed to convert  every matrix model trace involved when comparing to  supergravity results.

The spectral density of the supergravity theory, defined from $Z(\beta)=\int_0^{\infty}dE\rho_0^{\rm SJT}(E)e^{-\beta E}$, is given on the left of equations~(\ref{eq:classical_densities}). Meanwhile, the spectral density $\rho_0(q)$  of the supercharge {\it via} the matrix model is given on the right of equations~(\ref{eq:classical_densities}).  While all of these results are surprisingly simple, we stress that they get corrected by important non-perturbative effects, which we will compute using the methods of this paper in Section~\ref{sec:4}.


\subsection{Matrix model matching}

Section \ref{sec:2} reviewed a general class of  double scaled matrix models, specified by the coefficients $t_{2k}$ that determine the string equation (\ref{eq:13}) satisfied by $r(x)$, as well as by $\mu$ appearing in the computation of observables~(\ref{eq:8}). Fixing to a specific $\lbrace t_{2k},\mu \rbrace$ determines a particular double scaled model. Matching the leading genus behavior of $\gamma\langle{\rm Tr}\,e^{-\beta Q^2}\rangle$ to the supergravity partition function $Z(\beta)$ in equation~(\ref{eq:17}) uniquely fixes these parameters. 

We start by writing a perturbative expansion in $\hbar$ for the function $r(x)$:
\begin{equation}\label{eq:26}
r(x)=r_0(x)+\sum_{n=1}^{\infty}r_n(x)\hbar^n\ .
\end{equation}
Inserting this into the  string equation (\ref{eq:13}), the leading contribution $r_0(x)$ is determined from the following simple algebraic constraint
\begin{equation}
r_0(x)
\bigg[\,
\sum_{k=1}^{\infty}t_{2k}r_0(x)^{2k}+x
\bigg]=0\ ,
\end{equation}
which admits two possible solutions: Either $r_0(x) = 0$ or the quantity in parentheses vanishes, giving $r_0(x) \neq 0$. To get a non-trivial function $r_0(x)$ defined over the whole real line $x \in \mathbb{R}$,  we  use the piecewise and continuous solution
\begin{equation}\label{eq:18}
r_0(x):
\begin{cases}
\displaystyle
\,\,\sum_{k=1}^{\infty}t_{2k}r_0(x)^{2k}+x=0
\quad \ , \quad x\le 0\ , \\[4pt]
\qquad  \, r_0(x)=0   \hskip1.70cm \ ,
\quad x\ge 0\ .
\end{cases}
\end{equation}
From this solution, let us compute the leading genus contribution to $\langle{\rm Tr}\,e^{-\beta Q^2} \rangle$ in equation~(\ref{eq:8}), using the following identity
\begin{equation}\label{eq:28}
\braket{x_2|e^{-\beta \mathcal{H}_s}|x_1}=
\frac{e^{-\beta r_0(x_2)^2-\frac{1}{\beta}\left(\frac{x_1-x_2}{2\hbar}\right)^2}}
{2\hbar\sqrt{\pi \beta}}
+\mathcal{O}(\hbar)\ ,
\end{equation}
which can be easily proven by inserting a complete set of eigenstates of the momentum operator $\hat{p}$ and solving the resulting $p$ integral. Using this in equation~(\ref{eq:8}) we find
\begin{equation}\label{eq:22}
\begin{aligned}
\big\langle
{\rm Tr}\,e^{-\beta Q^2}
\big\rangle&=
\frac{2}{2\hbar \sqrt{\pi \beta}}
\int_{-\infty}^{\mu}dx\,e^{-\beta r_0(x)^2}+
\mathcal{O}(\hbar)\\
&\hskip-1cm\simeq
\frac{1}{\hbar\sqrt{\pi \beta}}
\left[\mu+
\sum_{k=1}^{\infty}t_{2k}2k
\int^{\infty}_{0}dr_0\,
r_0^{2k-1}
e^{-\beta r_0^2}
\right]\\
&\hskip-1cm=
\frac{1}{\hbar\sqrt{\pi \beta}}
\left[\mu+
\sum_{k=1}^{\infty}
\frac{t_{2k}k!}{\beta^k}
\right]+
\mathcal{O}(\hbar)\ ,
\end{aligned}
\end{equation}
where in the second equality we changed the integration variable to $r_0$ and computed the Jacobian using equation~(\ref{eq:18}). Note that to leading order the operator~$\mathcal{H}_s$ in~(\ref{eq:34}) is independent of $s=\pm$. We have also assumed the boundary condition~${r_0(-\infty)=+\infty}$ that we shall verify shortly. 

\begin{figure}[ht]
\centering
\includegraphics[scale=0.35]{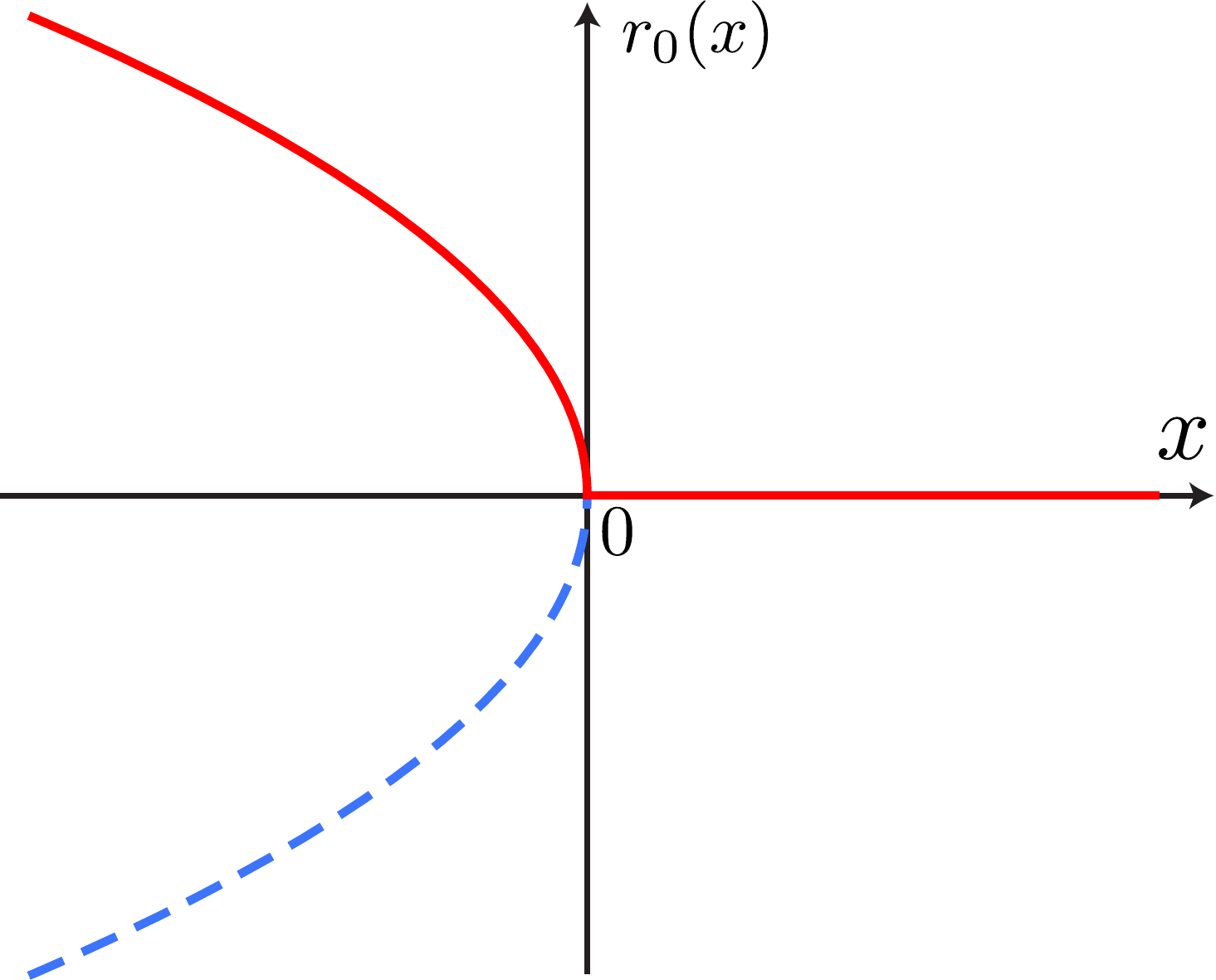}
\caption{Plot of the leading solution $r_0(x)$ obtained from (\ref{eq:18}) after fixing $t_{2k}$ according to (\ref{eq:25}). In the region $x<0$ there are two solutions to the implicit constraint, indicated in red (solid) and blue (dashed). We chose the red branch, which satisfies the boundary conditions assumed in (\ref{eq:22})}\label{fig:1}
\end{figure}

The leading $\hbar$ behavior in (\ref{eq:22}) depends on the parameters $\lbrace t_{2k},\mu\rbrace$, and it should  match with the supergravity observable $Z(\beta)/\gamma$ in equation~(\ref{eq:17}). Clearly, the matrix model result (\ref{eq:22}) has precisely the right structure, and we find agreement if
\begin{equation}\label{eq:25}
(t_{2k},\mu)=
\left(\frac{\pi^{2k}}{k!^2},1\right)\ .
\end{equation}
This unambiguously defines the multi-cut double scaled model.\footnote{Note that these agree with the values of the $t_k$ parameters found for the complex matrix model definition of the other supergravities discussed in refs.~\cite{Johnson:2020heh,Johnson:2020exp}.} With this choice, the spectral density $\rho_0(q)$ of the matrix model is the correct function  given in equation~(\ref{eq:classical_densities}) and the definition of $r_0(x)$ in the $x<0$ region of equation~(\ref{eq:18}) can be resummed and written in terms of a modified Bessel function. Specifically, the potential $r_{0}(x)$ satisfies the constraint
\beq 
I_{0}(2\pi r_{0})-1
+x=0\;.\eeq
Figure \ref{fig:1} shows $r_0(x)$ for $x\in \mathbb{R}$ (red curve), where we see that the assumed boundary condition $r_0(-\infty)={+}\infty$ is satisfied.

\subsection{Further comparison with JT supergravity}
\label{sec:sugra-matching}

Fixing the parameters $\lbrace t_{2k},\mu \rbrace$ of the multi-cut model according to equation~(\ref{eq:25}) only ensures the matching between $Z(\beta)$ and $\gamma\langle {\rm Tr}\,e^{-\beta Q^2}\rangle$ to leading order in $\hbar$. Now we must check whether higher trace operators and perturbative corrections reproduce the all-orders supergravity results~(\ref{eq:17}), after using the identification in equation~(\ref{eq:gravMMequiv}) where $H=Q^2$. 

\subsubsection{Leading genus multi-trace observables}

We start by extending the leading genus analysis in equation~(\ref{eq:22}) to multi-trace observables. For two and three ${\rm Tr}\,e^{-\beta Q^2}$ insertions the computation is reasonably straightforward, starting from the Appendix~\ref{zapp:1} general formulae~(\ref{eq:27}) and (\ref{eq:42}). Denoting $G_0(\beta_1,\dots,\beta_n)$ as the leading~$\hbar$ behavior of the connected expectation value of~$n$ insertions of ${\rm Tr}\,e^{-\beta_i Q^2}$, we derive:
\begin{equation}\label{eq:43}
\begin{aligned}
G_0(\beta_1,\beta_2)&=
2
\frac{\sqrt{\beta_1\beta_2}}{2\pi \beta_T}
e^{-\beta_Tr_0(\mu)^2}\ , \\
G_0(\beta_1,\beta_2,\beta_3)&=
2
\frac{\sqrt{\beta_1\beta_2\beta_3}}
{2\pi^{3/2}\beta_T}
\left[
(\hbar\partial_x)
e^{-\beta_Tr_0(x)^2}
\right]_{x=\mu}
\end{aligned}
\end{equation}
where $\beta_T{=}\sum_{i=1}^n\beta_i$. Using that $r_0(x)$ vanishes for positive $x$ and $\mu{>}0$, $G_0(\beta_1,\beta_2)$ precisely matches $ Z(\beta_1,\beta_2)/\gamma^2$ in equation~(\ref{eq:17}) and $G_0(\beta_1,\beta_2,\beta_3){=}0$, such that $Z_0(\beta_{1},\beta_{2},\beta_{3}){=}0$, again in agreement with the supergravity result.\footnote{Again, as explained in ref. \cite{Stanford:2019vob}, there is a factor $\gamma=\sqrt{2}$ for every
matrix model trace involved when comparing to supergravity results.} 

Since the procedure used in Appendix \ref{zapp:1} becomes increasingly tedious when computing higher trace observables, we can instead use a compact formula derived in ref.~\cite{Ambjorn:1990ji,Moore:1991ir} for the leading genus behavior of single-cut Hermitian matrix models. While single-cut matrix models are in a different universality class, observables are also computed from an almost identical effective quantum mechanical system \cite{DiFrancesco:1993cyw,Ginsparg:1993is}. To leading order, the only difference is that the potential $r_0(x)^2$ is positive definite and a factor of two coming from the summation over $s=\pm$ in (\ref{eq:8}). Taking this into account, we can apply the general formula of refs.~\cite{Ambjorn:1990ji,Moore:1991ir,Ginsparg:1993is} to the double-cut case and find
\begin{equation}\label{eq:45}
G_0(\beta_1,\dots,\beta_n)=
2
\frac{\sqrt{\beta_1\cdots\beta_n}}
{2\pi^{n/2}\beta_T}
\left[
(\hbar\partial_x)^{n-2}
e^{-\beta_T r_0(x)^2}
\right]_{x=\mu}
\end{equation}
For $n{=}2,3$ we recover the results in~(\ref{eq:43}). Applying this formula for the $n>3$ cases we find $G_0(\beta_1,\dots,\beta_n)$ vanishes for $n\ge 3$, given that $r_0(x)$ and all its derivatives vanish at $x=\mu>0$. In summary, we have shown how the double-cut matrix model exactly reproduces all the supergravity results to leading order in genus and arbitrary number of boundaries.

\subsubsection{Higher genus corrections}

Next, we compute $\hbar$ corrections to~$r_0(x)$ in equation~(\ref{eq:26}), by solving the string equation~(\ref{eq:13}) in perturbation theory. There are two different perturbative expansions, one valid for large positive $x$ and the other for large negative~$x$. Since $\mu$ is positive,  the relevant one for comparing with supergravity is the large positive~$x$ expansion. This is completely analogous to choices present when using  the type~0A models, as discussed in ref.~\cite{Johnson:2020heh}. It is convenient to introduce the parameter~$c\in \mathbb{R}_+$, and add $\hbar c$ to  the right-hand side of the string equation (\ref{eq:13}), giving:
\begin{equation}\label{eq:33}
\sum_{k=1}^{\infty}t_{2k}K_{2k}+r(x)x=\hbar c\ ,
\end{equation}
so that $c=0$ corresponds to the case  of interest. Inserting (\ref{eq:26}) in this differential equation it is straightforward to solve for the first few orders and find:
\begin{equation}\label{eq:44}
\begin{aligned}
r(x)=&r_0(x)+
\frac{\hbar c}{x}\left[
1+
t_2\frac{\hbar^2(1-c^2)}{x^3}+
\frac{\hbar^4(1-c^2)}{x^6}\right.\\[3pt]
&
\times
(t_2^2(10-3c^2)-t_4(4-c^2)x)
+\mathcal{O}(\hbar^6)
\bigg]\ .
\end{aligned}
\end{equation}
Notice that  the whole perturbative series vanishes when $c=0$, the case  of interest. This is no accident and follows from the observation that $r(x)=0$ is an exact solution to the full string equation when $c=0$. This means the function $r(x)$ for the model relevant to JT supergravity receives no perturbative corrections and is given by $r_0(x)$ of equation~(\ref{eq:18}) (with $t_{2k}$ as in~(\ref{eq:25}) and plotted in Figure~\ref{fig:1}) to all orders in perturbation theory.  We will  compute and display the non-perturbative contributions in Section~\ref{sec:4}.

To compute higher order corrections to $\langle {\rm Tr}\,e^{-\beta Q^2}\rangle$ using equation~(\ref{eq:8}) essentially involves computing the subleading terms in equation~(\ref{eq:28}). Building on some results of  Gel'fand and Dikii \cite{Gelfand:1975rn}, we derive the following expansion in Appendix \ref{zapp:2}
\begin{equation}\label{eq:32}
\begin{aligned}
&\braket{x|e^{-\beta \mathcal{H}_s}|x}\simeq
\frac{e^{-\beta r_0^2}}{2\hbar \sqrt{\pi \beta}}
\bigg\lbrace 
1+\hbar \beta(s r_0'-2r_0r_1)+\\
& \quad + \frac{\hbar^2\beta}{12}
\bigg[
\beta^2[(r_0^2)']^2
-12\left(
r_1^2+2r_0r_2-sr_1'
\right)+ \\
&\quad 
+2\beta\left(
12(r_0r_1)^2
+(r_0')^2
-2r_0r_0''
-12sr_0r_1r_0'
\right)
\bigg]
\bigg\rbrace\ ,
\end{aligned}
\end{equation}
where the~$r_n(x)$ are   corrections to~$r_0(x)$ defined in equation~(\ref{eq:26}), and  `$\simeq$' means that  terms of order $\hbar^{3}$ have been dropped. Integrating this expansion as in (\ref{eq:8}), we obtain higher~$\hbar$ corrections to~${\langle {\rm Tr}\,e^{-\beta Q^2} \rangle}$. In doing so, there is an important distinction that must be made regarding the leading and subleading terms in equation~(\ref{eq:32}).


When integrating (\ref{eq:32}) only the leading term must be integrated in the whole range $x\in(-\infty,\mu]$ indicated in equation~(\ref{eq:8}). All subleading $\hbar$ contributions can only be integrated on the region $x\in[0,\mu]$, where the expansion is meaningful (recall that we did a positive $x$ expansion, not a negative $x$ one, and it is meaningless to have both expansions present in the same expression). But we have already shown that the solution $r_0(x)=0$ for $x>0$ (\ref{eq:18}) receives no perturbative corrections, meaning $r_n(x)=0$ for $n\ge 0$. As a result, all  $\hbar$ corrections to (\ref{eq:32}) vanish and we conclude that
\begin{equation}\label{eq:46}
\langle{\rm Tr}\,e^{-\beta Q^2}  \rangle=
\frac{1}{\hbar\sqrt{\pi \beta}}
\int_{-\infty}^{\mu}dx\,e^{-\beta r_0(x)^2}=
\frac{e^{\pi^2/\beta}}{\hbar\sqrt{\pi \beta}}\ .
\end{equation}
The important difference with the previous computation in equation~(\ref{eq:22}), is that this is an exact result to all orders in perturbation theory, matching with the supergravity result in equation~(\ref{eq:17}). 

Let us emphasize that the vanishing of higher $\hbar$ corrections to $\langle {\rm Tr}\,e^{-\beta Q^2} \rangle$ does not depend on the special tuning of couplings that define the model, {\it  i.e.,} $\lbrace t_{2k},\mu \rbrace$ in equation (\ref{eq:25}). Instead, it is a consequence of having $r(x)=0$ for~${x>0}$ to all orders in perturbation theory which, as we can see from equation (\ref{eq:44}), holds for arbitrary $t_{2k}$.\footnote{This same feature (but in a different language) was also observed in Section~5.2.1 of ref.~\cite{Stanford:2019vob}, where the vanishing of the perturbation series was shown to be independent of the detailed structure of $\rho_0(q)$ (\ref{eq:classical_densities}) but relied on $\rho_0(q)$ being supported on the whole real line ${q\in \mathbb{R}}$, leaving no cut endpoints to source the relevant resolvents. An analogous situation also applies to other JT supergravity theories, as shown in ref.~\cite{Johnson:2020heh} for the cases $(\boldsymbol{\alpha},\boldsymbol{\beta})=(\lbrace 0,2 \rbrace,2)$.}

With this in  mind,  consider higher $\hbar$ corrections for multiple insertions of the macroscopic loop operator ${\rm Tr}\,e^{-\beta_i Q^2}$, which  \emph{via} the dictionary (\ref{eq:gravMMequiv}) will correspond to the JT supergravity partition function with higher genus corrections. For the supergravity of interest, all higher genus corrections vanish. It is interesting to see how this feature  emerges from  our definition of the double-scaled matrix model.   
Consider for example the two-point function, $\langle \text{Tr}(e^{-\beta_{1}Q^{2}})\text{Tr}(e^{-\beta_{2}Q^{2}})\rangle$.
The $\beta_2$ ``macroscopic loop''  can be expanded in terms of insertions of  point-like operators $\sigma_{2k}$, the  ``microscopic loops''~\cite{Banks:1990df,Ginsparg:1993is}, giving:
\begin{equation}
\langle 
\text{Tr}(e^{-\beta_{1}Q^{2}})
\text{Tr}(e^{-\beta_{2}Q^{2}})
\rangle=
\sum_{k=1}^{\infty}{\beta_{2}^{2k+\frac{1}{2}}}\langle 
\text{Tr}(e^{-\beta_{1}Q^{2}})
\sigma_{2k}
\rangle\, ,
\end{equation}
(in a particular normalization for the~$\sigma_{2k}$ that we will not need to specify here). However, a~$\sigma_{2k}$-insertion is equivalent~\cite{Gross:1990aw} (fully non-perturbatively) to differentiating the function~$r(x)$ with respect to the general coupling~$t_{2k}$, and so in practical terms, the right hand side of the above is computed by differentiating  the explicit~$r(x)$-dependence in~(\ref{eq:8}). In the genus expansion therefore, all contributions to the right hand side contain~$t_{2k}$ derivatives of~$r(x)$, which are given by the mKdV flows:
\begin{equation}
\frac{\partial r(x)}{\partial t_{2k}}\propto K^\prime_{2k}[r]\ , 
\end{equation} 
where the $K_{2k}[r]$, polynomials in $r(x)$ and its derivatives, are characterized in expressions~(\ref{eq:40}) and~(\ref{eq:40b}). Since it has been established that $r(x){=}0$ at every order in perturbation theory for arbitrary $t_{2k}$, we see that $\partial r(x)/\partial t_{2k}$ vanishes, and hence all higher genus contributions to this correlator. This procedure can be iterated to take care of correlators with higher numbers of ${\rm Tr}\,e^{-\beta Q^2}$ insertions, showing that they also vanish to all orders in perturbation theory. 

In summary, in this Section we have shown how the double-cut Hermitian matrix model  reproduces all of the perturbative results obtained from supergravity computations in ref.~\cite{Stanford:2019vob}.

\section{Non-perturbative physics}
\label{sec:4}

While the perturbative expansion of the supergravity observables is surprisingly simple, there can still be non-perturbative contributions that are not captured by the topological expansion. Although some important non-perturbative effects were discussed in ref.~\cite{Stanford:2019vob}, the main methods used there cannot derive generic  non-perturbative effects, whether using  supergravity or the recursive  loop equations technology which defines the matrix model order by order in the genus expansion. As emphasized in refs.~\cite{Johnson:2019eik,Johnson:2020heh,Johnson:2020exp}, the advantage of the alternative matrix model techniques used in this paper is that non-perturbative contributions can be explicitly computed because $r(x)$ is defined non-perturbatively by the string equation. 


\subsection{A toy model}
\label{sec:those-wonderful-toys}

Let us start by considering non-perturbative effects in a simple toy model that is relevant for JT supergravity and where everything can be computed analytically. This model is analogous to the Bessel model extensively studied in {\it e.g.} refs.~\cite{Stanford:2019vob,Johnson:2020heh,Johnson:2019eik}, and is:
\begin{equation}\label{eq:50}
r(x)=\frac{\hbar c}{x}\ .
\end{equation}
with $x>0$. When ${c=0,\pm 1}$ it is an exact solution to the  string equation (\ref{eq:33}). The eigenfunctions $\psi_{\mathcal{E},s}(x)$ of the operator $\mathcal{H}_s$ (\ref{eq:35}) can be computed exactly and written in terms of a Bessel function
\begin{equation}\label{eq:55}
\psi_{\mathcal{E},s}(x)=\frac{1}{\hbar}
\sqrt{\frac{x}{2}}J_{\alpha_s}(\sqrt{\mathcal{E}}x/\hbar)\ ,
\end{equation}
where $\alpha_s{=}c{+}s/2$ and the other independent solution is discarded since it is not regular at the origin. The normalization constant is fixed using the procedure described in Appendix \ref{zapp:3}. All observables can be constructed from the following kernel
\begin{equation}\label{eq:36}
\begin{aligned}
& L_s(\mathcal{E}_1,\mathcal{E}_2)  \equiv
\int_{0}^{\mu}
dx\,
\psi_{\mathcal{E}_1,s}(x)
\psi_{\mathcal{E}_2,s}^\ast(x) \\[4pt]
& = \hbar^2\left[
\frac{
\psi_{\mathcal{E}_1,s}(x)\psi'_{\mathcal{E}_2,s}(x)
-
\psi_{\mathcal{E}_1,s}'(x)\psi_{\mathcal{E}_2,s}(x)
}{\mathcal{E}_1-\mathcal{E}_2}
\right]_{x=\mu}\\[4pt]
& =
\frac{1}{2}\left(\frac{\mu}{\hbar}\right)^2
\left[
\frac{
\zeta_2 J_{\alpha_s}(\zeta_1) J_{\alpha_s}'(\zeta_2)
-\zeta_1 J_{\alpha_s}'(\zeta_1)J_{\alpha_s}(\zeta_2)}{\zeta_1^2-\zeta_2^2}
\right]\ ,
\end{aligned}
\end{equation}
where $\zeta_i=\mu \sqrt{\mathcal{E}_i}/\hbar$ and the integration region is ${x\in[0,\mu]}$ since this is the region where $r(x)$ in (\ref{eq:50}) is defined. The identity in the second line holds for any model and can be derived using that $\psi_{\mathcal{E},s}(x)$ are eigenfunctions of the operator $\mathcal{H}_s$.\footnote{In the random matrix theory context, this  is sometimes called a Christoffel-Darboux formula, an integral version of an identity satisfied by orthogonal polynomials~\cite{ForresterBook,andrews_askey_roy_1999}.} The full spectral density $\rho(q)$ computed from~(\ref{eq:38}) is given by
\begin{equation}\label{eq:51}
\begin{aligned}
\rho(q)&=
|q|\sum_{s=\pm}L_s(q^2,q^2) \\
&=
\sum_{s=\pm}
\frac{\mu \zeta }{4\hbar}
\left[
J_{\alpha_s}(\zeta)^2
-J_{\alpha_s-1}(\zeta)
J_{\alpha_s+1}(\zeta)
\right]\ ,
\end{aligned}
\end{equation}
where now $\zeta=\mu |q|/\hbar$. This exact result includes both perturbative and non-perturbative contributions. 

\begin{figure}
\centering
\includegraphics[scale=0.50]{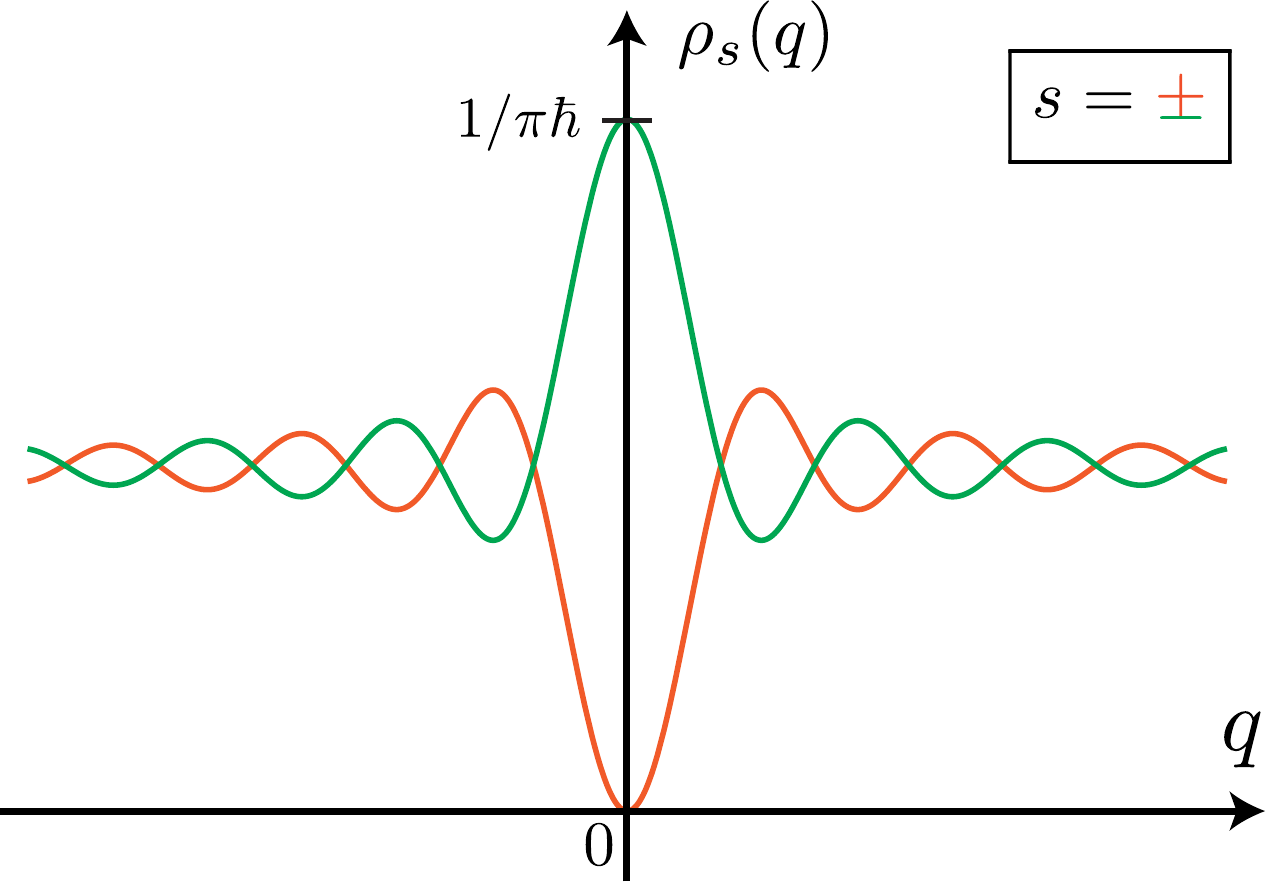}
\caption{Plot of the spectral density of the toy model with $(c,\mu)=(0,1)$ in (\ref{eq:52}), where $\rho_s(q)$ corresponds to the $s=\pm$ contribution to the spectral density. While for fixed $s$ we observe non-perturbative corrections, when summing both contribution we get a cancellation which ensures the full spectral density is exactly given by $\rho(q)=1/\pi \hbar$.}\label{fig:4}
\end{figure}

The case that is relevant for JT supergravity is when $(c,\mu)=(0,1)$, where the spectral density greatly simplifies to: 
\begin{equation}\label{eq:52}
\rho(q)\big|_{(c,\mu)=(0,1)}=
\sum_{s=\pm}\left[
\frac{1}{2\pi \hbar}-\frac{s}{4\pi q}\sin(2q/\hbar)
\right]\ .
\end{equation}
For each value of $s$ we observe there is a perturbative term in $\hbar$ as well as a single non-perturbative correction, given by a fast oscillatory function, see Figure \ref{fig:4}. A remarkable feature of the spectral density is that non-perturbative corrections coming from each $s$ sector precisely cancel each other, so that we are left with the exact result ${\rho(q)=1/\pi \hbar}$. Note that this captures the small $q$ behavior of the spectral density $\rho_0(q)$ in (\ref{eq:classical_densities}) required to describe the JT supergravity theory. 

The cancellation between non-perturbative corrections in this toy model is exclusive to the spectral density and single trace observable. Using the kernel (\ref{eq:36}) one can easily compute the expectation value of higher trace observables and explicitly show non-perturbative corrections are non-vanishing.

\subsection{The full model}

Let us now consider the full double scaled that is relevant for describing JT supergravity, defined from the values of $\mu$ and $t_{2k}$ indicated in (\ref{eq:25}). The analytic computation of non-perturbative contributions is much more challenging in this case, as it is not possible to write a simple solution $\psi_{\mathcal{E},s}(x)$ as in the toy model (\ref{eq:55}). However, we can make some progress by computing the eigenfunction $\psi_{\mathcal{E},s}(x)$ in the usual WKB approximation
\begin{equation}\label{eq:8.87}
\begin{aligned}
& \lim_{\hbar\rightarrow 0}
\psi_{\mathcal{E},s}(x)=
\frac{
A_+
\exp\left[\frac{i}{\hbar}\int^x_{x_{\rm min}}
d\bar{x}\sqrt{\mathcal{E}-r_0(\bar{x})^2}\right]}
{(\mathcal{E}-r_0(x)^2)^{1/4}} + \\[4pt]
& \qquad \qquad \,\,\,\,\, + 
\frac{
A_-\exp\left[-\frac{i}{\hbar}\int^x_{x_{\rm min}} d\bar{x}\sqrt{\mathcal{E}-r_0(\bar{x})^2}\right]}
{(\mathcal{E}-r_0(x)^2)^{1/4}}\ ,
\end{aligned}
\end{equation}
where the parameter $x_{\rm min}$ is defined from $r_0(x_{\rm min})^2=\mathcal{E}$, so that this is the solution in the classically allowed region. The challenge is fixing the undetermined constants $A_\pm$. To do so, we note that for any double scaled model at sufficiently large $x$, $r(x)$ approaches $r(x)=\hbar c/x$ of the toy model (\ref{eq:50}). Therefore, for large $x$ the general WKB approximation in (\ref{eq:8.87}) should match the classical limit $\hbar\rightarrow 0$ of the eigenfunctions of the toy model (\ref{eq:55}). In this way, we can fix the constants $A_\pm$ and obtain the WKB approximation for an arbitrary model
\begin{equation}
\lim_{\hbar\rightarrow 0}
\psi_{\mathcal{E},s}(x)=
\frac{\cos\left[
\frac{1}{\hbar}
\int_{x_{\rm min}}^x
d\bar{x}\sqrt{\mathcal{E}-r_0(\bar{x})^2}
-\frac{\pi}{4}(1+2\alpha_s)
\right]}
{\sqrt{\pi \hbar}(\mathcal{E}-r_0(x)^2)^{1/4}}
\end{equation}
Setting $x_{\rm min}=0$ and $r_0(x)=0$ we can easily check this expression agrees with the classical limit of (\ref{eq:55}). For the general case, $x_{\rm min}$ is a complicated function of $\mathcal{E}$ and $\psi_{\mathcal{E},s}(x)$ has a different classical limit.

From this we can compute the leading pertubative and non-perturbative contributions to the spectral density $\rho(q)$. Using the expression given in the second line of (\ref{eq:36}) (which holds for arbitrary $r(x)$) as well as the first line in (\ref{eq:51}) we find
\begin{equation}\label{eq:8.88}
\lim_{\hbar\rightarrow 0}
\rho(q)=
\rho_0(q)-
\sum_{s=\pm}
\frac{s}{4\pi |q|}
\sin\left[
2\pi
\int^{|q|}
d\bar{q}\,\rho_0(\bar{q})
-\pi c
\right]
\end{equation} 
where we have identified $\rho_0(q)$ as the leading perturbative contribution
\begin{equation}\label{eq:123}
\rho_0(q)=
\frac{|q|}{2\pi \hbar}\sum_{s=\pm}
\int_{x_{\rm min}}^{\mu}
\frac{dx}{\sqrt{q^2-r_0(x)^2}}\ ,
\end{equation}
and used the following identity
\begin{equation}
\int^{|q|}
d\bar{q}\,\rho_0(\bar{q})=
\frac{1}{\pi \hbar}\int_{x_{\rm min}}^\mu 
dx\sqrt{q^2-r_0(x)^2}\  .
\end{equation}
The second term in (\ref{eq:8.88}) is the interesting one, which gives the leading non-perturbative contribution to $\rho(q)$.\footnote{A different calculation in appendix E of \cite{Stanford:2019vob} gives similar (and in some sense equivalent) result for other random matrix ensembles.} The peculiarity is that the contributions from $s=\pm$ precisely cancel each other, so that there is effectively no leading non-perturbative term for the spectral density $\rho(q)$. There can still be subleading contributions, schematically given by
\begin{equation}\label{eq:8.89}
\rho(q)=\rho_0(q)
+\hbar^{n}\sin(\#/\hbar)+\dots\ , 
\end{equation} 
with $n>0$. For the toy model we have seen in (\ref{eq:52}) that all non-perturbative terms in $\rho(q)$ vanish. It is natural to ask if  this is also the case for the matrix model that describes JT supergravity. 

\begin{figure}
\centering
\includegraphics[scale=0.65]{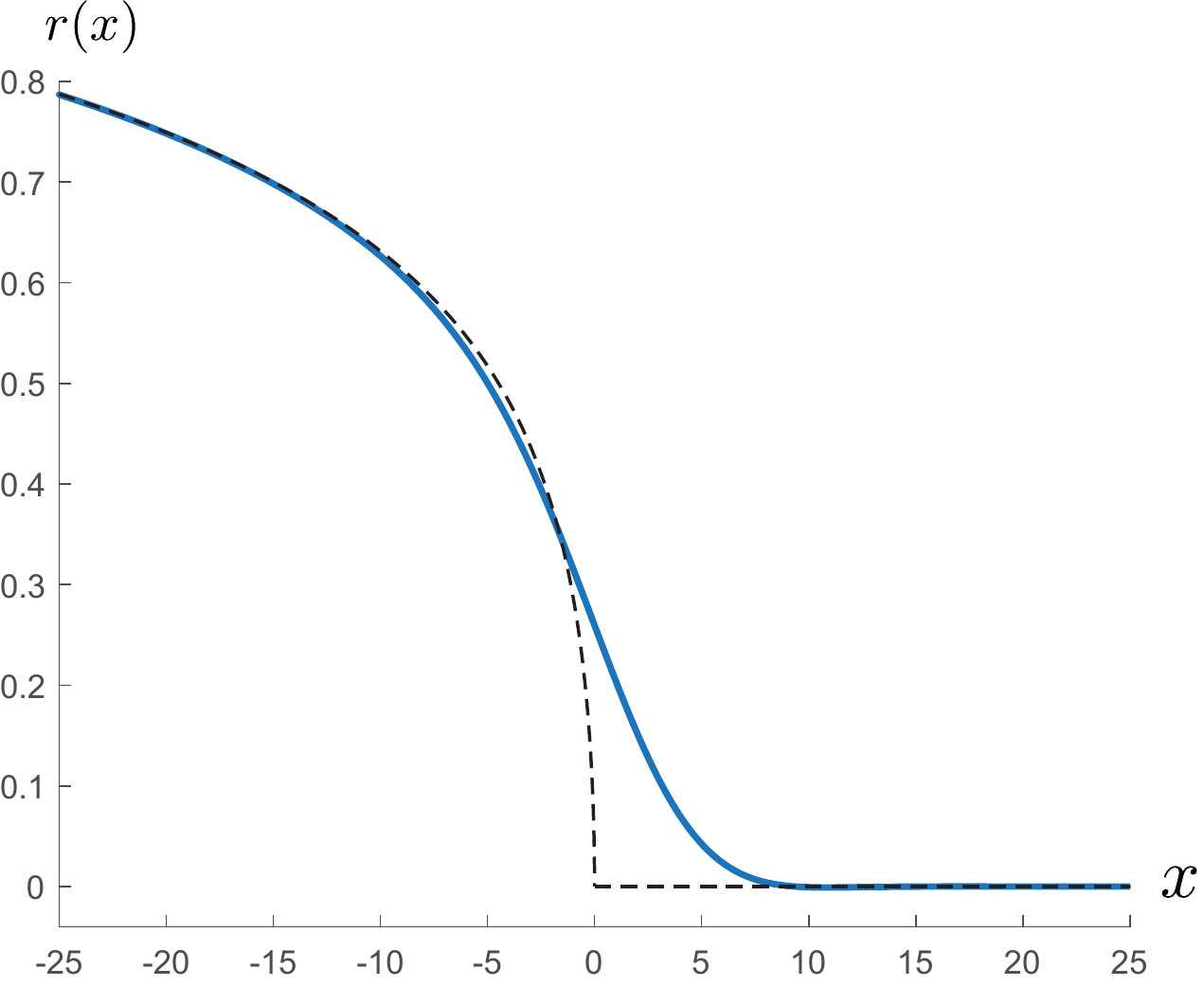}
\caption{The solid blue curve corresponds to the full non-perturbative numerical solution $r(x)$ to the string equation~(\ref{eq:13}) in the $k_{\rm max}{=}6$ truncation with $\hbar{=}1$. The dashed red curve is the leading genus solution $r_0(x)$ in (\ref{eq:18}).}\label{fig:2}
\end{figure}

To answer this, we need to compute the full eigenvalue spectral density $\rho(q)$, without making any approximation. Since there is no way of performing the calculation analytically, we proceed numerically. The first step is to solve the string equation~(\ref{eq:13}) for~$r(x)$. To make sense of it as a finite order differential equation, we follow \cite{Johnson:2020exp} and introduce a truncation by only including contributions from~$t_{2k}$ up to some maximum value~$k_{\rm max}$
\begin{equation}\label{eq:8.81}
\sum_{k=1}^{k_{\rm max}}
t_{2k}K_{2k}+r(x)x=0\ ,
\end{equation}
where $(t_{2k},\mu)$ are given in (\ref{eq:25}). For high enough values of $k_{\rm max}$ any artifacts due to the truncation are, at low enough energies, indistinguishable from other numerical errors due to discretization of~$r(x)$ or when subsequently solving the spectral problem numerically. From the general structure of $K_{2k}$ indicated in (\ref{eq:40b}) we see the differential equation for $r(x)$ is of order $2k_{\rm max}$. The boundary conditions at large values of $|x|$ are fixed by the leading~$\hbar$ solution $r_0(x)$ in (\ref{eq:18})
\begin{equation}
\lim_{|x|\rightarrow \infty}
\frac{\partial^n r(x)}{\partial x}
=
\lim_{|x|\rightarrow \infty}
\frac{\partial^n r_0(x)}{\partial x}\ ,
\end{equation}
with $n=0,1,\dots,k_{\rm max}-1$. Similar truncation procedures have been successfully applied to other double scaled matrix models in relation to JT gravity and supergravity~\cite{Johnson:2020mwi,Johnson:2020lns}.

\begin{figure}
\centering
\includegraphics[scale=0.70]{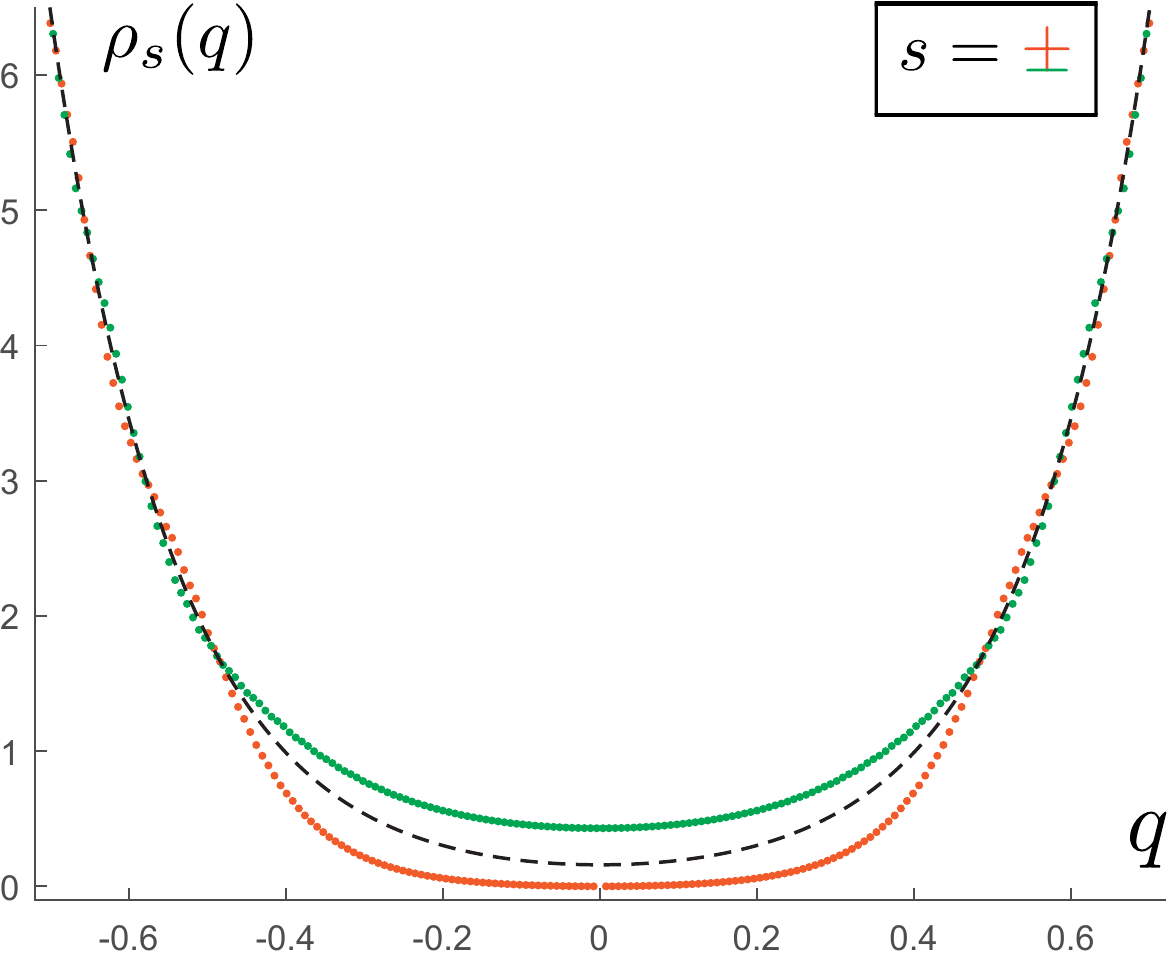}
\caption{Plot of each spin contribution $s=\pm$ to the spectral density for $\mu=1$, with the dashed line corresponding to the perturbative answer $\rho_0(q)/2$.}\label{fig:8.7}
\end{figure} 

In the diagram of Figure \ref{fig:2}, the blue solid line corresponds to the numerical solution for $r(x)$ with $\hbar=1$ and a truncation with $k_{\rm max}=6$. Working with higher truncations does not generate any substantial difference, at least for the numerical precision required for the computations here. The dashed line corresponds to $r_0(x)$ in (\ref{eq:18}), that is actually the exact solution to all orders in perturbation theory. The substantial difference between~$r(x)$ and~$r_0(x)$ for small values of $x$ is entirely due to non-perturbative effects.

Using this solution we can compute $r(x)^2-s\hbar r'(x)$, build the operator $\mathcal{H}_s$ and numerically compute the corresponding eigenfunctions $\psi_{\mathcal{E},s}(x)$, similarly as previously done in \cite{Johnson:2019eik,Johnson:2020heh,Johnson:2020exp,Johnson:2020lns}. Although the eigenfunctions are not normalizable, the normalization constant is fixed from the procedure explained in appendix \ref{zapp:3}. In this way, we can use (\ref{eq:38}) and compute the full spectral density $\rho(q)$ that is relevant for JT supergravity, including all perturbative and non-perturbative contributions. For each $s=\pm$ sector we plot the spectral density in Figure~\ref{fig:8.7}. Comparing with the perturbative answer shown in the dashed line, we observe non-perturbative corrections are present and dominate the spectral density for each value of $s=\pm$ at low values of $q$. The actual spectral density of the matrix model $\rho(q)$ is obtained from summing the two $s=\pm$ contributions, shown in the plot of Figure~\ref{fig:7}. Comparing with the perturbative answer shown in the dashed line, we observe there are indeed small subleading non-perturbative corrections generated by contributions of the schematic form given in (\ref{eq:8.89}).

\begin{figure}
\centering
\includegraphics[scale=0.65]{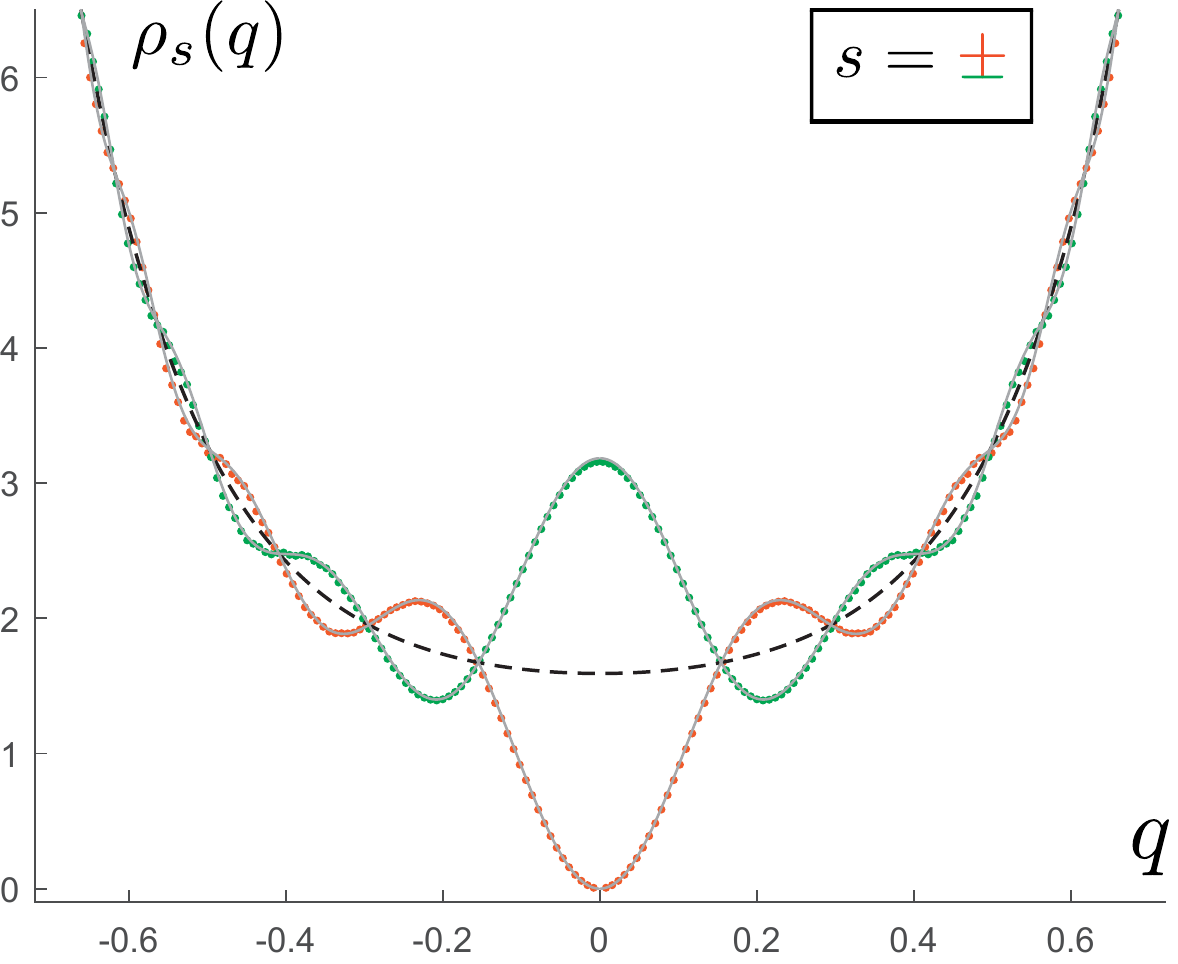}
\caption{Plot of each spin contribution $s=\pm$ to the spectral density for $\mu=10$. In this regime, there is good matching with the WKB expression in (\ref{eq:8.88}), shown as  the grey curves.}\label{fig:8}
\end{figure}

Since the difference between the perturbative and full spectral density in Figure \ref{fig:7} is quite small, we want to make sure it is not generated by numerical uncertainties but real physics. To do that, we can take smaller values of $\hbar$, where we expect non-perturbative effects to disappear. As noticed in \cite{Johnson:2020exp}, instead of taking $\hbar$ small, it is numerically more convenient to increase $\mu$. This has the same effect, as larger values of $\mu$ means the $x$ integral in (\ref{eq:38}) gets larger contribution from the classical region of the potential in $\mathcal{H}_s$, suppressing the quantum effects. In Figure \ref{fig:8} we plot each spin contribution $s=\pm$ to the spectral density when $\mu=10$, instead of $\mu=1$ as in (\ref{eq:25}). The grey curves in that diagram correspond to the WKB approximation in (\ref{eq:8.88}), where we find good agreement.\footnote{A similar matching for a different class of double scaled matrix models that are relevant for other JT supergravity theories was performed in ref.~\cite{Johnson:2020exp}.}

The full spectral density at $\mu=10$ is obtained by summing these contributions. Instead of doing that it is convenient to isolate the non-perturbative effects by plotting $\rho(q)-\rho_0(q)$, shown in Figure \ref{fig:8} for $\mu=10$ (red) and $\mu=1$ (blue). As expected, we observe the non-perturbative corrections that are present for $\mu=1$ are suppressed as~$\mu$ increases and the answer is dominated by the WKB approximation (\ref{eq:8.88}). This provides good evidence that  the non-perturbative effects seen for $\mu=1$ in Figure \ref{fig:7} are real physics and not  artifacts of the numerics.


Having the non-perturbatively computed eigenfunctions $\psi_{\mathcal{E},s}(x)$ allows for the  computation of  many other observables of interest, including all perturbative and non-perturbative corrections. A quantity of particular interest is the spectral form factor, defined as the two-point correlator involving  two asymptotic boundaries, written as a sum of disconnected and connected pieces:
\begin{equation}
\langle Z(\beta+it)Z(\beta-it) \rangle
\equiv Z(\beta+it)Z(\beta-it) +Z(\beta+it,\beta-it)\ ,
\end{equation}
where the two boundary lengths $\beta_1$ and $\beta_2$ have been analytically continued to $\beta{+}it$ and $\beta{-}it$. This quantity is useful for diagnosing certain  universal aspects of quantum chaotic behavior \cite{Cotler:2016fpe,Guhr:1997ve,Liu:2018hlr}. Random matrix theory predicts the spectral form factor to exhibit   ``dip'', ``ramp" and ``plateau" behavior,  describing the interaction of eigenvalues. In the context of the JT gravity/SYK correspondence, these  features of SYK describe aspects of the information scrambling properties of black holes~\cite{Cotler:2016fpe}.

\begin{figure}
\centering
\includegraphics[scale=0.63]{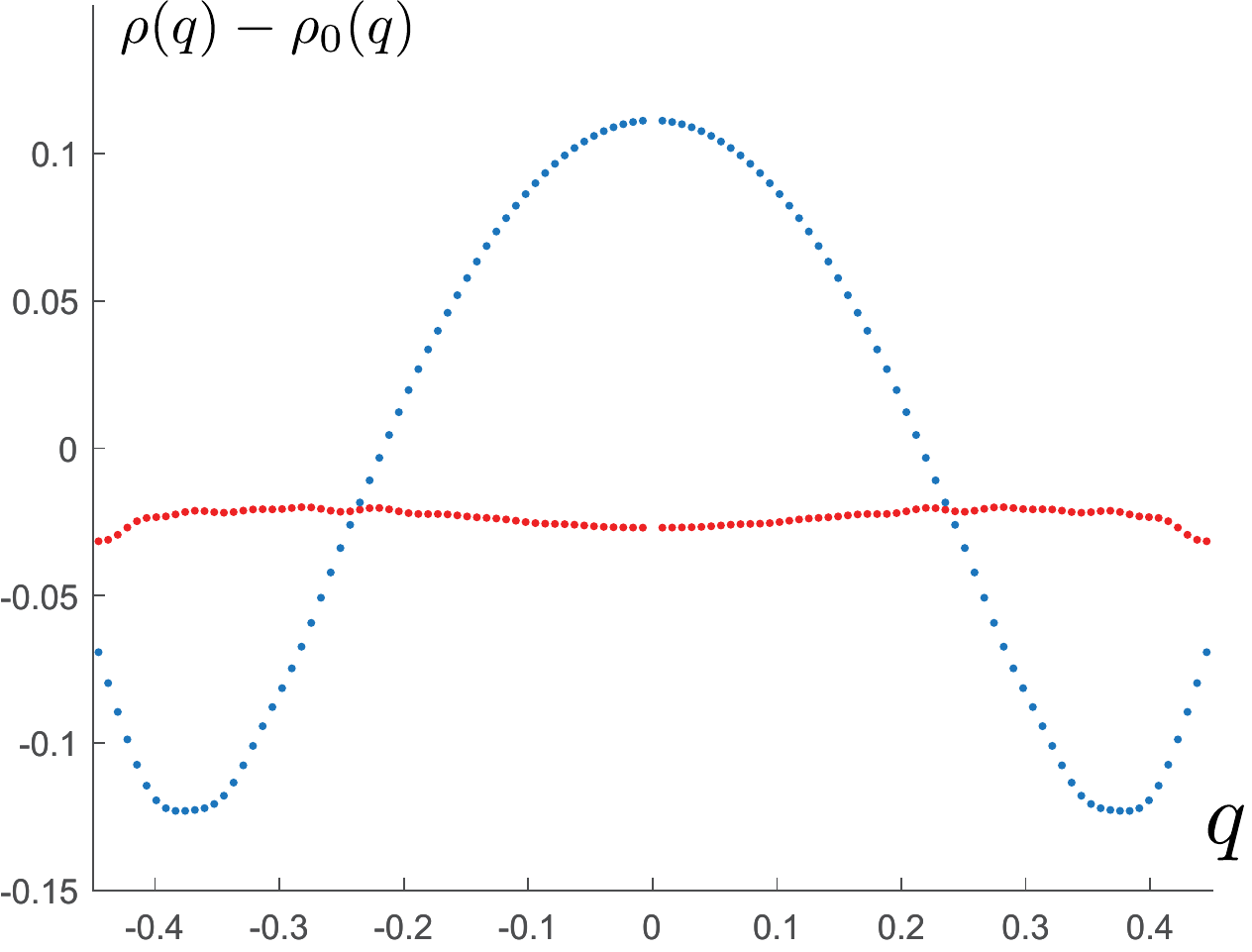}
\caption{Difference between the full spectral density $\rho(q)$ and the perturbative result $\rho_0(q)$ for $\mu=1$ (blue) and $\mu=10$ (red). We explicitly observe how the non-perturbative corrections are supressed as the value of $\mu$ increase.}
\end{figure}

\begin{figure}
\centering
\includegraphics[scale=0.46]{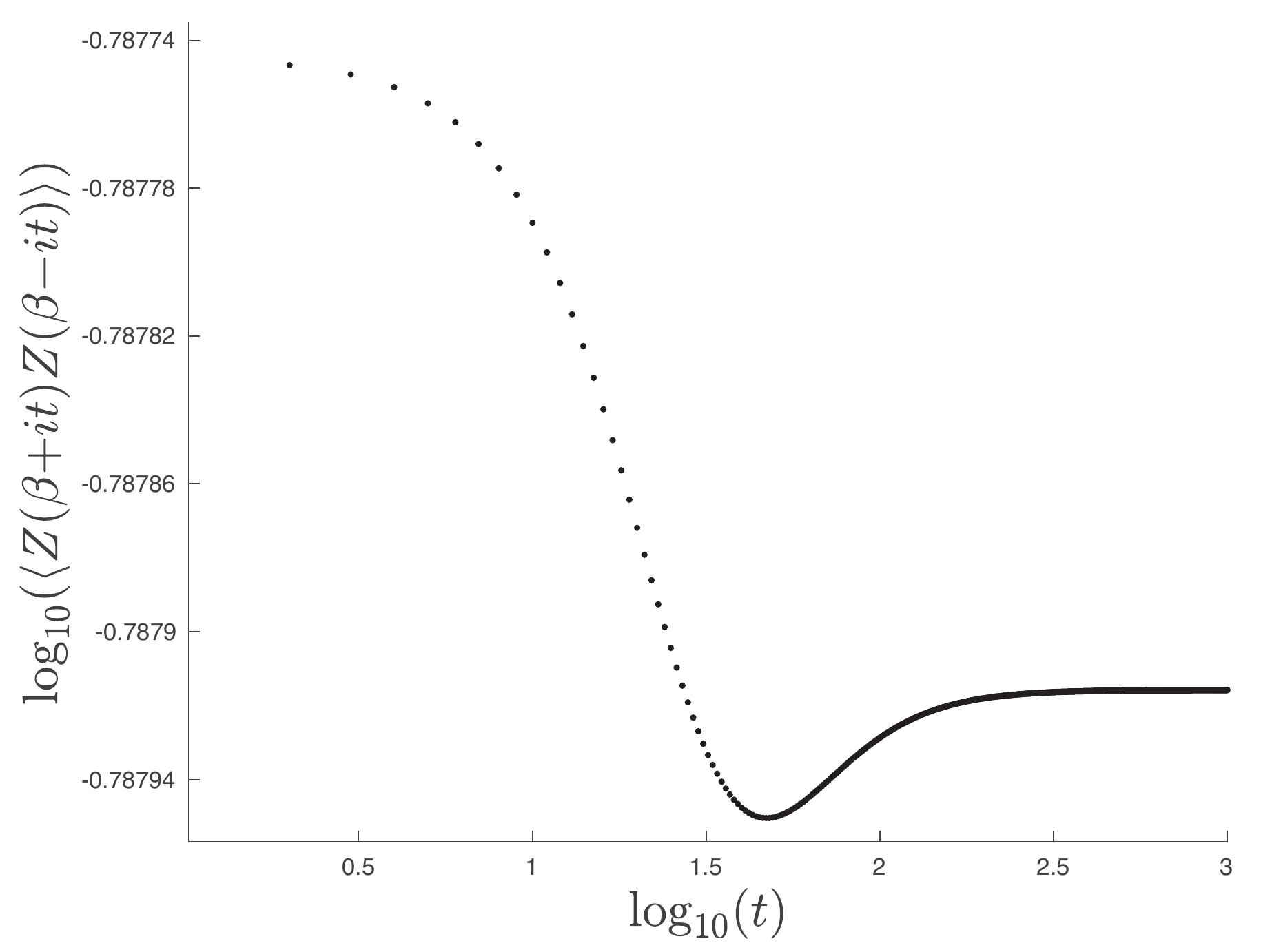}
\caption{Full non-perturbative spectral form factor {\it vs.} time~$t$, computed from the matrix model side with $\beta{=}50$ and $\hbar{=}1$, using (\ref{eq:8}), (\ref{eq:27}), and the methods described in refs.~\cite{Johnson:2020exp,Johnson:2020mwi}.}\label{fig:6}
\end{figure}

Using the identification in equation~(\ref{eq:gravMMequiv}), the spectral form factor can be computed from the matrix model side by inserting two operators ${\rm Tr}\,e^{-\beta Q^2}$ and performing the analytic continuation. For other JT supergravity theories this has been recently computed in refs.~\cite{Johnson:2020exp,Johnson:2020mwi} (see equation (26) of ref.~\cite{Johnson:2020exp}). Using the eigenfunctions $\psi_{\mathcal{E},s}(x)$ relevant to this case, we compute an example and obtain the result shown in Figure~\ref{fig:6}, showing the expected features. As in the previous work, it is observed that non-perturbative effects again play a crucial role in developing the plateau, and (since $\hbar{=}1$ in the figure) rapidly take over and hide the usually linear part of the ramp. For brevity, we have omitted showing plots of the individual disconnected portions of the spectral form factor, but there is a great deal of similarity to the (0,2) case studied in ref.~\cite{Johnson:2020exp}, and so we refer the reader there for results and extended discussion. In particular, the initial slope of the fall-off of the disconnected piece is unity, as it should be for a JT supergravity (this follows from the leading  small $\beta$ dependence of $Z(\beta)$). Also,  following from the fact that  the spectral density $\rho(E)$ diverges (even after non-perturbative corrections)
as $E{\to}0$, the connected piece takes much longer to saturate to the plateau value, just as happened for the (0,2) model.
%

\section{Closing remarks}
\label{sec:6}

We have shown how a particular JT supergravity theory---one  which sums only over orientable surfaces and equally weights spin structures in the Euclidean path integral---is given a complete non-perturbative description by a double-cut Hermitian matrix model ({\it i.e.} a $\boldsymbol{\beta}{=}2$ Dyson-Wigner ensemble). While generally the double-cut matrix model is characterized by two string equations associated to  the ZS integrable hierarchy, here the two string equations collapse to a single string equation,  the Painlev\'e~II/mKdV integrable hierarchy, also known from unitary matrix models. These equations are known to describe  type~0B minimal strings~\cite{Klebanov:2003wg}, and we showed a precise combination of such minimal models reproduces the JT supergravity. This is completely analogous to the work of refs.~\cite{Johnson:2020heh,Johnson:2020exp,Johnson:2020mwi},  which produced similar results for the Altland-Zirnbauer $(\boldsymbol{\alpha},\boldsymbol{\beta}){=}(\lbrace 0,1,2\rbrace,2)$ JT supergravities, showing how to  construct them  from type~0A minimal strings.

It was observed long ago in ref.~\cite{Dalley:1992br} that there is a non-perturbative map between the string equations of the complex matrix models (type 0A minimal strings) and those of the unitary matrix models (the type 0B minimal strings). It is in fact the  Miura transformation~\cite{doi:10.1063/1.1664700,doi:10.1063/1.1664701} that links the parent KdV and mKdV hierarchies: $u(x){=}r(x)^2{+}\hbar r'(x)$, where the $v(x)$ of ref.~\cite{Dalley:1992br} is   $r(x)$ here. (This is different from the map for the $k{=}1$ case noticed in ref.~\cite{Morris:1990bw}, or the connection more recently discussed for unscaled models in ref.~\cite{Mizoguchi:2004ne}.) In this sense,  the $(\boldsymbol{\alpha},\boldsymbol{\beta}{=}2)$ JT supergravities, non-perturbatively defined as  complex matrix models in refs.~\cite{Johnson:2020heh,Johnson:2020exp}, can straightforwardly be cast in terms of unitary matrix model language.\footnote{In fact, this was done later   in ref.~\cite{Okuyama:2020qpm}.} This does {\it not} mean that such a recasting captures the Dyson-Wigner $\boldsymbol{\beta}{=}2$ JT supergravity discussed in this paper however. The key new ingredient, arising from  using the  loop operator in equation~(\ref{eq:8}), is that for the $\boldsymbol{\beta}{=}2$ case, the physics comes from summing the results of the two separate Schr\"odinger systems with potentials $r(x)^2\pm\hbar r^\prime(x)$, and {\it not} just a single copy of one or (equivalently) the other, which is appropriate for the other family of JT supergravities. It is a subtle, but crucial difference.  

In a sense,  from a non-perturbative (geometry-independent)  high-altitude perspective, the fundamental variable is really $r(x)$, and the two different (inequivalent) uses of it to define potentials amounts to  the two (inequivalent)  JT supergravity families. This perspective deserves further exploration, since this simple difference in choice  changes major features of the theory, such as whether it has non-orientable geometries (preserves ${\textsf T}$), respects $(-1)^{\textsf F}$, and so on. 

It is worth noting that there is a  language in which the two different uses of $r(x)$ are  both natural. The  integrable hierarchies in question have an underlying $sl(2,\mathbb{C})$ structure~\cite{Drinfeld:1984qv,Date:1982yeu,Jimbo:1983if,Segal:1985aga,KacWakimoto,Hollowood:1992xq} from which naturally arises two  ``$\tau$-functions'', $\tau_0$ and $\tau_1$. In terms of these, define $u_{0,1}{=}2 \hbar^2\partial^2_x\ln \tau_{0,1}$ where either choice $u_{0,1}{=}v^2{\pm}\hbar v'$ gives a KdV-type function $u(x)$, and recall that $v(x)$ is  the $r(x)$ of this paper. Their difference gives $v{=}\hbar\partial_x\ln(\tau_0/\tau_1)$ while their sum gives $v^2{=}\hbar^2\partial^2_x\ln (\tau_0\tau_1)$. Perhaps this $\tau$-function language can be aligned with the supergravity symmetries ${\textsf T}$ and $(-1)^{\textsf F}$.

There are a number of other interesting potential avenues for future work. For example, as discussed in ref.~\cite{Stanford:2019vob}, there are  several other JT supergravity theories that are ${\textsf T}$--invariant  (so they include non-orientable surfaces in the Euclidean path integral) and have not yet been studied non-perturbatively. For example, two of them are classified as   double scaled $\boldsymbol{\beta}\!=\!1,4$ Dyson-Wigner ensembles \cite{Harris:1990kc,Brezin:1990xr,Brezin:1990dk}, and ought to share many of the qualitative features discussed in this paper. In particular, the leading spectral density $\rho_{0}$ for these models  has a branch cut comprising the entire real line. It is natural to  conjecture (and we do) that these SJT models may be similarly described by  double-cut matrix models. It is a natural guess that the string equations for those systems will define a function analogous to $r(x)$ that has a regime where it vanishes to all orders in perturbation theory. The details will be different however, with important non--perturbative differences, since the underlying orthogonal polynomials (and hence the string equations) arising from double scaling will be different.

As already  noted, the larger system of equations describing double-cut Hermitian matrix models are from the Zakharov-Shabat hierarchy, where the function $\alpha(x)$ and the parameter $\eta$ (see Subsection~\ref{sec:double-scaling}) are turned on. In our matching to the JT supergravity we restricted to the symmetric sector and turned those off. However, ref.~\cite{Klebanov:2003wg} gave a type~0B minimal string interpretation to some of the more general solutions in terms of symmetry breaking R-R fluxes. It would be interesting to explore if those can be understood in terms of the JT supergravity.

Notably, the  KdV, mKdV,  and ZS hierarchies, are all naturally embedded in a larger hierarchy of differential equations called the dispersive water wave hierarchy (see ref.~\cite{1985CMaPh..99...51K} and references therein). It was noticed~\cite{Iyer:2010ss,Iyer:2010ex} that it is possible to derive string-equation-like ODEs from that larger system that have many properties of minimal string theories, suggesting a large interconnected web of minimal string theories of which  type~0A and type~0B are merely a small part. It is tantalising to imagine that (by using the procedures used here and elsewhere for building JT gravity models out of minimal strings) an interconnected web of JT gravity and supergravity theories can similarly be defined.

Lastly, it would be worthwhile to use complex matrix models and double-cut hermitian matrix models to describe  non-trivial deformations of  JT supergravity analogous to those described in refs.~\cite{Maxfield:2020ale,Witten:2020wvy},  and also to JT gravity in de Sitter space \cite{Maldacena:2019cbz,Cotler:2019nbi}. Non-perturbative insights into such deformations would be useful. Progress has already been made for deformed JT gravity \cite{Johnson:2020lns}, and is currently underway for JT supergravity \cite{toappear}.

\acknowledgements

CVJ and FR are partially supported by the DOE grant DE-SC0011687. AS is supported by the Simons Foundation through \emph{It from Qubit: Simons collaboration on Quantum Fields, Gravity, and Information}. CVJ  thanks Amelia for her patience and support, especially during the long lockdown due to the pandemic.

\appendix

\onecolumngrid
\section{Higher trace operators}
\label{zapp:1}

In this appendix we show how to compute the expectation value of higher trace operators in the double scaled multi-cut model. Starting from the single trace formula in equation~(\ref{eq:8}) one way of deriving the analogous higher trace formulas is to use the free fermion formalism \cite{Banks:1989df,Ginsparg:1993is}. While this formalism was originally developed for single cut matrix models, the effective quantum mechanical system that arises when computing observables is basically the same after some simple redefinitions. 

\vspace{5pt}

\noindent \textbf{Double trace operators:} Starting from equation~(\ref{eq:8}), we obtain the following formula for two insertions of ${\rm Tr}\,e^{-\beta Q^2}$ \cite{toappear}
\begin{equation}\label{eq:27}
\big\langle 
{\rm Tr}\big(e^{-\beta_1 Q^2}\big)
{\rm Tr}\big(e^{-\beta_2 Q^2}\big)
\big\rangle_c=
\sum_{s=\pm}
\int_{-\infty}^{\mu}dx_1
\int_{\mu}^{+\infty}dx_2
\braket{x_1|
e^{-\beta_1\mathcal{H}_s}
|x_2}
\braket{x_2|e^{-\beta_2\mathcal{H}_s}|x_1}\ .
\end{equation}
Using equation~(\ref{eq:28}), it is quite simple to evaluate its leading $\hbar$ behavior and find
\begin{equation}
\big\langle 
{\rm Tr}\big(e^{-\beta_1 Q^2}\big)
{\rm Tr}\big(e^{-\beta_2 Q^2}\big)
\big\rangle_c=
\frac{1}{2\pi\hbar^2\sqrt{\beta_1\beta_2}}
\int_{-\infty}^{\mu}dx_1
\int_{\mu}^{+\infty}dx_2
e^{-\beta_1 r_0(x_1)^2-\beta_2 r_0(x_2)^2-
\frac{\beta_1+\beta_2}{\beta_1\beta_2}
\left(\frac{x_1-x_2}{2\hbar}\right)^2}+
\mathcal{O}(1)\ .
\end{equation}
Changing the integration variables to $x_1=\hbar(w_1-w_2)+\mu$ and $x_2=\hbar(w_1+w_2)+\mu$ the integrals decouple after we expand the exponentials $e^{-\beta_i r_0(x_i)^2}\simeq e^{-\beta_i r_0(\mu)^2}$ to leading order in $\hbar$. The remaining $(w_1,w_2)$ integrals can be easily solved and we find
\begin{equation}
\begin{aligned}
\big\langle 
{\rm Tr}\big(e^{-\beta_1 Q^2}\big)
{\rm Tr}\big(e^{-\beta_2 Q^2}\big)
\big\rangle_c=
2
\frac{\sqrt{\beta_1\beta_2}}{2\pi\beta_T}
e^{-\beta_T r_0(\mu)^2}
+\mathcal{O}(\hbar^2)\ ,
\end{aligned}
\end{equation}
where we have defined $\beta_T=\sum_{i=1}^n\beta_i$, in this case with $n=2$. Using $r_0(\mu)=r_0(1)=0$, the leading behavior reproduces the supergravity result in equation~(\ref{eq:17}).

\vspace{5pt}

\noindent \textbf{Triple trace operators:} Let us now consider three insertions of the matrix operator ${\rm Tr}\,e^{-\beta Q^2}$. The general formula in this case can be obtain from the neat general expression derived in ref.~\cite{Okuyama:2018aij}, and written in equation (3.28) of ref.~\cite{Okuyama:2020ncd}. In our conventions, it is given by
\begin{equation}\label{eq:42}
\begin{aligned}
\big\langle 
{\rm Tr}\big(&e^{-\beta_1 Q^2}\big)
{\rm Tr}\big(e^{-\beta_2 Q^2}\big)
{\rm Tr}\big(e^{-\beta_3 Q^2}\big)
\big\rangle_c  =
\sum_{s=\pm}
\int_{-\infty}^{\mu}dx_1
\int_{\mu}^{+\infty}dx_2
\braket{x_1|
e^{-\beta_1\mathcal{H}_s}
|x_2}
 \\
\times 
&\left[
\int_{\mu}^{+\infty}dx_3
\braket{x_2|e^{-\beta_2\mathcal{H}_s}|x_3}
\braket{x_3|e^{-\beta_3\mathcal{H}_s}|x_1}
-
\int_{-\infty}^{\mu}dx_3
\braket{x_2|e^{-\beta_3\mathcal{H}_s}|x_3}
\braket{x_3|e^{-\beta_2\mathcal{H}_s}|x_1}
\right]
\ .
\end{aligned}
\end{equation}
To evaluate its leading $\hbar$ behavior, that we denote as $G_0(\beta_1,\beta_2,\beta_3)$, we set without loss of generality $\beta_2=\beta_3=\beta$, as the general answer is then obtained by requiring its symmetric under arbitrary exchanges of $\beta_i\leftrightarrow \beta_j$. Using (\ref{eq:28}) we find
\begin{equation}\label{eq:47}
\begin{split}
G_0(\beta_1,\beta,\beta)=
2
\frac{\sqrt{\beta_1\beta^2}}{2\pi^{3/2}}
\int_{-\infty}^{+\infty}\frac{dx_3}{4\beta_1\beta^2\hbar^3}
\,{\rm sign}(x_3-\mu)
&\int_{-\infty}^{\mu}\hspace{-2mm}dx_1 
\biggr(\int_{\mu}^{+\infty}\hspace{-2mm}dx_2\,
e^{-\beta_1r_0(x_1)^2-\frac{1}{\beta_1}(\frac{x_1-x_2}{2\hbar})^2}\\[4pt]
&\times e^{-\beta r_0(x_2)^2-\frac{1}{\beta}(\frac{x_2-x_3}{2\hbar})^2}
e^{-\beta r_0(x_3)^2-\frac{1}{\beta}(\frac{x_1-x_3}{2\hbar})^2}\biggr)\ .
\end{split}
\end{equation} 
Changing the integration variables to $x_i=\hbar \bar{x}_i+\mu$ we can use the following expansion in $\hbar$
\begin{equation}\label{eq:48}
\begin{split}
e^{-\beta_i r_0(x_i)^2}&=
e^{-\beta_ir_0(\mu)^2}\left[
1-\beta_i \left[(\hbar\partial_x) r_0(x)^2\right]_{x=\mu}\bar{x}_i+\mathcal{O}(\hbar^3)\right]\,.
\end{split}
\end{equation}
The first term contributes to (\ref{eq:47}) as an integral over the whole real line in $\bar{x}_3$ of the following function
\begin{equation}
\begin{split}
&F(\bar{x}_3)\equiv
{\rm sign}(\bar{x}_3)
\int_{-\infty}^{0}d\bar{x}_1 
\int_{0}^{+\infty}d\bar{x}_2
e^{
-\frac{(\bar{x}_1-\bar{x}_2)^2}{4\beta_1}
-\frac{(\bar{x}_2-\bar{x}_3)^2}{4\beta}
-\frac{(\bar{x}_1-\bar{x}_3)^2}{4\beta}}\ .
\end{split}
\end{equation} 
It is straightforward to show this is an odd function $F(\bar{x}_3)=-F(-\bar{x}_3)$, so that it vanishes when integrated over the whole real line. This means the leading contribution is given by the second term in (\ref{eq:48}), so that (\ref{eq:47}) can be written as
\begin{equation}\label{eq:49}
\begin{split}
G_0(\beta_1,\beta,\beta)&=
2
\frac{\sqrt{\beta_1\beta^2}}{2\pi^{3/2}(\beta_1+2\beta)}
\left[(\hbar \partial_x)
e^{-(\beta_1+2\beta)r_0(x)^2}
\right]_{x=\mu}
\left\lbrace
\frac{1}{4\beta_1\beta^2}
\sum_{i=1}^3 \beta_i I_i\right\rbrace\ ,
\end{split}
\end{equation} 
where we have defined
\begin{equation}
\begin{split}
I_i&=
\int_{-\infty}^{+\infty}d\bar{x}_3
\,{\rm sign}(\bar{x}_3)
\int_{-\infty}^{0}d\bar{x}_1
\int_{0}^{+\infty}d\bar{x}_2
\,
\bar{x}_i e^{
-\frac{ (\bar{x}_1-\bar{x}_2)^2 }{4\beta_1}
-\frac{(\bar{x}_2-\bar{x}_3)^2}{4\beta}
-\frac{(\bar{x}_1-\bar{x}_3)^2}{4\beta}}\ .
\end{split}
\end{equation}
To evaluate these triple integrals, we first apply a change of coordinates that decouples the exponents
\begin{equation}
\begin{cases}
\begin{aligned}
\,\,
3\bar{x}_1&=\sqrt{2\beta\beta_1}(2w_1+w_2+3w_3)
\ , \\
3\bar{x}_2&=\sqrt{2\beta\beta_1}(2w_1+w_2-3w_3) \ , \\
3\bar{x}_3&=2\sqrt{2\beta\beta_1}(w_1-w_2)\ ,
\end{aligned}
\end{cases}
\end{equation}
so that the integration region gets mapped to
\begin{equation}
\begin{cases}
\begin{aligned}
\,\,
\bar{x}_1
\in&(-\infty,0]
\quad \longrightarrow \quad 
2w_1\in[-w_2+3w_3,-w_2-3w_3]\ , \\
\bar{x}_2
\in&[0,+\infty)
\quad \longrightarrow \quad 
w_2\in(-\infty,+\infty)\ , \\
\bar{x}_3
\in& (-\infty,+\infty)
\quad \longrightarrow \quad 
w_3\in (-\infty,0]\,.
\end{aligned}
\end{cases}
\end{equation}
Applying this transformation we find
\begin{equation}
\begin{split}
I_i&=
\frac{\sqrt{2}}{3}
(4\beta\beta_1)^{3/2}
\int_{-\infty}^0dw_3
e^{-(\beta_1+2\beta)w_3^2}
\int_{-\infty}^{+\infty}dw_2
e^{-\beta_1w_2^2}\int_{\frac{-w_2+3w_3}{2}}^{\frac{-w_2-3w_3}{2}}dw_1
{\rm sign}(w_1-w_2)
\,\bar{x}_i(w_j)
\ ,
\end{split}
\end{equation}
where the prefactor comes from the Jacobian in the change of variables. Solving the integral as written in the $w_i$ variables is simpler, but still tedious. The final result in each case is given by
\begin{equation}
\begin{split}
&\frac{\beta_1 I_1}{4\beta_1 \beta^2}=\frac{\beta_1}{\beta_1+2\beta}-\frac{2\beta_1\beta}{\sqrt{\beta_1}(\beta_1+2\beta)^{3/2}}\tan^{-1}
\sqrt{\frac{\beta_1}{\beta_1+2\beta}}\ ,\\
&\frac{\beta I_2}{4\beta_1 \beta^2}=
\frac{\beta}{\beta_1+2\beta}
-\frac{2\beta^2}{\sqrt{\beta_1}(\beta_1+2\beta)^{3/2}}
\tan^{-1}\sqrt{
\frac{\beta_1}{\beta_1+2\beta}}\ ,\\
&\frac{\beta I_3}{4\beta_1 \beta^2}=
\frac{\beta}{\beta_1+2\beta}
+\frac{2\beta(\beta_1+\beta)}{\sqrt{\beta_1}(\beta_1+2\beta)^{3/2}}
\tan^{-1}\sqrt{
\frac{\beta_1}{\beta_1+2\beta}}\ .
\end{split}
\end{equation}
Summing these three terms as indicated in (\ref{eq:49}), we get a nice cancellation such that the factor between curly brackets in (\ref{eq:49}) is equal to one. 

This gives an explicit expression for $G_0(\beta_1,\beta,\beta)$. To get the result for arbitrary values of $\beta_i$ we note that the symmetry of the observable under arbitrary exchanges $\beta_i\leftrightarrow \beta_j$ uniquely fixes $(\beta_1+2\beta)\rightarrow (\beta_1+\beta_2+\beta_3)$ and $\beta_1\beta^2 \rightarrow\beta_1\beta_2\beta_3$. Putting everything together, we arrive at the final result in equation~(\ref{eq:43}).

\section{Higher genus perturbative corrections}
\label{zapp:2}

In this Appendix we compute perturbative $\hbar$ corrections to the one point function $\langle {\rm Tr}\,e^{-\beta Q^2}\rangle$ using (\ref{eq:8}), by calculating the subleading terms in (\ref{eq:28}). One way of doing this, is by using the expansion for the resolvent of the operator $\mathcal{H}_s=-(\hbar\partial_x)^2+r(x)^2-s\hbar r'(x)$, worked out long ago by Gel'fand and Dikii \cite{Gelfand:1975rn} and given by
\begin{equation}\label{eq:29}
\braket{x|\frac{1}{\mathcal{H}_s-\xi}|x}=
\frac{1}{\hbar}
\sum_{p=0}^{\infty}
\frac{1}{(-\xi)^{p+1/2}}
\frac{(2p-1)!\widetilde{R}_p[u_s(x)]}{(-4)^pp!(p-1)!}
\ ,
\end{equation}
where $\xi<0$ and we defined $u_s(x)=r(x)^2-s\hbar r'(x)$. The Gel'fand-Dikii functionals $\widetilde{R}_p[r(x)^2]$ are polynomials in $r(x)^2$ and its derivatives computed from the following recursion relation
\begin{equation}\label{eq:30}
\widetilde{R}_{p+1}=
\frac{2p+2}{2p+1}\left[
u_s(x)\widetilde{R}_p-\frac{\hbar^2}{4}
\widetilde{R}''_p-
\frac{1}{2}\int^x d\bar{x}\,u'_s(x)\widetilde{R}_p
\right]\ ,
\end{equation}
with $\widetilde{R}_0=1$. Our normalization of the functionals $\widetilde{R}_p$ is different from the one used in ref.~\cite{Gelfand:1975rn}, as we have defined things differently so that $\widetilde{R}_p=u_s(x)^{p}+\mathcal{O}(\hbar^2)$. Explicit expressions for the first few functionals are given in equation (10) of ref.~\cite{Gelfand:1975rn}. Applying an inverse Laplace transformation in $\xi$ to equation~(\ref{eq:29}) we obtain the analogous expansion for $\braket{x|e^{-\beta \mathcal{H}_s}|x}$
\begin{equation}\label{eq:31}
\braket{x|e^{-\beta \mathcal{Q}^2}|x}=
\frac{1}{2\hbar\sqrt{\pi \beta}}
\sum_{p=0}^{\infty}
\frac{(-\beta)^p}{p!}
\widetilde{R}_p[u_s(x)]\ .
\end{equation}

While this asymptotic series formula is valid to all orders in $\hbar$, we cannot exchange the infinite series with the $x$ integral in (\ref{eq:8}). Instead we must solve the series order by order in $\hbar$. To do so, we can use the recursion relation (\ref{eq:30}) to expand $\widetilde{R}_p$ in $\hbar$. The expansion becomes simpler when written in terms of $u(x)=r(x)^2$, so that we find
\begin{equation}
\begin{aligned}
\widetilde{R}_p[u_s]=&u_s^p-\frac{\hbar^2}{12}p(p-1)u_s^{p-3}\left[
2u_su_s''+(p-2)(u_s')^2
\right]+
\frac{\hbar^4}{1440}p(p-1)(p-2)u_s^{p	-6}\left[
24u_s^3u_s^{(4)}
+
48(p-3)u_s^2u_s'u_s'''\right.\\[8pt]
&+
36(p-3)u_s^2(u_s'')^2+
44(p-3)(p-4)u_s(u_s')^2u_s''+
5(p-3)(p-4)(p-5)(u_s')^4
\Big]+\mathcal{O}(\hbar^6)\ .
\end{aligned}
\end{equation}
Inserting this into (\ref{eq:31}) and solving the series order by order in $\hbar$ we find
\begin{equation}
\begin{split}
\braket{x|e^{-\beta \mathcal{H}_s}|x}&=
\frac{e^{-\beta u_s}}{2\hbar\sqrt{\pi \beta}}
\biggr[
1-\frac{\hbar^2\beta^2}{12}
\left(2u_s''-\beta(u_s')^2\right)\\
&-\frac{\hbar^4\beta^3}{1440}
\biggr(
24u_s^{(4)}
-48\beta u_s'u_s^{(3)}
-36\beta(u_s'')^2
+44\beta^2(u_s')^2u_s''-5\beta^3(u_s')^4
\biggr)+\mathcal{O}(\hbar^6)
\biggr]\ .
\end{split}
\end{equation}
The final expansion used in the main text (\ref{eq:32}) is obtained by replacing $u_s(x)=r(x)^2-s\hbar r'(x)$ with $r(x)$ in (\ref{eq:26}) and expanding in $\hbar$ one last time.

To check we did not miss any factor in this expansion, let us use it to compute higher $\hbar$ corrections for the toy model defined from $r(x)=\hbar c/x$. Using (\ref{eq:8}) we find
\begin{equation}
\begin{aligned}
\langle {\rm Tr}\,e^{-\beta Q^2} \rangle &=
\sum_{s=\pm}
\frac{1}{2\hbar\sqrt{\pi \beta}}
\int_{0}^{\mu}
dx
\left[
1-\frac{c(c+s)\beta \hbar^2}{x^2}
+
\frac{c(c+s)(c(c+s)-2)\beta^2\hbar^4}{2x^4}
+\mathcal{O}(\hbar^6)
\right] \\
&=\sum_{s=\pm}
\frac{\mu }{2\hbar\sqrt{\pi \beta}}
\left[
1+\frac{c(c+s)\beta\hbar^2}{\mu^2}
-\frac{c(c+s)(c(c+s)-2)\beta^2\hbar^4}{6\mu^4}
+\mathcal{O}(\hbar^6)
\right]\ ,
\end{aligned}
\end{equation}
where the integration region is given by $x\in[0,\mu]$ since it is for this range that $r(x)=\hbar c/x$ is well defined. There is a divergent contribution coming from the $x\rightarrow 0$ limit of the integral that we have disregarded. From this expression it is straightforward to compute the spectral density $\rho(q)$ and find
\begin{equation}
\rho(q)=
\sum_{s=\pm}
\frac{\mu}{2\pi \hbar}
\left[
1-\frac{c(c+s)\hbar^2}{2\mu^2q^2}
-\frac{c(c+s)(c(c+s)-2)\hbar^4}{8\mu^4q^4}+
\mathcal{O}(\hbar^6)
\right]\ .
\end{equation}
Comparing this with the first perturbative terms obtained from expanding the exact expression in (\ref{eq:51}), we find perfect agreement.

\section{Eigenfunctions normalization}
\label{zapp:3}

In this Appendix we show how to normalize the eigenfunctions $\mathcal{H}_s\psi_{\mathcal{E},s}(x)=\mathcal{E}\psi_{\mathcal{E},s}(x)$ from their classical $\hbar\rightarrow 0$ behavior. To do so, let us start by writing an expression for $\rho_0(q)$ from the first line in (\ref{eq:22}). After applying an inverse Laplace transform we obtain the following expression for $\rho_0(x)$ in terms of $r_0(x)$
\begin{equation}
\rho_0(q)=
\frac{|q|}{2\pi \hbar}
\int_{-\infty}^{\mu}
dx
\frac{\Theta[q^2-r_0(x)^2]}{\sqrt{q^2-r_0(x)^2}}\ ,
\end{equation}
which is precisely the perturbative term obtained from the WKB approximation (\ref{eq:123}). Comparing with the full expression for $\rho(q)$ in (\ref{eq:38}), we obtain the following condition satisfied by $|\psi_{\mathcal{E},s}(x)|^2$
\begin{equation}\label{eq:56}
\lim_{\hbar\rightarrow 0}
|\psi_{\mathcal{E},s}(x)|^2=
\frac{1}{2\pi \hbar\sqrt{\mathcal{E}}}
+{\rm oscillating}\ ,
\qquad \quad x>0\ ,
\end{equation}
where we have used that $r_0(x)=0$ for $x>0$. Since $\psi_{\mathcal{E},s}(x)$ for $x\rightarrow +\infty$ behaves like a free particle, we expect to have additional oscillating terms which average to zero. Using this condition we can unambiguously fix the normalization constant in the eigenfunctions $\psi_{\mathcal{E},s}(x)$. Let us see how this works for the toy model eigenfunctions (\ref{eq:55}). Computing the norm square and taking the classical limit we find
\begin{equation}
\lim_{\hbar\rightarrow 0}
|\psi_{\mathcal{E},s}(x)|^2= 
\frac{1}{2\pi \hbar \sqrt{\mathcal{E}}}-
\frac{1}{2\pi \hbar \sqrt{\mathcal{E}}}
\cos\left[
\frac{2x \sqrt{\mathcal{E}}}{\hbar}
-\frac{\pi}{2}(2c+s-1)
\right]
\ ,
\end{equation}
which is precisely the normalization required by equation~(\ref{eq:56}).



\twocolumngrid

\bibliography{SJT_doublecut,NP_super_JT_gravity,sample}

\begin{thebibliography}{106}%
\makeatletter
\providecommand \@ifxundefined [1]{%
 \@ifx{#1\undefined}
}%
\providecommand \@ifnum [1]{%
 \ifnum #1\expandafter \@firstoftwo
 \else \expandafter \@secondoftwo
 \fi
}%
\providecommand \@ifx [1]{%
 \ifx #1\expandafter \@firstoftwo
 \else \expandafter \@secondoftwo
 \fi
}%
\providecommand \natexlab [1]{#1}%
\providecommand \enquote  [1]{``#1''}%
\providecommand \bibnamefont  [1]{#1}%
\providecommand \bibfnamefont [1]{#1}%
\providecommand \citenamefont [1]{#1}%
\providecommand \href@noop [0]{\@secondoftwo}%
\providecommand \href [0]{\begingroup \@sanitize@url \@href}%
\providecommand \@href[1]{\@@startlink{#1}\@@href}%
\providecommand \@@href[1]{\endgroup#1\@@endlink}%
\providecommand \@sanitize@url [0]{\catcode `\\12\catcode `\$12\catcode
  `\&12\catcode `\#12\catcode `\^12\catcode `\_12\catcode `\%12\relax}%
\providecommand \@@startlink[1]{}%
\providecommand \@@endlink[0]{}%
\providecommand \url  [0]{\begingroup\@sanitize@url \@url }%
\providecommand \@url [1]{\endgroup\@href {#1}{\urlprefix }}%
\providecommand \urlprefix  [0]{URL }%
\providecommand \Eprint [0]{\href }%
\providecommand \doibase [0]{http://dx.doi.org/}%
\providecommand \selectlanguage [0]{\@gobble}%
\providecommand \bibinfo  [0]{\@secondoftwo}%
\providecommand \bibfield  [0]{\@secondoftwo}%
\providecommand \translation [1]{[#1]}%
\providecommand \BibitemOpen [0]{}%
\providecommand \bibitemStop [0]{}%
\providecommand \bibitemNoStop [0]{.\EOS\space}%
\providecommand \EOS [0]{\spacefactor3000\relax}%
\providecommand \BibitemShut  [1]{\csname bibitem#1\endcsname}%
\let\auto@bib@innerbib\@empty
\bibitem [{\citenamefont {Jackiw}(1985)}]{Jackiw:1984je}%
  \BibitemOpen
  \bibfield  {author} {\bibinfo {author} {\bibfnamefont {R.}~\bibnamefont
  {Jackiw}},\ }\href {\doibase 10.1016/0550-3213(85)90448-1} {\bibfield
  {journal} {\bibinfo  {journal} {Nucl. Phys. B}\ }\textbf {\bibinfo {volume}
  {252}},\ \bibinfo {pages} {343} (\bibinfo {year} {1985})}\BibitemShut
  {NoStop}%
\bibitem [{\citenamefont {Teitelboim}(1983)}]{Teitelboim:1983ux}%
  \BibitemOpen
  \bibfield  {author} {\bibinfo {author} {\bibfnamefont {C.}~\bibnamefont
  {Teitelboim}},\ }\href {\doibase 10.1016/0370-2693(83)90012-6} {\bibfield
  {journal} {\bibinfo  {journal} {Phys. Lett. B}\ }\textbf {\bibinfo {volume}
  {126}},\ \bibinfo {pages} {41} (\bibinfo {year} {1983})}\BibitemShut
  {NoStop}%
\bibitem [{\citenamefont {Achucarro}\ and\ \citenamefont
  {Ortiz}(1993)}]{Achucarro:1993fd}%
  \BibitemOpen
  \bibfield  {author} {\bibinfo {author} {\bibfnamefont {A.}~\bibnamefont
  {Achucarro}}\ and\ \bibinfo {author} {\bibfnamefont {M.~E.}\ \bibnamefont
  {Ortiz}},\ }\href {\doibase 10.1103/PhysRevD.48.3600} {\bibfield  {journal}
  {\bibinfo  {journal} {Phys. Rev. D}\ }\textbf {\bibinfo {volume} {48}},\
  \bibinfo {pages} {3600} (\bibinfo {year} {1993})},\ \Eprint
  {http://arxiv.org/abs/hep-th/9304068} {arXiv:hep-th/9304068} \BibitemShut
  {NoStop}%
\bibitem [{\citenamefont {Fabbri}\ \emph {et~al.}(2001)\citenamefont {Fabbri},
  \citenamefont {Navarro},\ and\ \citenamefont
  {Navarro-Salas}}]{Fabbri:2000xh}%
  \BibitemOpen
  \bibfield  {author} {\bibinfo {author} {\bibfnamefont {A.}~\bibnamefont
  {Fabbri}}, \bibinfo {author} {\bibfnamefont {D.}~\bibnamefont {Navarro}}, \
  and\ \bibinfo {author} {\bibfnamefont {J.}~\bibnamefont {Navarro-Salas}},\
  }\href {\doibase 10.1016/S0550-3213(00)00661-1} {\bibfield  {journal}
  {\bibinfo  {journal} {Nucl. Phys. B}\ }\textbf {\bibinfo {volume} {595}},\
  \bibinfo {pages} {381} (\bibinfo {year} {2001})},\ \Eprint
  {http://arxiv.org/abs/hep-th/0006035} {arXiv:hep-th/0006035} \BibitemShut
  {NoStop}%
\bibitem [{\citenamefont {Nayak}\ \emph {et~al.}(2018)\citenamefont {Nayak},
  \citenamefont {Shukla}, \citenamefont {Soni}, \citenamefont {Trivedi},\ and\
  \citenamefont {Vishal}}]{Nayak:2018qej}%
  \BibitemOpen
  \bibfield  {author} {\bibinfo {author} {\bibfnamefont {P.}~\bibnamefont
  {Nayak}}, \bibinfo {author} {\bibfnamefont {A.}~\bibnamefont {Shukla}},
  \bibinfo {author} {\bibfnamefont {R.~M.}\ \bibnamefont {Soni}}, \bibinfo
  {author} {\bibfnamefont {S.~P.}\ \bibnamefont {Trivedi}}, \ and\ \bibinfo
  {author} {\bibfnamefont {V.}~\bibnamefont {Vishal}},\ }\href {\doibase
  10.1007/JHEP09(2018)048} {\bibfield  {journal} {\bibinfo  {journal} {JHEP}\
  }\textbf {\bibinfo {volume} {09}},\ \bibinfo {pages} {048} (\bibinfo {year}
  {2018})},\ \Eprint {http://arxiv.org/abs/1802.09547} {arXiv:1802.09547
  [hep-th]} \BibitemShut {NoStop}%
\bibitem [{\citenamefont {Kolekar}\ and\ \citenamefont
  {Narayan}(2018)}]{Kolekar:2018sba}%
  \BibitemOpen
  \bibfield  {author} {\bibinfo {author} {\bibfnamefont {K.~S.}\ \bibnamefont
  {Kolekar}}\ and\ \bibinfo {author} {\bibfnamefont {K.}~\bibnamefont
  {Narayan}},\ }\href {\doibase 10.1103/PhysRevD.98.046012} {\bibfield
  {journal} {\bibinfo  {journal} {Phys. Rev. D}\ }\textbf {\bibinfo {volume}
  {98}},\ \bibinfo {pages} {046012} (\bibinfo {year} {2018})},\ \Eprint
  {http://arxiv.org/abs/1803.06827} {arXiv:1803.06827 [hep-th]} \BibitemShut
  {NoStop}%
\bibitem [{\citenamefont {Ghosh}\ \emph {et~al.}(2020)\citenamefont {Ghosh},
  \citenamefont {Maxfield},\ and\ \citenamefont {Turiaci}}]{Ghosh:2019rcj}%
  \BibitemOpen
  \bibfield  {author} {\bibinfo {author} {\bibfnamefont {A.}~\bibnamefont
  {Ghosh}}, \bibinfo {author} {\bibfnamefont {H.}~\bibnamefont {Maxfield}}, \
  and\ \bibinfo {author} {\bibfnamefont {G.~J.}\ \bibnamefont {Turiaci}},\
  }\href {\doibase 10.1007/JHEP05(2020)104} {\bibfield  {journal} {\bibinfo
  {journal} {JHEP}\ }\textbf {\bibinfo {volume} {05}},\ \bibinfo {pages} {104}
  (\bibinfo {year} {2020})},\ \Eprint {http://arxiv.org/abs/1912.07654}
  {arXiv:1912.07654 [hep-th]} \BibitemShut {NoStop}%
\bibitem [{\citenamefont {Sachdev}\ and\ \citenamefont
  {Ye}(1993)}]{Sachdev:1992fk}%
  \BibitemOpen
  \bibfield  {author} {\bibinfo {author} {\bibfnamefont {S.}~\bibnamefont
  {Sachdev}}\ and\ \bibinfo {author} {\bibfnamefont {J.}~\bibnamefont {Ye}},\
  }\href {\doibase 10.1103/PhysRevLett.70.3339} {\bibfield  {journal} {\bibinfo
   {journal} {Phys. Rev. Lett.}\ }\textbf {\bibinfo {volume} {70}},\ \bibinfo
  {pages} {3339} (\bibinfo {year} {1993})},\ \Eprint
  {http://arxiv.org/abs/cond-mat/9212030} {arXiv:cond-mat/9212030} \BibitemShut
  {NoStop}%
\bibitem [{\citenamefont {Kitaev}(2015)}]{Kitaev:talks}%
  \BibitemOpen
  \bibfield  {author} {\bibinfo {author} {\bibfnamefont {A.}~\bibnamefont
  {Kitaev}},\ }\href@noop {} {\bibfield  {journal} {\bibinfo  {journal} {KITP
  seminars, April 7th and May 27th}\ } (\bibinfo {year} {2015})}\BibitemShut
  {NoStop}%
\bibitem [{\citenamefont {Jensen}(2016)}]{Jensen:2016pah}%
  \BibitemOpen
  \bibfield  {author} {\bibinfo {author} {\bibfnamefont {K.}~\bibnamefont
  {Jensen}},\ }\href {\doibase 10.1103/PhysRevLett.117.111601} {\bibfield
  {journal} {\bibinfo  {journal} {Phys. Rev. Lett.}\ }\textbf {\bibinfo
  {volume} {117}},\ \bibinfo {pages} {111601} (\bibinfo {year} {2016})},\
  \Eprint {http://arxiv.org/abs/1605.06098} {arXiv:1605.06098 [hep-th]}
  \BibitemShut {NoStop}%
\bibitem [{\citenamefont {Maldacena}\ and\ \citenamefont
  {Stanford}(2016)}]{Maldacena:2016hyu}%
  \BibitemOpen
  \bibfield  {author} {\bibinfo {author} {\bibfnamefont {J.}~\bibnamefont
  {Maldacena}}\ and\ \bibinfo {author} {\bibfnamefont {D.}~\bibnamefont
  {Stanford}},\ }\href {\doibase 10.1103/PhysRevD.94.106002} {\bibfield
  {journal} {\bibinfo  {journal} {Phys. Rev. D}\ }\textbf {\bibinfo {volume}
  {94}},\ \bibinfo {pages} {106002} (\bibinfo {year} {2016})},\ \Eprint
  {http://arxiv.org/abs/1604.07818} {arXiv:1604.07818 [hep-th]} \BibitemShut
  {NoStop}%
\bibitem [{\citenamefont {Maldacena}\ \emph {et~al.}(2016)\citenamefont
  {Maldacena}, \citenamefont {Stanford},\ and\ \citenamefont
  {Yang}}]{Maldacena:2016upp}%
  \BibitemOpen
  \bibfield  {author} {\bibinfo {author} {\bibfnamefont {J.}~\bibnamefont
  {Maldacena}}, \bibinfo {author} {\bibfnamefont {D.}~\bibnamefont {Stanford}},
  \ and\ \bibinfo {author} {\bibfnamefont {Z.}~\bibnamefont {Yang}},\ }\href
  {\doibase 10.1093/ptep/ptw124} {\bibfield  {journal} {\bibinfo  {journal}
  {PTEP}\ }\textbf {\bibinfo {volume} {2016}},\ \bibinfo {pages} {12C104}
  (\bibinfo {year} {2016})},\ \Eprint {http://arxiv.org/abs/1606.01857}
  {arXiv:1606.01857 [hep-th]} \BibitemShut {NoStop}%
\bibitem [{\citenamefont {Engels{\"o}y}\ \emph {et~al.}(2016)\citenamefont
  {Engels{\"o}y}, \citenamefont {Mertens},\ and\ \citenamefont
  {Verlinde}}]{Engelsoy:2016xyb}%
  \BibitemOpen
  \bibfield  {author} {\bibinfo {author} {\bibfnamefont {J.}~\bibnamefont
  {Engels{\"o}y}}, \bibinfo {author} {\bibfnamefont {T.~G.}\ \bibnamefont
  {Mertens}}, \ and\ \bibinfo {author} {\bibfnamefont {H.}~\bibnamefont
  {Verlinde}},\ }\href {\doibase 10.1007/JHEP07(2016)139} {\bibfield  {journal}
  {\bibinfo  {journal} {JHEP}\ }\textbf {\bibinfo {volume} {07}},\ \bibinfo
  {pages} {139} (\bibinfo {year} {2016})},\ \Eprint
  {http://arxiv.org/abs/1606.03438} {arXiv:1606.03438 [hep-th]} \BibitemShut
  {NoStop}%
\bibitem [{\citenamefont {Penington}(2020)}]{Penington:2019npb}%
  \BibitemOpen
  \bibfield  {author} {\bibinfo {author} {\bibfnamefont {G.}~\bibnamefont
  {Penington}},\ }\href {\doibase 10.1007/JHEP09(2020)002} {\bibfield
  {journal} {\bibinfo  {journal} {JHEP}\ }\textbf {\bibinfo {volume} {09}},\
  \bibinfo {pages} {002} (\bibinfo {year} {2020})},\ \Eprint
  {http://arxiv.org/abs/1905.08255} {arXiv:1905.08255 [hep-th]} \BibitemShut
  {NoStop}%
\bibitem [{\citenamefont {Almheiri}\ \emph {et~al.}(2019)\citenamefont
  {Almheiri}, \citenamefont {Engelhardt}, \citenamefont {Marolf},\ and\
  \citenamefont {Maxfield}}]{Almheiri:2019psf}%
  \BibitemOpen
  \bibfield  {author} {\bibinfo {author} {\bibfnamefont {A.}~\bibnamefont
  {Almheiri}}, \bibinfo {author} {\bibfnamefont {N.}~\bibnamefont
  {Engelhardt}}, \bibinfo {author} {\bibfnamefont {D.}~\bibnamefont {Marolf}},
  \ and\ \bibinfo {author} {\bibfnamefont {H.}~\bibnamefont {Maxfield}},\
  }\href {\doibase 10.1007/JHEP12(2019)063} {\bibfield  {journal} {\bibinfo
  {journal} {JHEP}\ }\textbf {\bibinfo {volume} {12}},\ \bibinfo {pages} {063}
  (\bibinfo {year} {2019})},\ \Eprint {http://arxiv.org/abs/1905.08762}
  {arXiv:1905.08762 [hep-th]} \BibitemShut {NoStop}%
\bibitem [{\citenamefont {Almheiri}\ \emph {et~al.}(2020)\citenamefont
  {Almheiri}, \citenamefont {Hartman}, \citenamefont {Maldacena}, \citenamefont
  {Shaghoulian},\ and\ \citenamefont {Tajdini}}]{Almheiri:2019qdq}%
  \BibitemOpen
  \bibfield  {author} {\bibinfo {author} {\bibfnamefont {A.}~\bibnamefont
  {Almheiri}}, \bibinfo {author} {\bibfnamefont {T.}~\bibnamefont {Hartman}},
  \bibinfo {author} {\bibfnamefont {J.}~\bibnamefont {Maldacena}}, \bibinfo
  {author} {\bibfnamefont {E.}~\bibnamefont {Shaghoulian}}, \ and\ \bibinfo
  {author} {\bibfnamefont {A.}~\bibnamefont {Tajdini}},\ }\href {\doibase
  10.1007/JHEP05(2020)013} {\bibfield  {journal} {\bibinfo  {journal} {JHEP}\
  }\textbf {\bibinfo {volume} {05}},\ \bibinfo {pages} {013} (\bibinfo {year}
  {2020})},\ \Eprint {http://arxiv.org/abs/1911.12333} {arXiv:1911.12333
  [hep-th]} \BibitemShut {NoStop}%
\bibitem [{\citenamefont {Penington}\ \emph {et~al.}(2019)\citenamefont
  {Penington}, \citenamefont {Shenker}, \citenamefont {Stanford},\ and\
  \citenamefont {Yang}}]{Penington:2019kki}%
  \BibitemOpen
  \bibfield  {author} {\bibinfo {author} {\bibfnamefont {G.}~\bibnamefont
  {Penington}}, \bibinfo {author} {\bibfnamefont {S.~H.}\ \bibnamefont
  {Shenker}}, \bibinfo {author} {\bibfnamefont {D.}~\bibnamefont {Stanford}}, \
  and\ \bibinfo {author} {\bibfnamefont {Z.}~\bibnamefont {Yang}},\ }\href@noop
  {} {\  (\bibinfo {year} {2019})},\ \Eprint {http://arxiv.org/abs/1911.11977}
  {arXiv:1911.11977 [hep-th]} \BibitemShut {NoStop}%
\bibitem [{\citenamefont {Saad}\ \emph {et~al.}(2019)\citenamefont {Saad},
  \citenamefont {Shenker},\ and\ \citenamefont {Stanford}}]{Saad:2019lba}%
  \BibitemOpen
  \bibfield  {author} {\bibinfo {author} {\bibfnamefont {P.}~\bibnamefont
  {Saad}}, \bibinfo {author} {\bibfnamefont {S.~H.}\ \bibnamefont {Shenker}}, \
  and\ \bibinfo {author} {\bibfnamefont {D.}~\bibnamefont {Stanford}},\
  }\href@noop {} {\  (\bibinfo {year} {2019})},\ \Eprint
  {http://arxiv.org/abs/1903.11115} {arXiv:1903.11115 [hep-th]} \BibitemShut
  {NoStop}%
\bibitem [{\citenamefont {Brezin}\ and\ \citenamefont
  {Kazakov}(1990)}]{Brezin:1990rb}%
  \BibitemOpen
  \bibfield  {author} {\bibinfo {author} {\bibfnamefont {E.}~\bibnamefont
  {Brezin}}\ and\ \bibinfo {author} {\bibfnamefont {V.~A.}\ \bibnamefont
  {Kazakov}},\ }\href@noop {} {\bibfield  {journal} {\bibinfo  {journal} {Phys.
  Lett.}\ }\textbf {\bibinfo {volume} {B236}},\ \bibinfo {pages} {144}
  (\bibinfo {year} {1990})}\BibitemShut {NoStop}%
\bibitem [{\citenamefont {Douglas}\ and\ \citenamefont
  {Shenker}(1990)}]{Douglas:1990ve}%
  \BibitemOpen
  \bibfield  {author} {\bibinfo {author} {\bibfnamefont {M.~R.}\ \bibnamefont
  {Douglas}}\ and\ \bibinfo {author} {\bibfnamefont {S.~H.}\ \bibnamefont
  {Shenker}},\ }\href@noop {} {\bibfield  {journal} {\bibinfo  {journal} {Nucl.
  Phys.}\ }\textbf {\bibinfo {volume} {B335}},\ \bibinfo {pages} {635}
  (\bibinfo {year} {1990})}\BibitemShut {NoStop}%
\bibitem [{\citenamefont {Gross}\ and\ \citenamefont
  {Migdal}(1990{\natexlab{a}})}]{Gross:1990vs}%
  \BibitemOpen
  \bibfield  {author} {\bibinfo {author} {\bibfnamefont {D.~J.}\ \bibnamefont
  {Gross}}\ and\ \bibinfo {author} {\bibfnamefont {A.~A.}\ \bibnamefont
  {Migdal}},\ }\href@noop {} {\bibfield  {journal} {\bibinfo  {journal} {Phys.
  Rev. Lett.}\ }\textbf {\bibinfo {volume} {64}},\ \bibinfo {pages} {127}
  (\bibinfo {year} {1990}{\natexlab{a}})}\BibitemShut {NoStop}%
\bibitem [{\citenamefont {Gross}\ and\ \citenamefont
  {Migdal}(1990{\natexlab{b}})}]{Gross:1990aw}%
  \BibitemOpen
  \bibfield  {author} {\bibinfo {author} {\bibfnamefont {D.~J.}\ \bibnamefont
  {Gross}}\ and\ \bibinfo {author} {\bibfnamefont {A.~A.}\ \bibnamefont
  {Migdal}},\ }\href@noop {} {\bibfield  {journal} {\bibinfo  {journal} {Nucl.
  Phys.}\ }\textbf {\bibinfo {volume} {B340}},\ \bibinfo {pages} {333}
  (\bibinfo {year} {1990}{\natexlab{b}})}\BibitemShut {NoStop}%
\bibitem [{\citenamefont {Mirzakhani}(2006)}]{Mirzakhani:2006fta}%
  \BibitemOpen
  \bibfield  {author} {\bibinfo {author} {\bibfnamefont {M.}~\bibnamefont
  {Mirzakhani}},\ }\href {\doibase 10.1007/s00222-006-0013-2} {\bibfield
  {journal} {\bibinfo  {journal} {Invent. Math.}\ }\textbf {\bibinfo {volume}
  {167}},\ \bibinfo {pages} {179} (\bibinfo {year} {2006})}\BibitemShut
  {NoStop}%
\bibitem [{\citenamefont {Eynard}(2004)}]{Eynard:2004mh}%
  \BibitemOpen
  \bibfield  {author} {\bibinfo {author} {\bibfnamefont {B.}~\bibnamefont
  {Eynard}},\ }\href {\doibase 10.1088/1126-6708/2004/11/031} {\bibfield
  {journal} {\bibinfo  {journal} {JHEP}\ }\textbf {\bibinfo {volume} {11}},\
  \bibinfo {pages} {031} (\bibinfo {year} {2004})},\ \Eprint
  {http://arxiv.org/abs/hep-th/0407261} {arXiv:hep-th/0407261} \BibitemShut
  {NoStop}%
\bibitem [{\citenamefont {Eynard}\ and\ \citenamefont
  {Orantin}(2007)}]{Eynard:2007kz}%
  \BibitemOpen
  \bibfield  {author} {\bibinfo {author} {\bibfnamefont {B.}~\bibnamefont
  {Eynard}}\ and\ \bibinfo {author} {\bibfnamefont {N.}~\bibnamefont
  {Orantin}},\ }\href {\doibase 10.4310/CNTP.2007.v1.n2.a4} {\bibfield
  {journal} {\bibinfo  {journal} {Commun. Num. Theor. Phys.}\ }\textbf
  {\bibinfo {volume} {1}},\ \bibinfo {pages} {347} (\bibinfo {year} {2007})},\
  \Eprint {http://arxiv.org/abs/math-ph/0702045} {arXiv:math-ph/0702045}
  \BibitemShut {NoStop}%
\bibitem [{\citenamefont {Johnson}(2020{\natexlab{a}})}]{Johnson:2019eik}%
  \BibitemOpen
  \bibfield  {author} {\bibinfo {author} {\bibfnamefont {C.~V.}\ \bibnamefont
  {Johnson}},\ }\href {\doibase 10.1103/PhysRevD.101.106023} {\bibfield
  {journal} {\bibinfo  {journal} {Phys. Rev. D}\ }\textbf {\bibinfo {volume}
  {101}},\ \bibinfo {pages} {106023} (\bibinfo {year} {2020}{\natexlab{a}})},\
  \Eprint {http://arxiv.org/abs/1912.03637} {arXiv:1912.03637 [hep-th]}
  \BibitemShut {NoStop}%
\bibitem [{\citenamefont {Stanford}\ and\ \citenamefont
  {Witten}(2019)}]{Stanford:2019vob}%
  \BibitemOpen
  \bibfield  {author} {\bibinfo {author} {\bibfnamefont {D.}~\bibnamefont
  {Stanford}}\ and\ \bibinfo {author} {\bibfnamefont {E.}~\bibnamefont
  {Witten}},\ }\href@noop {} {\  (\bibinfo {year} {2019})},\ \Eprint
  {http://arxiv.org/abs/1907.03363} {arXiv:1907.03363 [hep-th]} \BibitemShut
  {NoStop}%
\bibitem [{\citenamefont {Dyson}(1962)}]{Dyson:1962es}%
  \BibitemOpen
  \bibfield  {author} {\bibinfo {author} {\bibfnamefont {F.}~\bibnamefont
  {Dyson}},\ }\href {\doibase 10.1063/1.1703773} {\bibfield  {journal}
  {\bibinfo  {journal} {J. Math. Phys.}\ }\textbf {\bibinfo {volume} {3}},\
  \bibinfo {pages} {140} (\bibinfo {year} {1962})}\BibitemShut {NoStop}%
\bibitem [{\citenamefont {Altland}\ and\ \citenamefont
  {Zirnbauer}(1997)}]{Altland:1997zz}%
  \BibitemOpen
  \bibfield  {author} {\bibinfo {author} {\bibfnamefont {A.}~\bibnamefont
  {Altland}}\ and\ \bibinfo {author} {\bibfnamefont {M.~R.}\ \bibnamefont
  {Zirnbauer}},\ }\href {\doibase 10.1103/PhysRevB.55.1142} {\bibfield
  {journal} {\bibinfo  {journal} {Phys. Rev. B}\ }\textbf {\bibinfo {volume}
  {55}},\ \bibinfo {pages} {1142} (\bibinfo {year} {1997})},\ \Eprint
  {http://arxiv.org/abs/cond-mat/9602137} {arXiv:cond-mat/9602137} \BibitemShut
  {NoStop}%
\bibitem [{\citenamefont {Fu}\ \emph {et~al.}(2017)\citenamefont {Fu},
  \citenamefont {Gaiotto}, \citenamefont {Maldacena},\ and\ \citenamefont
  {Sachdev}}]{Fu:2016vas}%
  \BibitemOpen
  \bibfield  {author} {\bibinfo {author} {\bibfnamefont {W.}~\bibnamefont
  {Fu}}, \bibinfo {author} {\bibfnamefont {D.}~\bibnamefont {Gaiotto}},
  \bibinfo {author} {\bibfnamefont {J.}~\bibnamefont {Maldacena}}, \ and\
  \bibinfo {author} {\bibfnamefont {S.}~\bibnamefont {Sachdev}},\ }\href
  {\doibase 10.1103/PhysRevD.95.026009} {\bibfield  {journal} {\bibinfo
  {journal} {Phys. Rev. D}\ }\textbf {\bibinfo {volume} {95}},\ \bibinfo
  {pages} {026009} (\bibinfo {year} {2017})},\ \bibinfo {note} {[Addendum:
  Phys.Rev.D 95, 069904 (2017)]},\ \Eprint {http://arxiv.org/abs/1610.08917}
  {arXiv:1610.08917 [hep-th]} \BibitemShut {NoStop}%
\bibitem [{\citenamefont {Stanford}\ and\ \citenamefont
  {Witten}(2017)}]{Stanford:2017thb}%
  \BibitemOpen
  \bibfield  {author} {\bibinfo {author} {\bibfnamefont {D.}~\bibnamefont
  {Stanford}}\ and\ \bibinfo {author} {\bibfnamefont {E.}~\bibnamefont
  {Witten}},\ }\href {\doibase 10.1007/JHEP10(2017)008} {\bibfield  {journal}
  {\bibinfo  {journal} {JHEP}\ }\textbf {\bibinfo {volume} {10}},\ \bibinfo
  {pages} {008} (\bibinfo {year} {2017})},\ \Eprint
  {http://arxiv.org/abs/1703.04612} {arXiv:1703.04612 [hep-th]} \BibitemShut
  {NoStop}%
\bibitem [{\citenamefont {Li}\ \emph {et~al.}(2017)\citenamefont {Li},
  \citenamefont {Liu}, \citenamefont {Xin},\ and\ \citenamefont
  {Zhou}}]{Li:2017hdt}%
  \BibitemOpen
  \bibfield  {author} {\bibinfo {author} {\bibfnamefont {T.}~\bibnamefont
  {Li}}, \bibinfo {author} {\bibfnamefont {J.}~\bibnamefont {Liu}}, \bibinfo
  {author} {\bibfnamefont {Y.}~\bibnamefont {Xin}}, \ and\ \bibinfo {author}
  {\bibfnamefont {Y.}~\bibnamefont {Zhou}},\ }\href {\doibase
  10.1007/JHEP06(2017)111} {\bibfield  {journal} {\bibinfo  {journal} {JHEP}\
  }\textbf {\bibinfo {volume} {06}},\ \bibinfo {pages} {111} (\bibinfo {year}
  {2017})},\ \Eprint {http://arxiv.org/abs/1702.01738} {arXiv:1702.01738
  [hep-th]} \BibitemShut {NoStop}%
\bibitem [{\citenamefont {Kanazawa}\ and\ \citenamefont
  {Wettig}(2017)}]{Kanazawa:2017dpd}%
  \BibitemOpen
  \bibfield  {author} {\bibinfo {author} {\bibfnamefont {T.}~\bibnamefont
  {Kanazawa}}\ and\ \bibinfo {author} {\bibfnamefont {T.}~\bibnamefont
  {Wettig}},\ }\href {\doibase 10.1007/JHEP09(2017)050} {\bibfield  {journal}
  {\bibinfo  {journal} {JHEP}\ }\textbf {\bibinfo {volume} {09}},\ \bibinfo
  {pages} {050} (\bibinfo {year} {2017})},\ \Eprint
  {http://arxiv.org/abs/1706.03044} {arXiv:1706.03044 [hep-th]} \BibitemShut
  {NoStop}%
\bibitem [{\citenamefont {Sun}\ and\ \citenamefont {Ye}(2019)}]{Sun:2019yqp}%
  \BibitemOpen
  \bibfield  {author} {\bibinfo {author} {\bibfnamefont {F.}~\bibnamefont
  {Sun}}\ and\ \bibinfo {author} {\bibfnamefont {J.}~\bibnamefont {Ye}},\
  }\href@noop {} {\  (\bibinfo {year} {2019})},\ \Eprint
  {http://arxiv.org/abs/1905.07694} {arXiv:1905.07694 [cond-mat.str-el]}
  \BibitemShut {NoStop}%
\bibitem [{\citenamefont {Forste}\ and\ \citenamefont
  {Golla}(2017)}]{Forste:2017kwy}%
  \BibitemOpen
  \bibfield  {author} {\bibinfo {author} {\bibfnamefont {S.}~\bibnamefont
  {Forste}}\ and\ \bibinfo {author} {\bibfnamefont {I.}~\bibnamefont {Golla}},\
  }\href {\doibase 10.1016/j.physletb.2017.05.039} {\bibfield  {journal}
  {\bibinfo  {journal} {Phys. Lett. B}\ }\textbf {\bibinfo {volume} {771}},\
  \bibinfo {pages} {157} (\bibinfo {year} {2017})},\ \Eprint
  {http://arxiv.org/abs/1703.10969} {arXiv:1703.10969 [hep-th]} \BibitemShut
  {NoStop}%
\bibitem [{\citenamefont {Murugan}\ \emph {et~al.}(2017)\citenamefont
  {Murugan}, \citenamefont {Stanford},\ and\ \citenamefont
  {Witten}}]{Murugan:2017eto}%
  \BibitemOpen
  \bibfield  {author} {\bibinfo {author} {\bibfnamefont {J.}~\bibnamefont
  {Murugan}}, \bibinfo {author} {\bibfnamefont {D.}~\bibnamefont {Stanford}}, \
  and\ \bibinfo {author} {\bibfnamefont {E.}~\bibnamefont {Witten}},\ }\href
  {\doibase 10.1007/JHEP08(2017)146} {\bibfield  {journal} {\bibinfo  {journal}
  {JHEP}\ }\textbf {\bibinfo {volume} {08}},\ \bibinfo {pages} {146} (\bibinfo
  {year} {2017})},\ \Eprint {http://arxiv.org/abs/1706.05362} {arXiv:1706.05362
  [hep-th]} \BibitemShut {NoStop}%
\bibitem [{\citenamefont {Johnson}(2020{\natexlab{b}})}]{Johnson:2020heh}%
  \BibitemOpen
  \bibfield  {author} {\bibinfo {author} {\bibfnamefont {C.~V.}\ \bibnamefont
  {Johnson}},\ }\href@noop {} {\  (\bibinfo {year} {2020}{\natexlab{b}})},\
  \Eprint {http://arxiv.org/abs/2005.01893} {arXiv:2005.01893 [hep-th]}
  \BibitemShut {NoStop}%
\bibitem [{\citenamefont {Johnson}(2020{\natexlab{c}})}]{Johnson:2020exp}%
  \BibitemOpen
  \bibfield  {author} {\bibinfo {author} {\bibfnamefont {C.~V.}\ \bibnamefont
  {Johnson}},\ }\href@noop {} {\  (\bibinfo {year} {2020}{\natexlab{c}})},\
  \Eprint {http://arxiv.org/abs/2006.10959} {arXiv:2006.10959 [hep-th]}
  \BibitemShut {NoStop}%
\bibitem [{\citenamefont {Johnson}(2020{\natexlab{d}})}]{Johnson:2020mwi}%
  \BibitemOpen
  \bibfield  {author} {\bibinfo {author} {\bibfnamefont {C.~V.}\ \bibnamefont
  {Johnson}},\ }\href@noop {} {\  (\bibinfo {year} {2020}{\natexlab{d}})},\
  \Eprint {http://arxiv.org/abs/2008.13120} {arXiv:2008.13120 [hep-th]}
  \BibitemShut {NoStop}%
\bibitem [{\citenamefont {Morris}(1991)}]{Morris:1991cq}%
  \BibitemOpen
  \bibfield  {author} {\bibinfo {author} {\bibfnamefont {T.~R.}\ \bibnamefont
  {Morris}},\ }\href@noop {} {\bibfield  {journal} {\bibinfo  {journal} {Nucl.
  Phys.}\ }\textbf {\bibinfo {volume} {B356}},\ \bibinfo {pages} {703}
  (\bibinfo {year} {1991})}\BibitemShut {NoStop}%
\bibitem [{\citenamefont {Dalley}\ \emph
  {et~al.}(1992{\natexlab{a}})\citenamefont {Dalley}, \citenamefont {Johnson},\
  and\ \citenamefont {Morris}}]{Dalley:1992qg}%
  \BibitemOpen
  \bibfield  {author} {\bibinfo {author} {\bibfnamefont {S.}~\bibnamefont
  {Dalley}}, \bibinfo {author} {\bibfnamefont {C.~V.}\ \bibnamefont {Johnson}},
  \ and\ \bibinfo {author} {\bibfnamefont {T.}~\bibnamefont {Morris}},\
  }\href@noop {} {\bibfield  {journal} {\bibinfo  {journal} {Nucl. Phys.}\
  }\textbf {\bibinfo {volume} {B368}},\ \bibinfo {pages} {625} (\bibinfo {year}
  {1992}{\natexlab{a}})}\BibitemShut {NoStop}%
\bibitem [{\citenamefont {Dalley}\ \emph
  {et~al.}(1992{\natexlab{b}})\citenamefont {Dalley}, \citenamefont {Johnson},
  \citenamefont {Morris},\ and\ \citenamefont {Watterstam}}]{Dalley:1992br}%
  \BibitemOpen
  \bibfield  {author} {\bibinfo {author} {\bibfnamefont {S.}~\bibnamefont
  {Dalley}}, \bibinfo {author} {\bibfnamefont {C.}~\bibnamefont {Johnson}},
  \bibinfo {author} {\bibfnamefont {T.}~\bibnamefont {Morris}}, \ and\ \bibinfo
  {author} {\bibfnamefont {A.}~\bibnamefont {Watterstam}},\ }\href {\doibase
  10.1142/S0217732392002226} {\bibfield  {journal} {\bibinfo  {journal} {Mod.
  Phys. Lett. A}\ }\textbf {\bibinfo {volume} {7}},\ \bibinfo {pages} {2753}
  (\bibinfo {year} {1992}{\natexlab{b}})},\ \Eprint
  {http://arxiv.org/abs/hep-th/9206060} {arXiv:hep-th/9206060} \BibitemShut
  {NoStop}%
\bibitem [{\citenamefont {Dalley}\ \emph
  {et~al.}(1992{\natexlab{c}})\citenamefont {Dalley}, \citenamefont {Johnson},\
  and\ \citenamefont {Morris}}]{Dalley:1992vr}%
  \BibitemOpen
  \bibfield  {author} {\bibinfo {author} {\bibfnamefont {S.}~\bibnamefont
  {Dalley}}, \bibinfo {author} {\bibfnamefont {C.~V.}\ \bibnamefont {Johnson}},
  \ and\ \bibinfo {author} {\bibfnamefont {T.}~\bibnamefont {Morris}},\
  }\href@noop {} {\bibfield  {journal} {\bibinfo  {journal} {Nucl. Phys.}\
  }\textbf {\bibinfo {volume} {B368}},\ \bibinfo {pages} {655} (\bibinfo {year}
  {1992}{\natexlab{c}})}\BibitemShut {NoStop}%
\bibitem [{\citenamefont {Dalley}\ \emph
  {et~al.}(1992{\natexlab{d}})\citenamefont {Dalley}, \citenamefont {Johnson},\
  and\ \citenamefont {Morris}}]{Dalley:1992yi}%
  \BibitemOpen
  \bibfield  {author} {\bibinfo {author} {\bibfnamefont {S.}~\bibnamefont
  {Dalley}}, \bibinfo {author} {\bibfnamefont {C.~V.}\ \bibnamefont {Johnson}},
  \ and\ \bibinfo {author} {\bibfnamefont {T.}~\bibnamefont {Morris}},\
  }\href@noop {} {\bibfield  {journal} {\bibinfo  {journal} {Nucl. Phys. Proc.
  Suppl.}\ }\textbf {\bibinfo {volume} {25A}},\ \bibinfo {pages} {87} (\bibinfo
  {year} {1992}{\natexlab{d}})},\ \Eprint {http://arxiv.org/abs/hep-th/9108016}
  {hep-th/9108016} \BibitemShut {NoStop}%
\bibitem [{\citenamefont {Klebanov}\ \emph {et~al.}(2004)\citenamefont
  {Klebanov}, \citenamefont {Maldacena},\ and\ \citenamefont
  {Seiberg}}]{Klebanov:2003wg}%
  \BibitemOpen
  \bibfield  {author} {\bibinfo {author} {\bibfnamefont {I.~R.}\ \bibnamefont
  {Klebanov}}, \bibinfo {author} {\bibfnamefont {J.~M.}\ \bibnamefont
  {Maldacena}}, \ and\ \bibinfo {author} {\bibfnamefont {N.}~\bibnamefont
  {Seiberg}},\ }\href {\doibase 10.1007/s00220-004-1183-7} {\bibfield
  {journal} {\bibinfo  {journal} {Commun. Math. Phys.}\ }\textbf {\bibinfo
  {volume} {252}},\ \bibinfo {pages} {275} (\bibinfo {year} {2004})},\ \Eprint
  {http://arxiv.org/abs/hep-th/0309168} {arXiv:hep-th/0309168} \BibitemShut
  {NoStop}%
\bibitem [{\citenamefont {Mertens}\ and\ \citenamefont
  {Turiaci}(2020)}]{Mertens:2020hbs}%
  \BibitemOpen
  \bibfield  {author} {\bibinfo {author} {\bibfnamefont {T.~G.}\ \bibnamefont
  {Mertens}}\ and\ \bibinfo {author} {\bibfnamefont {G.~J.}\ \bibnamefont
  {Turiaci}},\ }\href@noop {} {\  (\bibinfo {year} {2020})},\ \Eprint
  {http://arxiv.org/abs/2006.07072} {arXiv:2006.07072 [hep-th]} \BibitemShut
  {NoStop}%
\bibitem [{\citenamefont {Mertens}\ and\ \citenamefont
  {Turiaci}(2019)}]{Mertens:2019tcm}%
  \BibitemOpen
  \bibfield  {author} {\bibinfo {author} {\bibfnamefont {T.~G.}\ \bibnamefont
  {Mertens}}\ and\ \bibinfo {author} {\bibfnamefont {G.~J.}\ \bibnamefont
  {Turiaci}},\ }\href {\doibase 10.1007/JHEP08(2019)127} {\bibfield  {journal}
  {\bibinfo  {journal} {JHEP}\ }\textbf {\bibinfo {volume} {08}},\ \bibinfo
  {pages} {127} (\bibinfo {year} {2019})},\ \Eprint
  {http://arxiv.org/abs/1904.05228} {arXiv:1904.05228 [hep-th]} \BibitemShut
  {NoStop}%
\bibitem [{\citenamefont {Turiaci}\ \emph {et~al.}(2020)\citenamefont
  {Turiaci}, \citenamefont {Usatyuk},\ and\ \citenamefont
  {Weng}}]{Turiaci:2020fjj}%
  \BibitemOpen
  \bibfield  {author} {\bibinfo {author} {\bibfnamefont {G.~J.}\ \bibnamefont
  {Turiaci}}, \bibinfo {author} {\bibfnamefont {M.}~\bibnamefont {Usatyuk}}, \
  and\ \bibinfo {author} {\bibfnamefont {W.~W.}\ \bibnamefont {Weng}},\
  }\href@noop {} {\  (\bibinfo {year} {2020})},\ \Eprint
  {http://arxiv.org/abs/2011.06038} {arXiv:2011.06038 [hep-th]} \BibitemShut
  {NoStop}%
\bibitem [{\citenamefont {Mertens}(2020)}]{Mertens:2020pfe}%
  \BibitemOpen
  \bibfield  {author} {\bibinfo {author} {\bibfnamefont {T.~G.}\ \bibnamefont
  {Mertens}},\ }\href@noop {} {\  (\bibinfo {year} {2020})},\ \Eprint
  {http://arxiv.org/abs/2007.00998} {arXiv:2007.00998 [hep-th]} \BibitemShut
  {NoStop}%
\bibitem [{\citenamefont {Betzios}\ and\ \citenamefont
  {Papadoulaki}(2020)}]{Betzios:2020nry}%
  \BibitemOpen
  \bibfield  {author} {\bibinfo {author} {\bibfnamefont {P.}~\bibnamefont
  {Betzios}}\ and\ \bibinfo {author} {\bibfnamefont {O.}~\bibnamefont
  {Papadoulaki}},\ }\href@noop {} {\  (\bibinfo {year} {2020})},\ \Eprint
  {http://arxiv.org/abs/2004.00002} {arXiv:2004.00002 [hep-th]} \BibitemShut
  {NoStop}%
\bibitem [{\citenamefont {Polchinski}()}]{Polchinski:1998rr}%
  \BibitemOpen
  \bibfield  {author} {\bibinfo {author} {\bibfnamefont {J.}~\bibnamefont
  {Polchinski}},\ }\href@noop {} {\ }\bibinfo {note} {Cambridge, UK: Univ. Pr.
  (1998) 531 p}\BibitemShut {NoStop}%
\bibitem [{\citenamefont {Ginsparg}\ and\ \citenamefont
  {Moore}(1993)}]{Ginsparg:1993is}%
  \BibitemOpen
  \bibfield  {author} {\bibinfo {author} {\bibfnamefont {P.~H.}\ \bibnamefont
  {Ginsparg}}\ and\ \bibinfo {author} {\bibfnamefont {G.~W.}\ \bibnamefont
  {Moore}},\ }in\ \href@noop {} {\emph {\bibinfo {booktitle} {{Theoretical
  Advanced Study Institute (TASI 92): From Black Holes and Strings to
  Particles}}}}\ (\bibinfo {year} {1993})\ pp.\ \bibinfo {pages} {277--469},\
  \Eprint {http://arxiv.org/abs/hep-th/9304011} {arXiv:hep-th/9304011}
  \BibitemShut {NoStop}%
\bibitem [{\citenamefont {Periwal}\ and\ \citenamefont
  {Shevitz}(1990{\natexlab{a}})}]{Periwal:1990gf}%
  \BibitemOpen
  \bibfield  {author} {\bibinfo {author} {\bibfnamefont {V.}~\bibnamefont
  {Periwal}}\ and\ \bibinfo {author} {\bibfnamefont {D.}~\bibnamefont
  {Shevitz}},\ }\href {\doibase 10.1103/PhysRevLett.64.1326} {\bibfield
  {journal} {\bibinfo  {journal} {Phys. Rev. Lett.}\ }\textbf {\bibinfo
  {volume} {64}},\ \bibinfo {pages} {1326} (\bibinfo {year}
  {1990}{\natexlab{a}})}\BibitemShut {NoStop}%
\bibitem [{\citenamefont {Periwal}\ and\ \citenamefont
  {Shevitz}(1990{\natexlab{b}})}]{Periwal:1990qb}%
  \BibitemOpen
  \bibfield  {author} {\bibinfo {author} {\bibfnamefont {V.}~\bibnamefont
  {Periwal}}\ and\ \bibinfo {author} {\bibfnamefont {D.}~\bibnamefont
  {Shevitz}},\ }\href {\doibase 10.1016/0550-3213(90)90676-5} {\bibfield
  {journal} {\bibinfo  {journal} {Nucl. Phys. B}\ }\textbf {\bibinfo {volume}
  {344}},\ \bibinfo {pages} {731} (\bibinfo {year}
  {1990}{\natexlab{b}})}\BibitemShut {NoStop}%
\bibitem [{\citenamefont {Nappi}(1990)}]{Nappi:1990bi}%
  \BibitemOpen
  \bibfield  {author} {\bibinfo {author} {\bibfnamefont {C.~R.}\ \bibnamefont
  {Nappi}},\ }\href {\doibase 10.1142/S0217732390003243} {\bibfield  {journal}
  {\bibinfo  {journal} {Mod. Phys. Lett. A}\ }\textbf {\bibinfo {volume} {5}},\
  \bibinfo {pages} {2773} (\bibinfo {year} {1990})}\BibitemShut {NoStop}%
\bibitem [{\citenamefont {Crnkovic}\ and\ \citenamefont
  {Moore}(1991)}]{Crnkovic:1990mr}%
  \BibitemOpen
  \bibfield  {author} {\bibinfo {author} {\bibfnamefont {C.}~\bibnamefont
  {Crnkovic}}\ and\ \bibinfo {author} {\bibfnamefont {G.~W.}\ \bibnamefont
  {Moore}},\ }\href {\doibase 10.1016/0370-2693(91)91900-G} {\bibfield
  {journal} {\bibinfo  {journal} {Phys. Lett. B}\ }\textbf {\bibinfo {volume}
  {257}},\ \bibinfo {pages} {322} (\bibinfo {year} {1991})}\BibitemShut
  {NoStop}%
\bibitem [{\citenamefont {Hollowood}\ \emph
  {et~al.}(1992{\natexlab{a}})\citenamefont {Hollowood}, \citenamefont
  {Miramontes}, \citenamefont {Pasquinucci},\ and\ \citenamefont
  {Nappi}}]{Hollowood:1991xq}%
  \BibitemOpen
  \bibfield  {author} {\bibinfo {author} {\bibfnamefont {T.~J.}\ \bibnamefont
  {Hollowood}}, \bibinfo {author} {\bibfnamefont {L.}~\bibnamefont
  {Miramontes}}, \bibinfo {author} {\bibfnamefont {A.}~\bibnamefont
  {Pasquinucci}}, \ and\ \bibinfo {author} {\bibfnamefont {C.}~\bibnamefont
  {Nappi}},\ }\href {\doibase 10.1016/0550-3213(92)90457-M} {\bibfield
  {journal} {\bibinfo  {journal} {Nucl. Phys. B}\ }\textbf {\bibinfo {volume}
  {373}},\ \bibinfo {pages} {247} (\bibinfo {year} {1992}{\natexlab{a}})},\
  \Eprint {http://arxiv.org/abs/hep-th/9109046} {arXiv:hep-th/9109046}
  \BibitemShut {NoStop}%
\bibitem [{\citenamefont {Crnkovic}\ \emph {et~al.}(1992)\citenamefont
  {Crnkovic}, \citenamefont {Douglas},\ and\ \citenamefont
  {Moore}}]{Crnkovic:1992wd}%
  \BibitemOpen
  \bibfield  {author} {\bibinfo {author} {\bibfnamefont {C.}~\bibnamefont
  {Crnkovic}}, \bibinfo {author} {\bibfnamefont {M.~R.}\ \bibnamefont
  {Douglas}}, \ and\ \bibinfo {author} {\bibfnamefont {G.~W.}\ \bibnamefont
  {Moore}},\ }\href {\doibase 10.1142/S0217751X92003483} {\bibfield  {journal}
  {\bibinfo  {journal} {Int. J. Mod. Phys. A}\ }\textbf {\bibinfo {volume}
  {7}},\ \bibinfo {pages} {7693} (\bibinfo {year} {1992})},\ \Eprint
  {http://arxiv.org/abs/hep-th/9108014} {arXiv:hep-th/9108014} \BibitemShut
  {NoStop}%
\bibitem [{\citenamefont {Douglas}\ \emph {et~al.}(1990)\citenamefont
  {Douglas}, \citenamefont {Seiberg},\ and\ \citenamefont
  {Shenker}}]{Douglas:1990xv}%
  \BibitemOpen
  \bibfield  {author} {\bibinfo {author} {\bibfnamefont {M.~R.}\ \bibnamefont
  {Douglas}}, \bibinfo {author} {\bibfnamefont {N.}~\bibnamefont {Seiberg}}, \
  and\ \bibinfo {author} {\bibfnamefont {S.~H.}\ \bibnamefont {Shenker}},\
  }\href {\doibase 10.1016/0370-2693(90)90333-2} {\bibfield  {journal}
  {\bibinfo  {journal} {Phys. Lett. B}\ }\textbf {\bibinfo {volume} {244}},\
  \bibinfo {pages} {381} (\bibinfo {year} {1990})}\BibitemShut {NoStop}%
\bibitem [{\citenamefont {Gross}\ and\ \citenamefont
  {Witten}(1980)}]{Gross:1980he}%
  \BibitemOpen
  \bibfield  {author} {\bibinfo {author} {\bibfnamefont {D.~J.}\ \bibnamefont
  {Gross}}\ and\ \bibinfo {author} {\bibfnamefont {E.}~\bibnamefont {Witten}},\
  }\href {\doibase 10.1103/PhysRevD.21.446} {\bibfield  {journal} {\bibinfo
  {journal} {Phys. Rev. D}\ }\textbf {\bibinfo {volume} {21}},\ \bibinfo
  {pages} {446} (\bibinfo {year} {1980})}\BibitemShut {NoStop}%
\bibitem [{\citenamefont {Cicuta}\ \emph {et~al.}(1986)\citenamefont {Cicuta},
  \citenamefont {Molinari},\ and\ \citenamefont {Montaldi}}]{Cicuta:1986pu}%
  \BibitemOpen
  \bibfield  {author} {\bibinfo {author} {\bibfnamefont {G.~M.}\ \bibnamefont
  {Cicuta}}, \bibinfo {author} {\bibfnamefont {L.}~\bibnamefont {Molinari}}, \
  and\ \bibinfo {author} {\bibfnamefont {E.}~\bibnamefont {Montaldi}},\ }\href
  {\doibase 10.1142/S021773238600018X} {\bibfield  {journal} {\bibinfo
  {journal} {Mod. Phys. Lett. A}\ }\textbf {\bibinfo {volume} {1}},\ \bibinfo
  {pages} {125} (\bibinfo {year} {1986})}\BibitemShut {NoStop}%
\bibitem [{\citenamefont {Akemann}\ \emph {et~al.}(2011)\citenamefont
  {Akemann}, \citenamefont {Baik},\ and\ \citenamefont
  {Di~Francesco}}]{Akemann:2011csh}%
  \BibitemOpen
  \bibfield  {author} {\bibinfo {author} {\bibfnamefont {G.}~\bibnamefont
  {Akemann}}, \bibinfo {author} {\bibfnamefont {J.}~\bibnamefont {Baik}}, \
  and\ \bibinfo {author} {\bibfnamefont {P.}~\bibnamefont {Di~Francesco}},\
  }\href@noop {} {\emph {\bibinfo {title} {{The Oxford Handbook of Random
  Matrix Theory}}}},\ Oxford Handbooks in Mathematics\ (\bibinfo  {publisher}
  {Oxford University Press},\ \bibinfo {year} {2011})\BibitemShut {NoStop}%
\bibitem [{\citenamefont {Di~Francesco}\ \emph {et~al.}(1995)\citenamefont
  {Di~Francesco}, \citenamefont {Ginsparg},\ and\ \citenamefont
  {Zinn-Justin}}]{DiFrancesco:1993cyw}%
  \BibitemOpen
  \bibfield  {author} {\bibinfo {author} {\bibfnamefont {P.}~\bibnamefont
  {Di~Francesco}}, \bibinfo {author} {\bibfnamefont {P.~H.}\ \bibnamefont
  {Ginsparg}}, \ and\ \bibinfo {author} {\bibfnamefont {J.}~\bibnamefont
  {Zinn-Justin}},\ }\href {\doibase 10.1016/0370-1573(94)00084-G} {\bibfield
  {journal} {\bibinfo  {journal} {Phys. Rept.}\ }\textbf {\bibinfo {volume}
  {254}},\ \bibinfo {pages} {1} (\bibinfo {year} {1995})},\ \Eprint
  {http://arxiv.org/abs/hep-th/9306153} {arXiv:hep-th/9306153} \BibitemShut
  {NoStop}%
\bibitem [{\citenamefont {Kazakov}(1989)}]{Kazakov:1989bc}%
  \BibitemOpen
  \bibfield  {author} {\bibinfo {author} {\bibfnamefont {V.~A.}\ \bibnamefont
  {Kazakov}},\ }\href@noop {} {\bibfield  {journal} {\bibinfo  {journal} {Mod.
  Phys. Lett.}\ }\textbf {\bibinfo {volume} {A4}},\ \bibinfo {pages} {2125}
  (\bibinfo {year} {1989})}\BibitemShut {NoStop}%
\bibitem [{\citenamefont {Eynard}\ \emph {et~al.}(2015)\citenamefont {Eynard},
  \citenamefont {Kimura},\ and\ \citenamefont {Ribault}}]{Eynard:2015aea}%
  \BibitemOpen
  \bibfield  {author} {\bibinfo {author} {\bibfnamefont {B.}~\bibnamefont
  {Eynard}}, \bibinfo {author} {\bibfnamefont {T.}~\bibnamefont {Kimura}}, \
  and\ \bibinfo {author} {\bibfnamefont {S.}~\bibnamefont {Ribault}},\
  }\href@noop {} {\  (\bibinfo {year} {2015})},\ \Eprint
  {http://arxiv.org/abs/1510.04430} {arXiv:1510.04430 [math-ph]} \BibitemShut
  {NoStop}%
\bibitem [{\citenamefont {Anninos}\ and\ \citenamefont
  {M\"uhlmann}(2020)}]{Anninos:2020ccj}%
  \BibitemOpen
  \bibfield  {author} {\bibinfo {author} {\bibfnamefont {D.}~\bibnamefont
  {Anninos}}\ and\ \bibinfo {author} {\bibfnamefont {B.}~\bibnamefont
  {M\"uhlmann}},\ }\href {\doibase 10.1088/1742-5468/aba499} {\bibfield
  {journal} {\bibinfo  {journal} {J. Stat. Mech.}\ }\textbf {\bibinfo {volume}
  {2008}},\ \bibinfo {pages} {083109} (\bibinfo {year} {2020})},\ \Eprint
  {http://arxiv.org/abs/2004.01171} {arXiv:2004.01171 [hep-th]} \BibitemShut
  {NoStop}%
\bibitem [{\citenamefont {Neuberger}(1991)}]{Neuberger:1990vu}%
  \BibitemOpen
  \bibfield  {author} {\bibinfo {author} {\bibfnamefont {H.}~\bibnamefont
  {Neuberger}},\ }\href {\doibase 10.1016/0550-3213(91)90104-6} {\bibfield
  {journal} {\bibinfo  {journal} {Nucl. Phys. B}\ }\textbf {\bibinfo {volume}
  {352}},\ \bibinfo {pages} {689} (\bibinfo {year} {1991})}\BibitemShut
  {NoStop}%
\bibitem [{\citenamefont {Brezin}\ \emph {et~al.}(1978)\citenamefont {Brezin},
  \citenamefont {Itzykson}, \citenamefont {Parisi},\ and\ \citenamefont
  {Zuber}}]{Brezin:1978sv}%
  \BibitemOpen
  \bibfield  {author} {\bibinfo {author} {\bibfnamefont {E.}~\bibnamefont
  {Brezin}}, \bibinfo {author} {\bibfnamefont {C.}~\bibnamefont {Itzykson}},
  \bibinfo {author} {\bibfnamefont {G.}~\bibnamefont {Parisi}}, \ and\ \bibinfo
  {author} {\bibfnamefont {J.~B.}\ \bibnamefont {Zuber}},\ }\href@noop {}
  {\bibfield  {journal} {\bibinfo  {journal} {Commun. Math. Phys.}\ }\textbf
  {\bibinfo {volume} {59}},\ \bibinfo {pages} {35} (\bibinfo {year}
  {1978})}\BibitemShut {NoStop}%
\bibitem [{\citenamefont {Bessis}\ \emph {et~al.}(1980)\citenamefont {Bessis},
  \citenamefont {Itzykson},\ and\ \citenamefont {Zuber}}]{Bessis:1980ss}%
  \BibitemOpen
  \bibfield  {author} {\bibinfo {author} {\bibfnamefont {D.}~\bibnamefont
  {Bessis}}, \bibinfo {author} {\bibfnamefont {C.}~\bibnamefont {Itzykson}}, \
  and\ \bibinfo {author} {\bibfnamefont {J.~B.}\ \bibnamefont {Zuber}},\
  }\href@noop {} {\bibfield  {journal} {\bibinfo  {journal} {Adv. Appl. Math.}\
  }\textbf {\bibinfo {volume} {1}},\ \bibinfo {pages} {109} (\bibinfo {year}
  {1980})}\BibitemShut {NoStop}%
\bibitem [{\citenamefont {Zakharov}\ and\ \citenamefont
  {Shabat}(1974)}]{zakharov1974scheme}%
  \BibitemOpen
  \bibfield  {author} {\bibinfo {author} {\bibfnamefont {V.}~\bibnamefont
  {Zakharov}}\ and\ \bibinfo {author} {\bibfnamefont {A.}~\bibnamefont
  {Shabat}},\ }\href@noop {} {\bibfield  {journal} {\bibinfo  {journal}
  {Funkts. Anal. Prilozhen.}\ }\textbf {\bibinfo {volume} {8}},\ \bibinfo
  {pages} {54} (\bibinfo {year} {1974})}\BibitemShut {NoStop}%
\bibitem [{toa()}]{toappear}%
  \BibitemOpen
  \href@noop {} {\bibinfo  {journal} {To appear}\ }\BibitemShut {NoStop}%
\bibitem [{\citenamefont {Banks}\ \emph
  {et~al.}(1990{\natexlab{a}})\citenamefont {Banks}, \citenamefont {Douglas},
  \citenamefont {Seiberg},\ and\ \citenamefont {Shenker}}]{Banks:1989df}%
  \BibitemOpen
\bibfield  {journal} {  }\bibfield  {author} {\bibinfo {author} {\bibfnamefont
  {T.}~\bibnamefont {Banks}}, \bibinfo {author} {\bibfnamefont {M.~R.}\
  \bibnamefont {Douglas}}, \bibinfo {author} {\bibfnamefont {N.}~\bibnamefont
  {Seiberg}}, \ and\ \bibinfo {author} {\bibfnamefont {S.~H.}\ \bibnamefont
  {Shenker}},\ }\href {\doibase 10.1016/0370-2693(90)91736-U} {\bibfield
  {journal} {\bibinfo  {journal} {Phys. Lett. B}\ }\textbf {\bibinfo {volume}
  {238}},\ \bibinfo {pages} {279} (\bibinfo {year}
  {1990}{\natexlab{a}})}\BibitemShut {NoStop}%
\bibitem [{\citenamefont {Johnson}\ and\ \citenamefont
  {Rosso}(2020)}]{Johnson:2020lns}%
  \BibitemOpen
  \bibfield  {author} {\bibinfo {author} {\bibfnamefont {C.~V.}\ \bibnamefont
  {Johnson}}\ and\ \bibinfo {author} {\bibfnamefont {F.}~\bibnamefont
  {Rosso}},\ }\href@noop {} {\  (\bibinfo {year} {2020})},\ \Eprint
  {http://arxiv.org/abs/2011.06026} {arXiv:2011.06026 [hep-th]} \BibitemShut
  {NoStop}%
\bibitem [{\citenamefont {Gross}\ and\ \citenamefont
  {Migdal}(1990{\natexlab{c}})}]{Gross:1989aw}%
  \BibitemOpen
  \bibfield  {author} {\bibinfo {author} {\bibfnamefont {D.~J.}\ \bibnamefont
  {Gross}}\ and\ \bibinfo {author} {\bibfnamefont {A.~A.}\ \bibnamefont
  {Migdal}},\ }\href {\doibase 10.1016/0550-3213(90)90450-R} {\bibfield
  {journal} {\bibinfo  {journal} {Nucl. Phys. B}\ }\textbf {\bibinfo {volume}
  {340}},\ \bibinfo {pages} {333} (\bibinfo {year}
  {1990}{\natexlab{c}})}\BibitemShut {NoStop}%
\bibitem [{\citenamefont {Ambjorn}\ \emph {et~al.}(1990)\citenamefont
  {Ambjorn}, \citenamefont {Jurkiewicz},\ and\ \citenamefont
  {Makeenko}}]{Ambjorn:1990ji}%
  \BibitemOpen
  \bibfield  {author} {\bibinfo {author} {\bibfnamefont {J.}~\bibnamefont
  {Ambjorn}}, \bibinfo {author} {\bibfnamefont {J.}~\bibnamefont {Jurkiewicz}},
  \ and\ \bibinfo {author} {\bibfnamefont {Y.}~\bibnamefont {Makeenko}},\
  }\href {\doibase 10.1016/0370-2693(90)90790-D} {\bibfield  {journal}
  {\bibinfo  {journal} {Phys. Lett. B}\ }\textbf {\bibinfo {volume} {251}},\
  \bibinfo {pages} {517} (\bibinfo {year} {1990})}\BibitemShut {NoStop}%
\bibitem [{\citenamefont {Moore}\ \emph {et~al.}(1991)\citenamefont {Moore},
  \citenamefont {Seiberg},\ and\ \citenamefont {Staudacher}}]{Moore:1991ir}%
  \BibitemOpen
  \bibfield  {author} {\bibinfo {author} {\bibfnamefont {G.~W.}\ \bibnamefont
  {Moore}}, \bibinfo {author} {\bibfnamefont {N.}~\bibnamefont {Seiberg}}, \
  and\ \bibinfo {author} {\bibfnamefont {M.}~\bibnamefont {Staudacher}},\
  }\href {\doibase 10.1016/0550-3213(91)90548-C} {\bibfield  {journal}
  {\bibinfo  {journal} {Nucl. Phys. B}\ }\textbf {\bibinfo {volume} {362}},\
  \bibinfo {pages} {665} (\bibinfo {year} {1991})}\BibitemShut {NoStop}%
\bibitem [{\citenamefont {Gelfand}\ and\ \citenamefont
  {Dikii}(1975)}]{Gelfand:1975rn}%
  \BibitemOpen
  \bibfield  {author} {\bibinfo {author} {\bibfnamefont {I.}~\bibnamefont
  {Gelfand}}\ and\ \bibinfo {author} {\bibfnamefont {L.}~\bibnamefont
  {Dikii}},\ }\href {\doibase 10.1070/RM1975v030n05ABEH001522} {\bibfield
  {journal} {\bibinfo  {journal} {Russ. Math. Surveys}\ }\textbf {\bibinfo
  {volume} {30}},\ \bibinfo {pages} {77} (\bibinfo {year} {1975})}\BibitemShut
  {NoStop}%
\bibitem [{\citenamefont {Banks}\ \emph
  {et~al.}(1990{\natexlab{b}})\citenamefont {Banks}, \citenamefont {Douglas},
  \citenamefont {Seiberg},\ and\ \citenamefont {Shenker}}]{Banks:1990df}%
  \BibitemOpen
  \bibfield  {author} {\bibinfo {author} {\bibfnamefont {T.}~\bibnamefont
  {Banks}}, \bibinfo {author} {\bibfnamefont {M.~R.}\ \bibnamefont {Douglas}},
  \bibinfo {author} {\bibfnamefont {N.}~\bibnamefont {Seiberg}}, \ and\
  \bibinfo {author} {\bibfnamefont {S.~H.}\ \bibnamefont {Shenker}},\
  }\href@noop {} {\bibfield  {journal} {\bibinfo  {journal} {Phys. Lett.}\
  }\textbf {\bibinfo {volume} {B238}},\ \bibinfo {pages} {279} (\bibinfo {year}
  {1990}{\natexlab{b}})}\BibitemShut {NoStop}%
\bibitem [{\citenamefont {Forrester}(2010)}]{ForresterBook}%
  \BibitemOpen
  \bibfield  {author} {\bibinfo {author} {\bibfnamefont {P.}~\bibnamefont
  {Forrester}},\ }\href {\doibase 10.1515/9781400835416} {\emph {\bibinfo
  {title} {Log-Gases and Random Matrices}}},\ London Mathematical Society
  monographs (LMS-34)\ (\bibinfo  {publisher} {Princeton University Press},\
  \bibinfo {year} {2010})\BibitemShut {NoStop}%
\bibitem [{\citenamefont {Andrews}\ \emph {et~al.}(1999)\citenamefont
  {Andrews}, \citenamefont {Askey},\ and\ \citenamefont
  {Roy}}]{andrews_askey_roy_1999}%
  \BibitemOpen
  \bibfield  {author} {\bibinfo {author} {\bibfnamefont {G.~E.}\ \bibnamefont
  {Andrews}}, \bibinfo {author} {\bibfnamefont {R.}~\bibnamefont {Askey}}, \
  and\ \bibinfo {author} {\bibfnamefont {R.}~\bibnamefont {Roy}},\ }\href
  {\doibase 10.1017/CBO9781107325937} {\emph {\bibinfo {title} {Special
  Functions}}},\ Encyclopedia of Mathematics and its Applications\ (\bibinfo
  {publisher} {Cambridge University Press},\ \bibinfo {year}
  {1999})\BibitemShut {NoStop}%
\bibitem [{\citenamefont {Cotler}\ \emph {et~al.}(2017)\citenamefont {Cotler},
  \citenamefont {Gur-Ari}, \citenamefont {Hanada}, \citenamefont {Polchinski},
  \citenamefont {Saad}, \citenamefont {Shenker}, \citenamefont {Stanford},
  \citenamefont {Streicher},\ and\ \citenamefont {Tezuka}}]{Cotler:2016fpe}%
  \BibitemOpen
  \bibfield  {author} {\bibinfo {author} {\bibfnamefont {J.~S.}\ \bibnamefont
  {Cotler}}, \bibinfo {author} {\bibfnamefont {G.}~\bibnamefont {Gur-Ari}},
  \bibinfo {author} {\bibfnamefont {M.}~\bibnamefont {Hanada}}, \bibinfo
  {author} {\bibfnamefont {J.}~\bibnamefont {Polchinski}}, \bibinfo {author}
  {\bibfnamefont {P.}~\bibnamefont {Saad}}, \bibinfo {author} {\bibfnamefont
  {S.~H.}\ \bibnamefont {Shenker}}, \bibinfo {author} {\bibfnamefont
  {D.}~\bibnamefont {Stanford}}, \bibinfo {author} {\bibfnamefont
  {A.}~\bibnamefont {Streicher}}, \ and\ \bibinfo {author} {\bibfnamefont
  {M.}~\bibnamefont {Tezuka}},\ }\href {\doibase 10.1007/JHEP05(2017)118}
  {\bibfield  {journal} {\bibinfo  {journal} {JHEP}\ }\textbf {\bibinfo
  {volume} {05}},\ \bibinfo {pages} {118} (\bibinfo {year} {2017})},\ \bibinfo
  {note} {[Erratum: JHEP 09, 002 (2018)]},\ \Eprint
  {http://arxiv.org/abs/1611.04650} {arXiv:1611.04650 [hep-th]} \BibitemShut
  {NoStop}%
\bibitem [{\citenamefont {Guhr}\ \emph {et~al.}(1998)\citenamefont {Guhr},
  \citenamefont {Muller-Groeling},\ and\ \citenamefont
  {Weidenmuller}}]{Guhr:1997ve}%
  \BibitemOpen
  \bibfield  {author} {\bibinfo {author} {\bibfnamefont {T.}~\bibnamefont
  {Guhr}}, \bibinfo {author} {\bibfnamefont {A.}~\bibnamefont
  {Muller-Groeling}}, \ and\ \bibinfo {author} {\bibfnamefont {H.~A.}\
  \bibnamefont {Weidenmuller}},\ }\href {\doibase
  10.1016/S0370-1573(97)00088-4} {\bibfield  {journal} {\bibinfo  {journal}
  {Phys. Rept.}\ }\textbf {\bibinfo {volume} {299}},\ \bibinfo {pages} {189}
  (\bibinfo {year} {1998})},\ \Eprint {http://arxiv.org/abs/cond-mat/9707301}
  {arXiv:cond-mat/9707301} \BibitemShut {NoStop}%
\bibitem [{\citenamefont {Liu}(2018)}]{Liu:2018hlr}%
  \BibitemOpen
  \bibfield  {author} {\bibinfo {author} {\bibfnamefont {J.}~\bibnamefont
  {Liu}},\ }\href {\doibase 10.1103/PhysRevD.98.086026} {\bibfield  {journal}
  {\bibinfo  {journal} {Phys. Rev. D}\ }\textbf {\bibinfo {volume} {98}},\
  \bibinfo {pages} {086026} (\bibinfo {year} {2018})},\ \Eprint
  {http://arxiv.org/abs/1806.05316} {arXiv:1806.05316 [hep-th]} \BibitemShut
  {NoStop}%
\bibitem [{\citenamefont {Miura}(1968)}]{doi:10.1063/1.1664700}%
  \BibitemOpen
  \bibfield  {author} {\bibinfo {author} {\bibfnamefont {R.~M.}\ \bibnamefont
  {Miura}},\ }\href {\doibase 10.1063/1.1664700} {\bibfield  {journal}
  {\bibinfo  {journal} {J. Math. Phys.}\ }\textbf {\bibinfo {volume} {9}},\
  \bibinfo {pages} {1202} (\bibinfo {year} {1968})}\BibitemShut {NoStop}%
\bibitem [{\citenamefont {Miura}\ \emph {et~al.}(1968)\citenamefont {Miura},
  \citenamefont {Gardner},\ and\ \citenamefont
  {Kruskal}}]{doi:10.1063/1.1664701}%
  \BibitemOpen
  \bibfield  {author} {\bibinfo {author} {\bibfnamefont {R.~M.}\ \bibnamefont
  {Miura}}, \bibinfo {author} {\bibfnamefont {C.~S.}\ \bibnamefont {Gardner}},
  \ and\ \bibinfo {author} {\bibfnamefont {M.~D.}\ \bibnamefont {Kruskal}},\
  }\href {\doibase 10.1063/1.1664701} {\bibfield  {journal} {\bibinfo
  {journal} {J. Math. Phys.}\ }\textbf {\bibinfo {volume} {9}},\ \bibinfo
  {pages} {1204} (\bibinfo {year} {1968})}\BibitemShut {NoStop}%
\bibitem [{\citenamefont {Morris}()}]{Morris:1990bw}%
  \BibitemOpen
  \bibfield  {author} {\bibinfo {author} {\bibfnamefont {T.~R.}\ \bibnamefont
  {Morris}},\ }\href@noop {} {\ }\bibinfo {note}
  {FERMILAB-PUB-90-136-T}\BibitemShut {NoStop}%
\bibitem [{\citenamefont {Mizoguchi}(2005)}]{Mizoguchi:2004ne}%
  \BibitemOpen
  \bibfield  {author} {\bibinfo {author} {\bibfnamefont {S.}~\bibnamefont
  {Mizoguchi}},\ }\href {\doibase 10.1016/j.nuclphysb.2005.03.035} {\bibfield
  {journal} {\bibinfo  {journal} {Nucl. Phys. B}\ }\textbf {\bibinfo {volume}
  {716}},\ \bibinfo {pages} {462} (\bibinfo {year} {2005})},\ \Eprint
  {http://arxiv.org/abs/hep-th/0411049} {arXiv:hep-th/0411049} \BibitemShut
  {NoStop}%
\bibitem [{\citenamefont {Okuyama}\ and\ \citenamefont
  {Sakai}(2020{\natexlab{a}})}]{Okuyama:2020qpm}%
  \BibitemOpen
  \bibfield  {author} {\bibinfo {author} {\bibfnamefont {K.}~\bibnamefont
  {Okuyama}}\ and\ \bibinfo {author} {\bibfnamefont {K.}~\bibnamefont
  {Sakai}},\ }\href {\doibase 10.1007/JHEP10(2020)160} {\bibfield  {journal}
  {\bibinfo  {journal} {JHEP}\ }\textbf {\bibinfo {volume} {10}},\ \bibinfo
  {pages} {160} (\bibinfo {year} {2020}{\natexlab{a}})},\ \Eprint
  {http://arxiv.org/abs/2007.09606} {arXiv:2007.09606 [hep-th]} \BibitemShut
  {NoStop}%
\bibitem [{\citenamefont {Drinfeld}\ and\ \citenamefont
  {Sokolov}(1984)}]{Drinfeld:1984qv}%
  \BibitemOpen
  \bibfield  {author} {\bibinfo {author} {\bibfnamefont {V.~G.}\ \bibnamefont
  {Drinfeld}}\ and\ \bibinfo {author} {\bibfnamefont {V.~V.}\ \bibnamefont
  {Sokolov}},\ }\href {\doibase 10.1007/BF02105860} {\bibfield  {journal}
  {\bibinfo  {journal} {J. Sov. Math.}\ }\textbf {\bibinfo {volume} {30}},\
  \bibinfo {pages} {1975} (\bibinfo {year} {1984})}\BibitemShut {NoStop}%
\bibitem [{\citenamefont {Date}\ \emph {et~al.}(1982)\citenamefont {Date},
  \citenamefont {Jimbo}, \citenamefont {Kashiwara},\ and\ \citenamefont
  {Miwa}}]{Date:1982yeu}%
  \BibitemOpen
  \bibfield  {author} {\bibinfo {author} {\bibfnamefont {E.}~\bibnamefont
  {Date}}, \bibinfo {author} {\bibfnamefont {M.}~\bibnamefont {Jimbo}},
  \bibinfo {author} {\bibfnamefont {M.}~\bibnamefont {Kashiwara}}, \ and\
  \bibinfo {author} {\bibfnamefont {T.}~\bibnamefont {Miwa}},\ }\href {\doibase
  10.2977/prims/1195183297} {\bibfield  {journal} {\bibinfo  {journal} {Publ.
  Res. Inst. Math. Sci. Kyoto}\ }\textbf {\bibinfo {volume} {18}},\ \bibinfo
  {pages} {1077} (\bibinfo {year} {1982})}\BibitemShut {NoStop}%
\bibitem [{\citenamefont {Jimbo}\ and\ \citenamefont
  {Miwa}(1983)}]{Jimbo:1983if}%
  \BibitemOpen
  \bibfield  {author} {\bibinfo {author} {\bibfnamefont {M.}~\bibnamefont
  {Jimbo}}\ and\ \bibinfo {author} {\bibfnamefont {T.}~\bibnamefont {Miwa}},\
  }\href {\doibase 10.2977/prims/1195182017} {\bibfield  {journal} {\bibinfo
  {journal} {Publ. Res. Inst. Math. Sci. Kyoto}\ }\textbf {\bibinfo {volume}
  {19}},\ \bibinfo {pages} {943} (\bibinfo {year} {1983})}\BibitemShut
  {NoStop}%
\bibitem [{\citenamefont {Segal}\ and\ \citenamefont
  {Wilson}(1985)}]{Segal:1985aga}%
  \BibitemOpen
  \bibfield  {author} {\bibinfo {author} {\bibfnamefont {G.}~\bibnamefont
  {Segal}}\ and\ \bibinfo {author} {\bibfnamefont {G.}~\bibnamefont {Wilson}},\
  }\href {\doibase 10.1007/BF02698802} {\bibfield  {journal} {\bibinfo
  {journal} {Inst. Hautes Etudes Sci. Publ. Math.}\ }\textbf {\bibinfo {volume}
  {61}},\ \bibinfo {pages} {5} (\bibinfo {year} {1985})}\BibitemShut {NoStop}%
\bibitem [{\citenamefont {Kac}\ and\ \citenamefont
  {Wakimoto}(1989)}]{KacWakimoto}%
  \BibitemOpen
  \bibfield  {author} {\bibinfo {author} {\bibfnamefont {V.~G.}\ \bibnamefont
  {Kac}}\ and\ \bibinfo {author} {\bibfnamefont {M.}~\bibnamefont {Wakimoto}},\
  }\href {\doibase 10.1103/PhysRevD.101.106023} {\bibfield  {journal} {\bibinfo
   {journal} {Proc. Symp. Pure Math.}\ }\textbf {\bibinfo {volume} {49}},\
  \bibinfo {pages} {191} (\bibinfo {year} {1989})}\BibitemShut {NoStop}%
\bibitem [{\citenamefont {Hollowood}\ \emph
  {et~al.}(1992{\natexlab{b}})\citenamefont {Hollowood}, \citenamefont
  {Miramontes}, \citenamefont {Pasquinucci},\ and\ \citenamefont
  {Nappi}}]{Hollowood:1992xq}%
  \BibitemOpen
  \bibfield  {author} {\bibinfo {author} {\bibfnamefont {T.~J.}\ \bibnamefont
  {Hollowood}}, \bibinfo {author} {\bibfnamefont {L.}~\bibnamefont
  {Miramontes}}, \bibinfo {author} {\bibfnamefont {A.}~\bibnamefont
  {Pasquinucci}}, \ and\ \bibinfo {author} {\bibfnamefont {C.}~\bibnamefont
  {Nappi}},\ }\href@noop {} {\bibfield  {journal} {\bibinfo  {journal} {Nucl.
  Phys.}\ }\textbf {\bibinfo {volume} {B373}},\ \bibinfo {pages} {247}
  (\bibinfo {year} {1992}{\natexlab{b}})},\ \Eprint
  {http://arxiv.org/abs/hep-th/9109046} {hep-th/9109046} \BibitemShut {NoStop}%
\bibitem [{\citenamefont {Harris}\ and\ \citenamefont
  {Martinec}(1990)}]{Harris:1990kc}%
  \BibitemOpen
  \bibfield  {author} {\bibinfo {author} {\bibfnamefont {G.~R.}\ \bibnamefont
  {Harris}}\ and\ \bibinfo {author} {\bibfnamefont {E.~J.}\ \bibnamefont
  {Martinec}},\ }\href {\doibase 10.1016/0370-2693(90)90663-Q} {\bibfield
  {journal} {\bibinfo  {journal} {Phys. Lett. B}\ }\textbf {\bibinfo {volume}
  {245}},\ \bibinfo {pages} {384} (\bibinfo {year} {1990})}\BibitemShut
  {NoStop}%
\bibitem [{\citenamefont {Brezin}\ and\ \citenamefont
  {Neuberger}(1990)}]{Brezin:1990xr}%
  \BibitemOpen
  \bibfield  {author} {\bibinfo {author} {\bibfnamefont {E.}~\bibnamefont
  {Brezin}}\ and\ \bibinfo {author} {\bibfnamefont {H.}~\bibnamefont
  {Neuberger}},\ }\href {\doibase 10.1103/PhysRevLett.65.2098} {\bibfield
  {journal} {\bibinfo  {journal} {Phys. Rev. Lett.}\ }\textbf {\bibinfo
  {volume} {65}},\ \bibinfo {pages} {2098} (\bibinfo {year}
  {1990})}\BibitemShut {NoStop}%
\bibitem [{\citenamefont {Brezin}\ and\ \citenamefont
  {Neuberger}(1991)}]{Brezin:1990dk}%
  \BibitemOpen
  \bibfield  {author} {\bibinfo {author} {\bibfnamefont {E.}~\bibnamefont
  {Brezin}}\ and\ \bibinfo {author} {\bibfnamefont {H.}~\bibnamefont
  {Neuberger}},\ }\href {\doibase 10.1016/0550-3213(91)90154-P} {\bibfield
  {journal} {\bibinfo  {journal} {Nucl. Phys. B}\ }\textbf {\bibinfo {volume}
  {350}},\ \bibinfo {pages} {513} (\bibinfo {year} {1991})}\BibitemShut
  {NoStop}%
\bibitem [{\citenamefont {{Kupershmidt}}(1985)}]{1985CMaPh..99...51K}%
  \BibitemOpen
  \bibfield  {author} {\bibinfo {author} {\bibfnamefont {B.~A.}\ \bibnamefont
  {{Kupershmidt}}},\ }\href {\doibase 10.1007/BF01466593} {\bibfield  {journal}
  {\bibinfo  {journal} {Commun. Math. Phys.}\ }\textbf {\bibinfo {volume}
  {99}},\ \bibinfo {pages} {51} (\bibinfo {year} {1985})}\BibitemShut {NoStop}%
\bibitem [{\citenamefont {Iyer}\ \emph
  {et~al.}(2011{\natexlab{a}})\citenamefont {Iyer}, \citenamefont {Johnson},\
  and\ \citenamefont {Pennington}}]{Iyer:2010ss}%
  \BibitemOpen
  \bibfield  {author} {\bibinfo {author} {\bibfnamefont {R.}~\bibnamefont
  {Iyer}}, \bibinfo {author} {\bibfnamefont {C.~V.}\ \bibnamefont {Johnson}}, \
  and\ \bibinfo {author} {\bibfnamefont {J.~S.}\ \bibnamefont {Pennington}},\
  }\href {\doibase 10.1088/1751-8113/44/1/015403} {\bibfield  {journal}
  {\bibinfo  {journal} {J. Phys. A}\ }\textbf {\bibinfo {volume} {44}},\
  \bibinfo {pages} {015403} (\bibinfo {year} {2011}{\natexlab{a}})},\ \Eprint
  {http://arxiv.org/abs/1002.1120} {arXiv:1002.1120 [hep-th]} \BibitemShut
  {NoStop}%
\bibitem [{\citenamefont {Iyer}\ \emph
  {et~al.}(2011{\natexlab{b}})\citenamefont {Iyer}, \citenamefont {Johnson},\
  and\ \citenamefont {Pennington}}]{Iyer:2010ex}%
  \BibitemOpen
  \bibfield  {author} {\bibinfo {author} {\bibfnamefont {R.}~\bibnamefont
  {Iyer}}, \bibinfo {author} {\bibfnamefont {C.~V.}\ \bibnamefont {Johnson}}, \
  and\ \bibinfo {author} {\bibfnamefont {J.~S.}\ \bibnamefont {Pennington}},\
  }\href {\doibase 10.1088/1751-8113/44/37/375401} {\bibfield  {journal}
  {\bibinfo  {journal} {J. Phys.}\ }\textbf {\bibinfo {volume} {A44}},\
  \bibinfo {pages} {375401} (\bibinfo {year} {2011}{\natexlab{b}})},\ \Eprint
  {http://arxiv.org/abs/1011.6354} {arXiv:1011.6354 [hep-th]} \BibitemShut
  {NoStop}%
\bibitem [{\citenamefont {Maxfield}\ and\ \citenamefont
  {Turiaci}(2020)}]{Maxfield:2020ale}%
  \BibitemOpen
  \bibfield  {author} {\bibinfo {author} {\bibfnamefont {H.}~\bibnamefont
  {Maxfield}}\ and\ \bibinfo {author} {\bibfnamefont {G.~J.}\ \bibnamefont
  {Turiaci}},\ }\href@noop {} {\  (\bibinfo {year} {2020})},\ \Eprint
  {http://arxiv.org/abs/2006.11317} {arXiv:2006.11317 [hep-th]} \BibitemShut
  {NoStop}%
\bibitem [{\citenamefont {Witten}(2020)}]{Witten:2020wvy}%
  \BibitemOpen
  \bibfield  {author} {\bibinfo {author} {\bibfnamefont {E.}~\bibnamefont
  {Witten}},\ }\href@noop {} {\  (\bibinfo {year} {2020})},\ \Eprint
  {http://arxiv.org/abs/2006.13414} {arXiv:2006.13414 [hep-th]} \BibitemShut
  {NoStop}%
\bibitem [{\citenamefont {Maldacena}\ \emph {et~al.}(2019)\citenamefont
  {Maldacena}, \citenamefont {Turiaci},\ and\ \citenamefont
  {Yang}}]{Maldacena:2019cbz}%
  \BibitemOpen
  \bibfield  {author} {\bibinfo {author} {\bibfnamefont {J.}~\bibnamefont
  {Maldacena}}, \bibinfo {author} {\bibfnamefont {G.~J.}\ \bibnamefont
  {Turiaci}}, \ and\ \bibinfo {author} {\bibfnamefont {Z.}~\bibnamefont
  {Yang}},\ }\href@noop {} {\  (\bibinfo {year} {2019})},\ \Eprint
  {http://arxiv.org/abs/1904.01911} {arXiv:1904.01911 [hep-th]} \BibitemShut
  {NoStop}%
\bibitem [{\citenamefont {Cotler}\ \emph {et~al.}(2020)\citenamefont {Cotler},
  \citenamefont {Jensen},\ and\ \citenamefont {Maloney}}]{Cotler:2019nbi}%
  \BibitemOpen
  \bibfield  {author} {\bibinfo {author} {\bibfnamefont {J.}~\bibnamefont
  {Cotler}}, \bibinfo {author} {\bibfnamefont {K.}~\bibnamefont {Jensen}}, \
  and\ \bibinfo {author} {\bibfnamefont {A.}~\bibnamefont {Maloney}},\ }\href
  {\doibase 10.1007/JHEP06(2020)048} {\bibfield  {journal} {\bibinfo  {journal}
  {JHEP}\ }\textbf {\bibinfo {volume} {06}},\ \bibinfo {pages} {048} (\bibinfo
  {year} {2020})},\ \Eprint {http://arxiv.org/abs/1905.03780} {arXiv:1905.03780
  [hep-th]} \BibitemShut {NoStop}%
\bibitem [{\citenamefont {Okuyama}(2018)}]{Okuyama:2018aij}%
  \BibitemOpen
  \bibfield  {author} {\bibinfo {author} {\bibfnamefont {K.}~\bibnamefont
  {Okuyama}},\ }\href {\doibase 10.1007/JHEP10(2018)037} {\bibfield  {journal}
  {\bibinfo  {journal} {JHEP}\ }\textbf {\bibinfo {volume} {10}},\ \bibinfo
  {pages} {037} (\bibinfo {year} {2018})},\ \Eprint
  {http://arxiv.org/abs/1808.10161} {arXiv:1808.10161 [hep-th]} \BibitemShut
  {NoStop}%
\bibitem [{\citenamefont {Okuyama}\ and\ \citenamefont
  {Sakai}(2020{\natexlab{b}})}]{Okuyama:2020ncd}%
  \BibitemOpen
  \bibfield  {author} {\bibinfo {author} {\bibfnamefont {K.}~\bibnamefont
  {Okuyama}}\ and\ \bibinfo {author} {\bibfnamefont {K.}~\bibnamefont
  {Sakai}},\ }\href {\doibase 10.1007/JHEP08(2020)126} {\bibfield  {journal}
  {\bibinfo  {journal} {JHEP}\ }\textbf {\bibinfo {volume} {08}},\ \bibinfo
  {pages} {126} (\bibinfo {year} {2020}{\natexlab{b}})},\ \Eprint
  {http://arxiv.org/abs/2004.07555} {arXiv:2004.07555 [hep-th]} \BibitemShut
  {NoStop}%
\end{thebibliography}%



\end{document}